\documentclass[usenatbib]{aa}

\usepackage{dirtytalk}
\usepackage{graphicx}
\usepackage{txfonts}
\usepackage{hyperref}
\usepackage{url}
\usepackage[dvipsnames]{xcolor}
\hypersetup{
colorlinks,
linkcolor={red!80!black},
citecolor={blue!80!black},
urlcolor={blue!80!black}
}

\newenvironment{myfont}{\fontfamily{ppl}\selectfont}{\par}
\DeclareTextFontCommand{\textmyfont}{\myfont}

\newcommand{\code}[1]{\texttt{#1}}

\def\nifs{\iso{56}Ni}

\def\cm3{cm$^{-3}$}
\def\kms{\mbox{km~s$^{-1}$}}

\def\msun{$M_{\odot}$}

\def\one{\ts {\,\sc i}}
\def\two{\ts {\,\sc ii}}

\def\beq{\begin{equation}}
\def\eeq{\end{equation}}

\def\lesssim{\mathrel{\hbox{\rlap{\hbox{\lower4pt\hbox{$\sim$}}}\hbox{$<$}}}}
\def\gtrsim{\mathrel{\hbox{\rlap{\hbox{\lower4pt\hbox{$\sim$}}}\hbox{$>$}}}}

\def\one{{\,\sc i}}
\def\two{{\,\sc ii}}

\def\v1d{{\code{V1D}}}
\def\sumo{{\code{SUMO}}}
\def\kepler{{\code{KEPLER}}}

\def\cmfgen{{\code{CMFGEN}}}

\def\oidoub{[O\one]\,$\lambda\lambda$\,$6300,\,6364$}
\def\caiidoub{[Ca\two]\,$\lambda\lambda$\,$7291,\,7323$}
\def\caiitrip{Ca\two\,$\lambda\lambda$\,$8498, 8542, 8662$}

\newcommand{\iso}[2]{\ensuremath{^{#1}\rm{#2}}}

\begin{document}

   \title{The explosion of 9\,$-$\,29\,\msun\ stars as Type II supernovae : results from radiative-transfer modeling at one year after explosion}

   \titlerunning{SN nebular-phase modeling}

\author{
   Luc Dessart\inst{\ref{inst1}}
  \and
   D. John Hillier\inst{\ref{inst2}}
   \and
   Tuguldur Sukhbold\inst{\ref{inst3}}
   \and
   S.E. Woosley\inst{\ref{inst4}}
   \and
   H.-T. Janka\inst{\ref{inst5}}
  }

\institute{
Institut d'Astrophysique de Paris, CNRS-Sorbonne Universit\'e, 98 bis boulevard Arago, F-75014 Paris, France.\label{inst1}
\and
    Department of Physics and Astronomy \& Pittsburgh Particle Physics,
    Astrophysics, and Cosmology Center (PITT PACC),  \hfill \\ University of Pittsburgh,
    3941 O'Hara Street, Pittsburgh, PA 15260, USA.\label{inst2}
\and
Department of Astronomy, Ohio State University, Columbus, Ohio, 43210, USA\label{inst3}
\and
Department of Astronomy and Astrophysics, University of California, Santa Cruz, CA 95064, USA\label{inst4}
\and
Max-Planck-Institut  f\"{u}r  Astrophysik,  Postfach  1317,  85741, Garching, Germany \label{inst5}
  }

   \date{}

  \abstract{
We present a set of nonlocal thermodynamic equilibrium steady-state calculations of radiative transfer for one-year old type II supernovae (SNe) starting from state-of-the-art explosion models computed with detailed nucleosynthesis. This grid covers single-star progenitors with initial masses between 9 and 29\,\msun, all evolved with \kepler\ at solar metallicity and ignoring rotation.  The \oidoub\ line flux generally grows with progenitor mass, and H$\alpha$ exhibits an equally strong and opposite trend.  The \caiidoub\ strength increases at low \nifs\ mass, low explosion energy, or with clumping. This Ca\two\ doublet, which forms primarily in the explosively-produced Si/S zones, depends little on the progenitor mass, but may strengthen if Ca$^+$ dominates in the H-rich emitting zones or if Ca is abundant in the O-rich zones. Indeed, Si-O shell merging prior to core collapse may boost the Ca\two\ doublet at the expense of the O\one\ doublet, and may thus mimic the metal line strengths of a lower mass progenitor. We find that the \nifs\ bubble effect has a weak impact, probably because it is too weak to induce much of an ionization shift in the various emitting zones. Our simulations compare favorably to observed SNe II, including SN\,2008bk (e.g., 9\,\msun\ model), SN\,2012aw  (12\,\msun\ model), SN\,1987A (15\,\msun\ model), or SN\,2015bs (25\,\msun\ model with no Si-O shell merging). SNe II with narrow lines and a low \nifs\ mass are well matched by the weak explosion of 9\,$-$\,11\,\msun\ progenitors. The nebular-phase spectra of standard SNe II can be explained with progenitors in the mass range 12\,$-$\,15\,\msun, with one notable exception for SN\,2015bs. In the intermediate mass range, these mass estimates may increase by a few \msun\ with allowance for clumping of the O-rich material or CO molecular cooling.
}

   \maketitle


\section{Introduction}

   The late time radiative properties of core-collapse supernovae (SNe) contain information on the progenitor star at death and the mechanism by which it exploded. This topic has been extensively discussed in the past (for a review, see \citealt{jerkstrand_rev_17}). In \citet{DH20_neb}, we presented a general study of nebular phase spectra of Type II SNe. Although this study conveyed some  insights, it suffered from the adoption of a simplistic-toy setup for the ejecta. The treatment of chemical mixing was inconsistent with current theoretical expectations since the need to account for macroscopic mixing generally meant that we also enforced microscopic mixing, which is unphysical. A trick to avoid this problem was to decouple the mixing of \nifs\ and that of other species, which was again inconsistent. In \citet{DH20_shuffle}, we introduced a new technique for treating chemical mixing of SN ejecta in grid-based radiative-transfer codes. The method is inherently 1D and consists in shuffling spherical shells of distinct composition in mass space  in an ejecta in homologous expansion (say at one year after explosion). At that time, the velocity is essentially proportional to radius so that this shuffling corresponds to a mixing of material in velocity space, while preserving the original chemical mixture for each mass shell. As an introduction to this research topic and a
discussion of our modeling technique are given in \citet{DH20_neb,DH20_shuffle}, they are not repeated here.

In this work we extend our previous exploration, that was limited to the s15A model of \citet[WH07]{WH07}, to study 23 explosion models taken from WH07 and \citet[S16]{sukhbold_ccsn_16}. This grid covers the range of progenitor masses most likely to explode (i.e., 9\,$-$\,26\,\msun) and has two different treatments of the explosion. In WH07, the explosion is generated with a piston whose trajectory is designed to deliver an ejecta kinetic energy of about 10$^{51}$\,erg at infinity. In contrast, S16 simulates the evolution of the collapsing star through core bounce, shock formation, the stalling of the shock, and its revival through a consistent handling of the neutrino-energy deposition in the infalling mantle. Although the method retains some shortcomings (spherical symmetry, calibration to a core-collapse explosion ``engine" meant to reproduce SN\,1987A), the approach gains considerably in physical consistency and delivers an unbiased prediction of Type SN II explosion properties. The goal of the present study is to document the nebular-phase spectral properties of these ab~initio explosion models from S16, while the WH07 models are used to test the sensitivity of results to the numerical handling of the explosion.

Our work provides a complementary picture to the results obtained, for the most part, with the code \sumo\ \citep{jerkstrand_87a_11,jerkstrand_04et_12}, which uses a Monte Carlo treatment of the radiative transfer. The probabilistic approach is well suited to handle the stochastic propagation of photons in a complex mixture of clumps whose distinct composition reflects the macroscopically mixed progenitor metal-rich core. Such studies of nearby Type II SNe suggest a lack of high-mass RSG progenitors \citep{jerkstrand_04et_12,maguire_2p_12,jerkstrand_12aw_14,jerkstrand_ni_15,yuan_13ej_16}.
These studies were  all targeted towards the modeling of a specific Type II SN. Using a given explosion model (usually taken from WH07), the ejecta was adjusted so that the expansion rate of the metal-rich layers was compatible with the width of \oidoub, H$\alpha$, or \caiidoub. The adjustment to the initial model was tailored independently for each progenitor mass and guided by the need to match the kinematics inferred from the observations. The present work is therefore complementary to this approach since we take the actual ejecta models of S16 and investigate the range of spectral properties they exhibit at nebular times.

\begin{figure*}
\centering
\includegraphics[width=0.495\hsize]{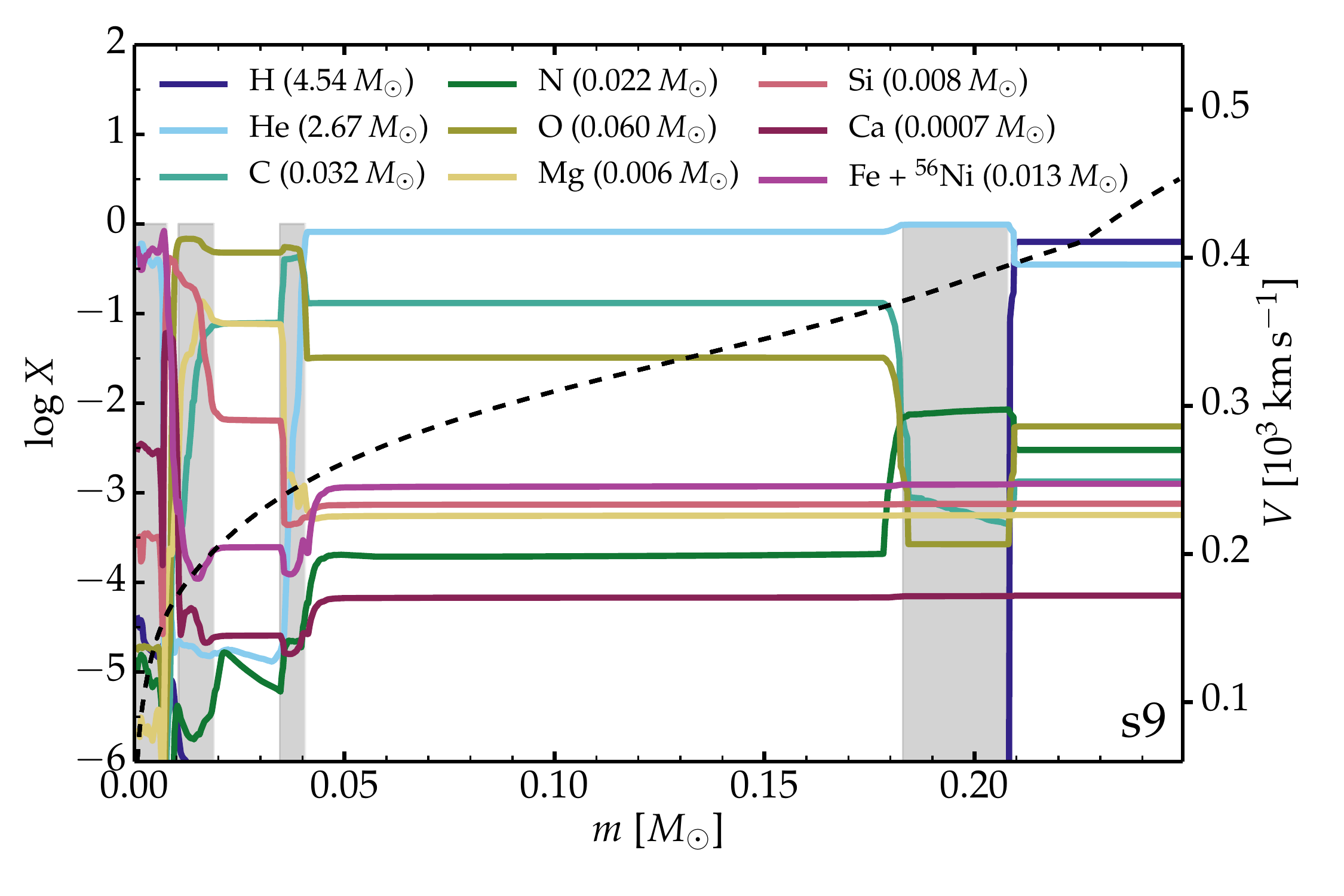}
\includegraphics[width=0.495\hsize]{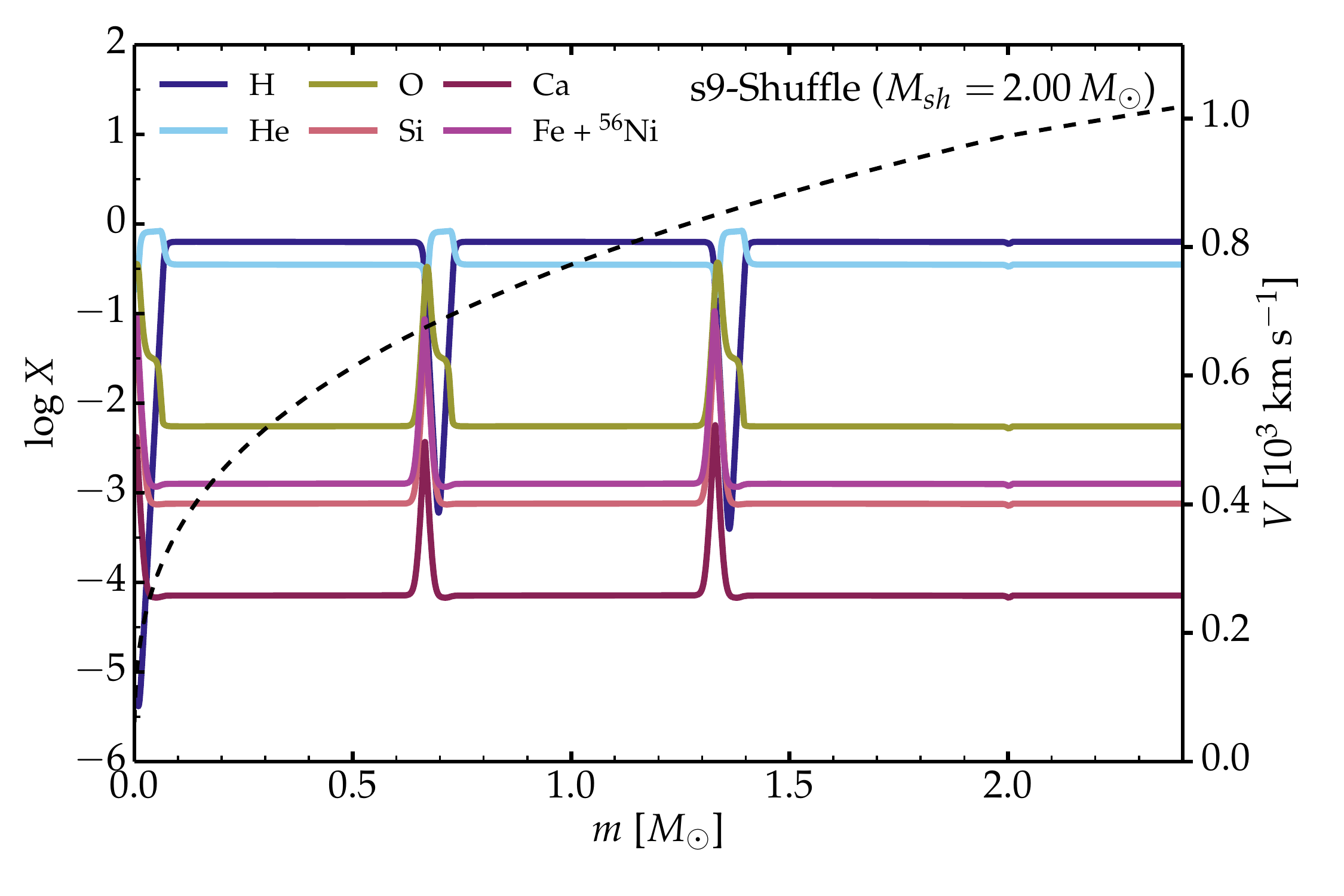}
\includegraphics[width=0.495\hsize]{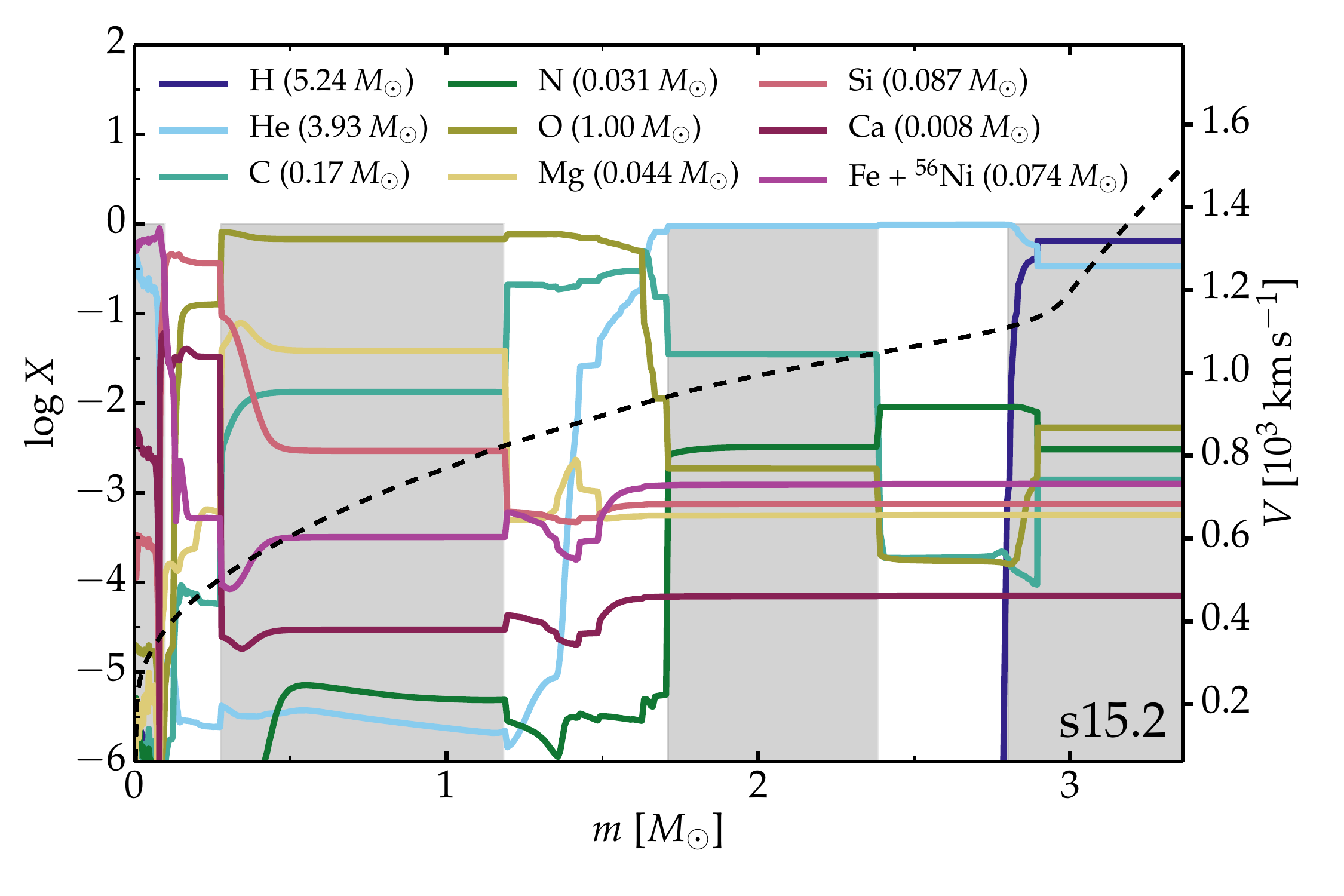}
\includegraphics[width=0.495\hsize]{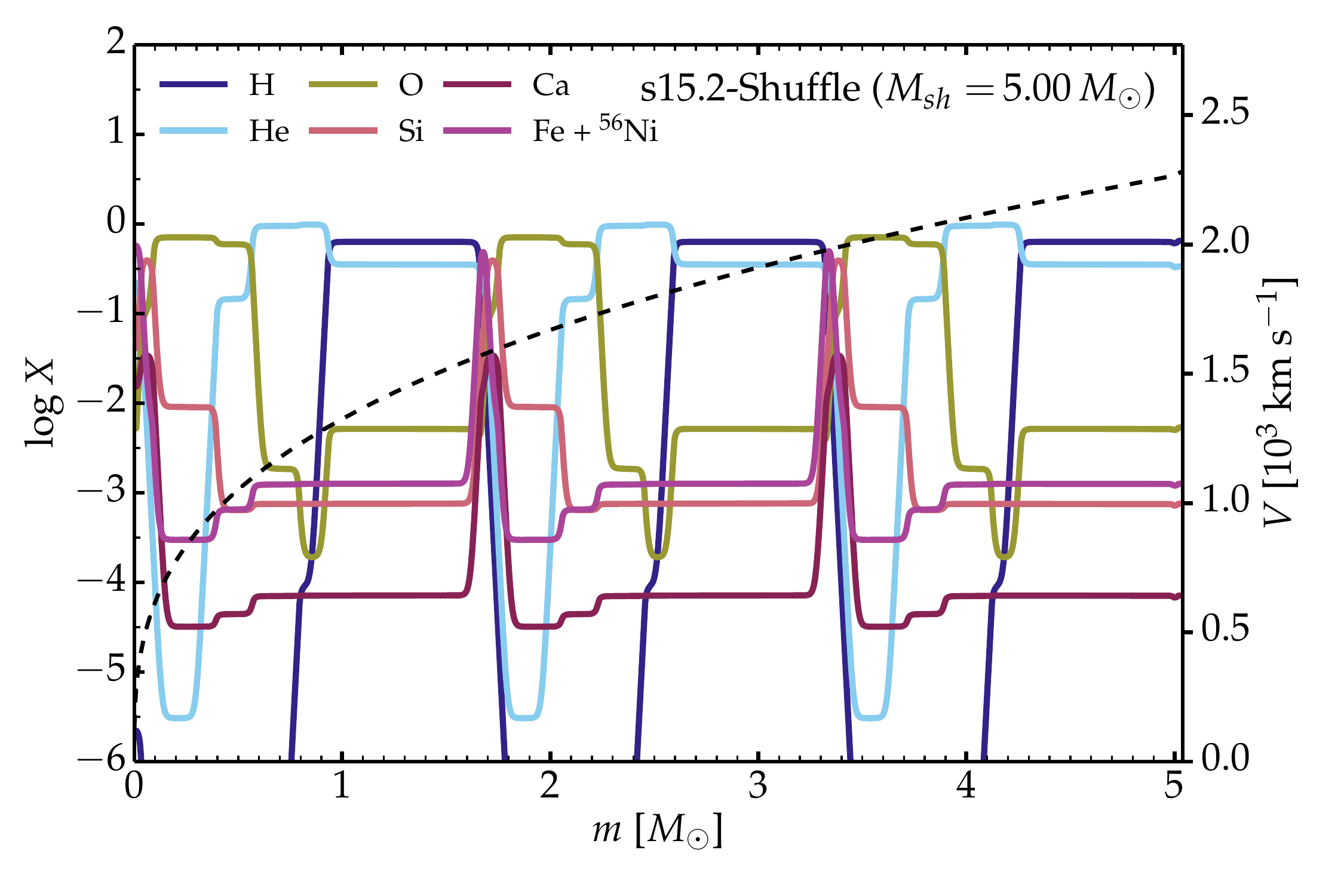}
\vspace{-0.2cm}
\caption{Left: Chemical composition versus Lagrangian mass in the s9 and s15p2 ejecta models of S16 at 350\,d (the original mass fraction is shown for \nifs) -- ejecta properties are summarized in Table~\ref{tab_prog}. The name and total yield of each selected species is labeled. An alternating white and grey background is used to distinguish between consecutive shells of distinct composition. Right: corresponding ejecta in which shells have been shuffled in mass space within the Lagrangian mass cut $M_{\rm sh}=\,$2.0 and 5.0\,\msun. Other ejecta models exhibit similar profiles, only modulated by the original composition (in particular the mass of each dominant shell) and the choice of  $M_{\rm sh}$ (details on the shuffling procedure are given in the appendix of \citealt{DH20_shuffle}).
\label{fig_prog_comp}
}
\end{figure*}

\begin{figure}
\centering
\includegraphics[width=\hsize]{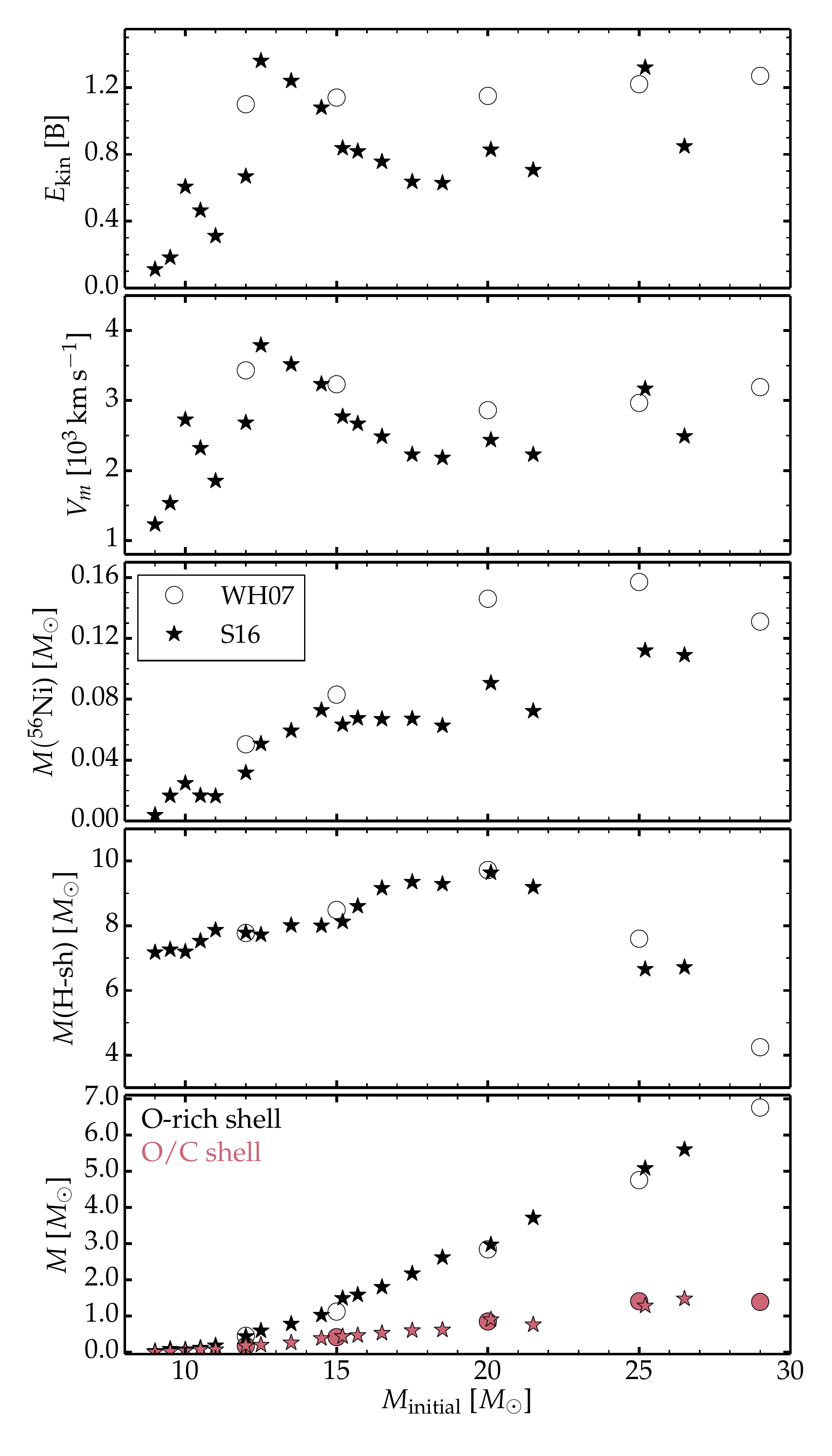}
\vspace{-0.6cm}
\caption{From top to bottom, this figures shows the ejecta kinetic energy, the mean expansion rate, the \nifs\ mass, the mass of the H-rich shell (i.e., corresponding to the progenitor H-rich envelope), and finally the masses of the O-rich shell (which includes the O/Si, the O/Ne/Mg, and the O/C shells) and of the O/C shell in our selection of WH07 (circles) and S16 (stars) models. The $E_{\rm kin}$ is prescribed in the WH07 models, but computed ab-initio in the S16 models. This explains the more complicated trends in the latter (e.g., the systematically lower  $E_{\rm kin}$ and $M$(\nifs) for masses above about 16\,\msun, with some exceptions with the  $E_{\rm kin}$ of s25p2), which includes regions in original mass that lead to no explosion at all (hence these models are not included in our sample; see S16 for discussion).
\label{fig_prop}
}
\end{figure}

In the next section, we present our numerical setup as well as the explosion models used as initial conditions for the radiative-transfer modeling with \cmfgen\ \citep{HD12,DH20_shuffle}. Section~\ref{sect_landscape} presents the qualitative and quantitative results for the simulations, with a particular focus on how the power absorbed in the various ejecta shells is radiated in nebular emission lines. Section~\ref{sect_ca2} discusses the origin of Ca\two\ emission in our Type II SN simulations and compares our results with those from previous work. The treatment of the \nifs-bubble effect, presented in Section~\ref{sect_bubble}, reveals only a minor impact on the SN radiative properties. Similarly, tests for material clumping, presented in Section~\ref{sect_fvol}, reveal a weak sensitivity over the parameter space explored. While our simulations adopt by default a complete macroscopic mixing of the metal-rich core, which is a conservative assumption, we find that the lack of such mixing, as would occur in a quasi-prompt explosion, would impact significantly the nebular spectral properties (Section~\ref{sect_mix}). Section~\ref{sect_prof} discusses the morphology of the strongest optical emission lines in nebular phase spectra of Type II SNe, in particular in connection to line overlap and optical depth effects. A comparison to a selection of observations is presented in Section~\ref{sect_obs}. We give our concluding remarks in Section~\ref{sect_conc}.

\section{Numerical setup}
\label{sect_setup}

Using the same numerical approach as in \citet{DH20_shuffle}, we study the radiative properties of a diverse set of explosion models derived from 9\,$-$\,29\,\msun\ stars on the main sequence. We consider two sets of explosion models that were taken from WH07 and S16. Both sets were evolved with \kepler\ at solar metallicity and without rotation until core collapse.  The physics in the \kepler\ code has been described by \citet{weaver_kepler_78} and \citet{whw02}. In the ``s-series'' of models used here, convection was included in the Ledoux formalism with a substantial amount of semiconvection  \citep{weaver_woosley_93}. The rate for \iso{12}C($\alpha, \gamma$)\iso{16}O corresponded to an $S$ factor at 300\,keV of 175 keV\,b, which is moderately high compared with some recent estimates \citep{deb17} but within experimental error bars. Nucleosynthesis was tracked using an adaptive network with all isotopes appropriate to a given stage \citep{rau02} - up to 2000 in the explosion itself.

Upon reaching core collapse, the WH07 models were exploded with a prescribed piston trajectory so that the asymptotic kinetic energy would be about 10$^{51}$\,erg. In contrast,  the S16 models were exploded with the P-HOTB code, which  uses a \say{neutrino engine} whose strength was calibrated to reproduce the elementary properties of the well studied SN\,1987A and SN\,1054 (the Crab explosion; \citealt{ugliano_ccsn_12}; \citealt{ertl_ccsn_16}; S16). Between these cases of a higher-mass and a lower-mass progenitor, respectively, the values of the engine parameters were interpolated as a function of stellar core compactness (see S16 for details). Effectively, the engine parameters determine the energy release of the cooling and accreting proto-neutron star (PNS), which, in turn, determines the  neutrino luminosity of the high-density PNS core. Neutrino transport in the accretion mantle of the PNS is treated by a gray approximation. The neutrino-driven supernova models employed in the present study are spherically symmetric. However, the neutrino-driven mechanism is known to be a multi-dimensional phenomenon. Although some multi-D effects may be accounted for in an effective way by using the engine calibration, other effects cannot be described by 1D models. For example, long-lasting simultaneous accretion downflows and re-ejected outflows around the PNS in 3D simulations (see \citealt{mueller_ccsn_3d_17}, \citealt{bollig_ccsn_3d_20}) are replaced by an extended post-bounce period of PNS accretion and, after a considerably delayed onset of the explosion, a subsequent neutrino-driven wind, whose strength is overestimated in order to provide a neutrino-heated mass with an energy that is needed to power the supernova. Clearly, explosion asymmetries and large-scale mixing processes, which are a natural consequence of nonradial hydrodynamic instabilities during explosions in 3D, cannot be described in spherical symmetry.

Neutrino-driven explosions modeled in 1D display a highly non-monotonic behavior of the explosion and remnant properties as a function of progenitor mass or compactness (see \citealt{ugliano_ccsn_12}, \citealt{ertl_ccsn_16}, S16). This behavior and most of its consequences were confirmed by independent numerical modeling approaches in 1D (e.g.  \citealt{pejcha_ccsn_15}; \citealt{ebinger_ccsn_19}) as well as a semi-analytic description that accounts for 3D effects in a parametrized way \citep{muller_prog_expl_16}. Although some authors express concerns that 3D effects in 1D models are lacking (e.g., \citealt{couch_stir_20}), the results of these models are compatible with a number of observational constraints, e.g., galactic chemical abundances, a rough correlation between explosion energy and \nifs\ mass \citep{muller_56ni_17}, the spread of observed neutron star masses (within uncertainties), and the observed rarity of Type II SN progenitors above 16\,$-$\,20\,\msun\ (e.g., \citealt{smartt_09}; \citealt{sukhbold_adams_20}, and references therein). Since the exact outcome of the 1D modeling depends on the strength of the neutrino engine, which was calibrated in the S16 explosion simulations for different suggestions of the SN\,1987A progenitor, further validation will have to come from observations. Ultimately 3D models will need to be employed. However,  3D explosions  are computationally expensive, are unsettled, for example,  due to uncertainties in nuclear and neutrino physics, and in the properties of the progenitor. Additionally,  there is a much larger parameter space making it more difficult to link any one model with an observed SN. Thus 1D models, guided by results from 3D models, still have an important role to play in advancing our understanding of core collapse SNe.

For the first set of models,  we selected the WH07 progenitors with an initial mass on the zero-age main sequence (ZAMS) of 12, 15, 20, 25, and 29\,\msun. In the nomenclature of WH07, these models are s12A, s15A, s20A, s25A, and s29A. The ``A'' suffix refers to models in which the ejecta kinetic energy at infinity is in the vicinity of $1.2 \times 10^{51}$\,erg and where the mass cut was placed at the location in the progenitor where the dimensionless entropy $S/N_{\rm A} k_{\rm B}=$\,4.\footnote{WH07 also computed a set of models with twice that energy at infinity and another set with the mass cut at the edge of the iron core. The models from the ``A'' series are considered the most realistic. The jump in entropy at $S/N_{\rm A} k_{\rm B}=$\,4 usually corresponds to the abrupt density decline at the base of the oxygen shell and modern calculations frequently find a mass cut there (see discussion, for example, in \citealt{ertl_ccsn_16}).} Simulations for other masses were not made because the explosion models of WH07 retain some level of arbitrariness -- the models are artificially exploded. For the second set of models, taken from S16, we selected 18 successful explosions (computed with the W18 engine) arising from stars with a ZAMS mass between 9 and 26.5\,\msun, only excluding a few with which we encountered convergence difficulties in the radiative-transfer calculation. In the nomenclature of S16, these models are s9, s9p5, s10 etc (where a \say{p} replaces the dot for non-integer masses). As is apparent from the next section, our sample of S16 models already exhibits some degeneracy in SN properties (i.e., nearly identical explosion and SN radiation properties), so we believe the present grid of models covers the range of nebular properties currently available from state-of-the-art explosion models. For the WH07 models s12A, s15A, s20A, and s25A, there is a model in the S16 sample with the same or a similar ZAMS mass (s12, s15p2, s20p1, s25p2; the models, however, differ in explosion energy, ejecta mass, or yields, including \nifs\ mass), which allows us to evaluate the impact of using different prescriptions for the explosion and pre-SN evolution (both sets were calculated with \kepler\ about ten years apart).

\begin{table*}
    \caption{Ejecta model properties from WH07 (upper part) and S16 (lower part;
    the selected models were exploded with the W18 engine), including the ejecta mass, ejecta kinetic energy, the cumulative
    yields of H, He, O, Mg, Si, Ca, and \nifs\ prior to decay. We specify whether the Si-rich and the O-rich shells merged before core collapse (this
    did not occur in the WH07 models, and only in three of the S16 models). The
    last two columns give the choice of mass cut $M_{\rm sh}$ for the shell shuffling, as well as the associated velocity that bounds 99\% of the
    total \nifs\ mass in the corresponding model.
\label{tab_prog}
}
\begin{center}
\begin{tabular}{l|cccccccccccc}
\hline
Model  &  $M_{\rm ej}$  &     $E_{\rm kin}$  &     $V_m$   &   \iso{1}H   & \iso{4}He &  \iso{16}O &  \iso{24}Mg &  \iso{28}Si &  \iso{40}Ca &  \iso{56}Ni$_{t=0}$ & $M_{\rm sh}$ & $V_{{\rm max},\nifs}$ \\
       &     [\msun]    &        [erg]    & [\kms]    &    [\msun] & [\msun] & [\msun] & [\msun] & [\msun] & [\msun] & [\msun] & [\msun] & [\kms] \\
\hline
\multicolumn{10}{c}{Models from \citet{WH07}}   \\
\hline
 s12A    &     9.39  &  1.10(51)   &  3432  &  5.27   &    3.40 &  3.45(-1) &  1.87(-2) &  4.53(-2) &  3.30(-3) &  5.04(-2)  &   3.8 & 1667 \\
 s15A    &    10.96  &  1.14(51)   &  3233  &  5.52   &    4.00 &  8.24(-1) &  3.80(-2) &  6.67(-2) &  4.41(-3) &  8.30(-2)   &  5.4 & 1808  \\
 s20A    &    14.11  &  1.15(51)   &  2862  &  5.86   &    4.96 &  1.94 &  7.54(-2) &  2.07(-1) &  9.20(-3) &  1.46(-1)   &      7.0 & 1588 \\
 s25A    &    13.94  &  1.22(51)   &  2966  &  4.03   &    4.74 &  3.33 &  1.74(-1) &  2.74(-1) &  1.29(-2) &  1.57(-1)   &      9.0 & 1981 \\
 s29A    &    12.53  &  1.27(51)   &  3192  &  1.97   &    3.52 &  4.79 &  1.61(-1) &  1.52(-1) &  9.68(-3) &  1.31(-1)   &     11.0 &  3343  \\
 \hline
\multicolumn{10}{c}{Models from \citet{sukhbold_ccsn_16}}   \\
\hline
   s9    &     7.38  &  1.11(50)   & 1229    &   4.54   &    2.67 &  6.04(-2) &  6.02(-3) &  7.84(-3) &  6.67(-4) &  3.88(-3)   & 2.0  &   851 \\
  s9p5   &     7.78  &  1.82(50)   & 1533    &    4.75    &    2.83   & 9.69(-2)  &  8.52(-3)  &  1.41(-2)  &  1.24(-3)  &   1.66(-2)   & 2.2 &   948   \\
  s10    &     8.19  &  6.06(50)   & 2727    &   4.94   &    2.97 &  9.65(-2) &  5.39(-3) &  1.33(-2) &  1.37(-3) &  2.48(-2)   & 3.0  &  1832 \\
 s10p5   &     8.67  &  4.64(50)   & 2320    &    5.13    &    3.19   & 1.41(-1)  &  6.92(-3)  &  1.49(-2)  &  1.24(-3)  &   1.67(-2)   & 3.0 &  1493   \\
  s11    &     9.12  &  3.11(50)   & 1851    &   5.32   &    3.37 &  1.97(-1) &  9.61(-3) &  2.06(-2) &  1.41(-3) &  1.63(-2)   & 3.0  &  1151 \\
  s12    &     9.32  &  6.68(50)   & 2684    &   5.27   &    3.41 &  3.29(-1) &  1.72(-2) &  2.89(-2) &  2.11(-3) &  3.17(-2)   & 4.0  &  1906 \\
 s12p5   &     9.51  &  1.36(51)   & 3791    &    5.22    &    3.45   & 4.56(-1)  &  3.51(-2)  &  4.58(-2)  &  3.03(-3)  &   5.07(-2)   & 4.0 &  2643   \\
 s13p5   &    10.07  &  1.24(51)   & 3518    &    5.32    &    3.69   & 5.90(-1)  &  4.04(-2)  &  5.18(-2)  &  3.58(-3)  &   5.93(-2)   & 4.0 &  2328   \\
 s14p5   &    10.37  &  1.08(51)   & 3235    &    5.23    &    3.81   & 7.63(-1)  &  3.87(-2)  &  5.69(-2)  &  3.83(-3)  &   7.28(-2)   & 4.5 &  2259   \\
s15p2    &    10.95  &  8.37(50)   & 2771    &   5.24   &    3.93 &  9.97(-1) &  4.41(-2) &  8.65(-2) &  7.77(-3) &  6.33(-2)   & 5.0  &  1997 \\
 s15p7   &    11.54  &  8.18(50)   & 2670    &    5.53    &    4.13   & 1.06      &  4.68(-2)  &  8.92(-2)  &  7.28(-3)  &   6.75(-2)   & 5.5 &  1988   \\
 s16p5   &    12.30  &  7.56(50)   & 2485    &    5.84    &    4.37   & 1.19      &  5.16(-2)  &  9.93(-2)  &  7.14(-3)  &   6.69(-2)   & 5.5 &  1770   \\
 s17p5   &    12.87  &  6.36(50)   & 2228    &    5.87    &    4.54   & 1.51      &  8.69(-2)  &  7.56(-2)  &  4.79(-3)  &   6.72(-2)   & 6.0 &  1629   \\
 s18p5   &    13.26  &  6.28(50)   & 2181    &    5.73    &    4.65   & 1.89      &  1.42(-1)  &  6.41(-2)  &  3.09(-3)  &   6.26(-2)   & 6.5 &  1638   \\
s20p1$^a$&    14.03  &  8.28(50)   & 2436    &   5.81   &    4.93 &  1.98 &  9.27(-2) &  2.33(-1) &  9.91(-3) &  9.06(-2)       & 7.0  &  1875 \\
 s21p5   &    14.30  &  7.06(50)   & 2227    &    5.38    &    4.94   & 2.49      &  7.98(-2)  &  1.18(-1)  &  1.09(-2)  &   7.22(-2)   & 8.5 &  1930   \\
s25p2$^a$&    13.22  &  1.32(51)   & 3169    &   3.42   &    4.43 &  3.53 &  6.83(-2) &  4.19(-1) &  1.27(-2) &  1.12(-1)      & 9.0  &  3168 \\
s26p5$^a$&    13.75  &  8.48(50)   & 2490    &    3.45    &    4.46   & 3.77      &  9.38(-2)  &  2.59(-1)  &  1.20(-2)  &   1.09(-1) & 9.5 &  2527   \\
\hline
\end{tabular}
\end{center}
    {\bf Notes:} $^a$: This indicates that the progenitor underwent Si-O shell merging prior to core collapse. The contamination of Si and Ca
    in the O-rich shell is strong in models s20p1 and s25p2 but moderate in model s26p5 (i.e., only the inner parts of the O-rich shell are contaminated by Si-rich material in model s26p5).
\end{table*}

A summary of model properties is given in Table~\ref{tab_prog},  with some properties also shown in Fig.~\ref{fig_prop}. Both WH07 and S16 models show the same trend of increasing O-rich shell mass (defined here as all ejecta depths with an O mass fraction greater than 10\%) with progenitor mass. Similar trends would be seen for the He-core or the CO-core mass. The S16 progenitors less massive than 12\,\msun\  on the ZAMS have an O-rich shell mass that drops down to about 0.1\,\msun\ for the lowest mass massive stars undergoing Fe-core collapse (the minimum mass is not firmly established; \citealt{poelarends_sAGB_08}). The explosions from these lower-mass stars eject mostly H and He, contribute little to the metal enrichment of the Universe, and thus may not exhibit strong metal lines in their associated SN nebular spectra.

The WH07 models have about the same ejecta kinetic energy of $1.2 \times 10^{51}$\,erg irrespective of ZAMS mass, with a monotonic increase of the \nifs\ yield with ZAMS mass that reflects the monotonic increase of the progenitor He-core mass or the compactness (this simply reflects the larger amount of dense material above the Fe core at collapse; see S16 for discussion). This uniform $E_{\rm kin}$ is not a prediction of the model but is imposed through the prescribed piston trajectory. In  practice, this trajectory could have been tuned further so that the $E_{\rm kin}$ were exactly the same for all models in the ``A'' series of WH07, irrespective of initial mass. This freedom is not ideal since the explosion energy is thought to depend sensitively on the structure in the inner $\sim$\,2\,\msun\ of the progenitor star at core collapse. This shortcoming is cured in the S16 models by means of a more consistent calculation of the collapse, bounce, and post-bounce evolution until explosion. In these calculations, the progenitor structure is an essential ingredient controlling the asymptotic ejecta kinetic energy (the S16 models still have shortcomings, such as the assumption of spherical symmetry or the calibration of the explosion engine; see earlier discussion in this section).

In the S16 models, a trend towards higher $E_{\rm kin}$ and \nifs\ yield with increasing ZAMS mass can be discerned. There is a modest scatter for the \nifs\ yield, but there are many outliers for the distribution in $E_{\rm kin}$. The non-monotonicity of $E_{\rm kin}$ with progenitor mass can be understood as follows. On the one hand the neutrino energy input to the explosion grows for more massive neutron stars, which have a tendency to be formed in the bigger He-cores of the most massive of the considered progenitors. On the other hand, the binding energy of the ejected stellar core material grows for stars with bigger He-cores. Because of the saturation of the explosion energy around $2 \times 10^{51}$\,erg (S16), the asymptotic ejecta kinetic energy attains a maximum at intermediate stellar masses.

While the models (in particular those drawn from the S16 set) have a higher physical consistency than previous ones, one must bear in mind that other explosion models could have been produced by varying the properties of the progenitor on the ZAMS (metallicity, rotation rate), or by adopting different numerical values for parameters that control the progenitor evolution (mixing length, mass loss prescription, overshoot etc). These variations will alter the progenitor properties,  such as the size of the CO and He cores and the extent of the H-rich envelope, at core collapse. Similarly, different explosion engines (see S16 for details) lead to different ejecta properties, most importantly $E_{\rm kin}$ and the \nifs\ mass.

Taking the selected WH07 and S16 models at 350\,d after explosion, we used the method discussed in \citet{DH20_shuffle} to introduce macroscopic mixing without any microscopic mixing. The technique amounts to shuffling shells of equal mass but located in different parts of the metal-rich unmixed ejecta at 350\,d. By design, this shuffling preserves the kinetic energy, the total mass, and all individual element masses. Only shells within a Lagrangian mass $M_{\rm sh}$ take part in this shuffling -- beyond $M_{\rm sh}$ the original ejecta composition from WH07 or S16 is left untouched. Because the ejecta is in homologous expansion at this late time, the shell shuffling in mass space is equivalent to a shell shuffling in velocity (or radial) space. Provided there is enough shuffling of the metal-rich shells relative to the \nifs-rich material, the exact shell arrangement in the inner metal-rich ejecta is unimportant (as demonstrated by \citealt{DH20_shuffle}). With this provision in mind, what counts is how far out in the ejecta we place the \nifs. To apply a consistent setup for the whole grid of models, we set $M_{\rm sh} \approx M_{\rm He,c} + 1.5\,$\,\msun.   This implies that the metal rich core is always \say{fully} mixed and that about 1.5\,\msun\ of material from the overlying H-rich shell is mixed into the metal-rich core. Because the He-core mass increases with progenitor mass, this criterion implies a greater mixing in velocity (or mass) space  (this statement would strictly hold if the final mass at core collapse was independent of ZAMS mass). It also means that in higher mass progenitors, the H-rich material mixed inwards represents a smaller fraction of the surrounding metal-rich material. The weakness of the method, which also applies to any similar works, is that we do not know the precise level of mixing in any given SN, nor how it varies with increasing ZAMS mass. The 1D approach also ignores the possibility of large-scale asymmetry, which probably exists at some level in all core-collapse SNe. We show the shuffled-shell composition structure of the s9 and s15p2 models of S16 in Fig.~\ref{fig_prog_comp}.

The \cmfgen\ simulations assume steady state, and typically use 350 points for the radial grid, uniformly spaced in velocity below  $M_{\rm sh}$ and uniformly spaced on an optical depth scale above that. We treat the \nifs\ two-step decay chain and ignore the contribution from other unstable isotopes.  Non-local energy deposition is computed by solving the grey radiative transfer equation (the $\gamma$-ray grey absorption-only opacity is set to 0.06 $Y_{\rm e}$ cm$^{2}$\,g$^{-1}$, where $Y_{\rm e}$ is the electron fraction).  Non-thermal processes are treated as per normal (for details, see \citealt{li_etal_12_nonte,d12_snibc}). All simulations presented  use a turbulent velocity of 10\,\kms. However, with the new shuffled-shell technique (which does not introduce any artificial microscopic mixing), we find that the choice of turbulent velocity is not as critical for the present simulations of Type II SNe at nebular times \citep{DH20_neb}. Indeed, tests show that the spectra obtained for 10\,\kms\ are essentially the same as those obtained for 50\,\kms\ for the present set of models.

\begin{figure*}
\centering
\includegraphics[width=0.95\hsize]{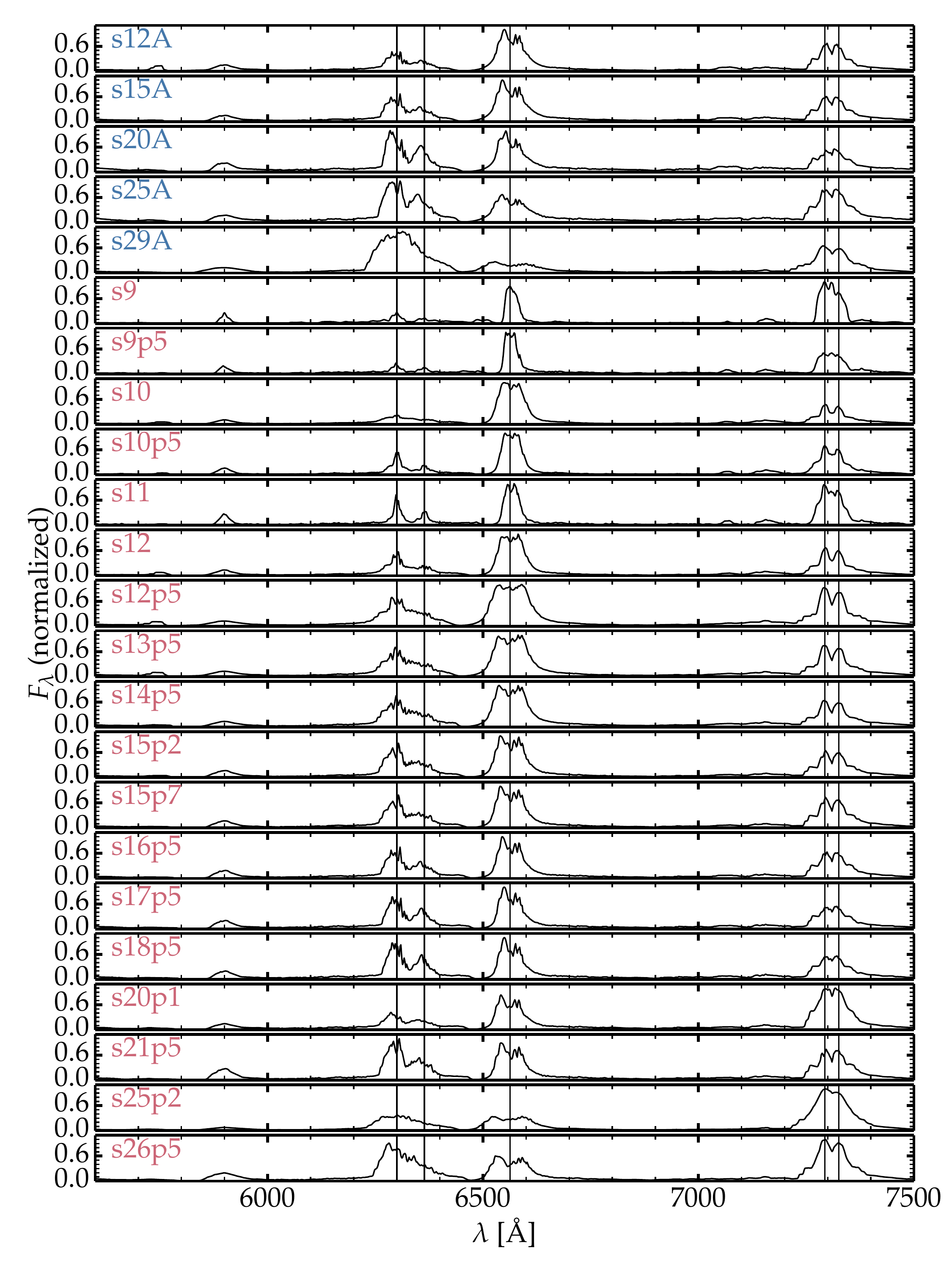}
\vspace{-0.3cm}
\caption{Montage of spectra for the set of WH07 models (blue label) and S16 models (red label) at 350\,d after explosion. The wavelength range is limited to the 5600\,$-$\,7500\,\AA\ region to reveal the variations in \oidoub, H$\alpha$, and \caiidoub. Note the non-monotonic evolution of these line fluxes with increasing initial mass in the S16 models, in contrast to the results obtained for the WH07 models.
\label{fig_montage_opt}
}
\end{figure*}

\section{Landscape of nebular phase properties from the S16 grid of explosion models}
\label{sect_landscape}

\subsection{Qualitative description}
\label{sect_qual}

Figure~\ref{fig_montage_opt} shows the optical spectra obtained with \cmfgen\ at an epoch of 350\,d after explosion. The spectra are normalized so that the maximum flux in the illustrated wavelength range is set to one. The models from the WH07 set are a good place to start the discussion since these models are characterized by a very similar ejecta kinetic energy of 1.1\,$-$\,1.3\,$\times$\,10$^{51}$\,erg, and a moderate range in \nifs\ mass between 0.05 and 0.13\,\msun. Hence, one expects that the key spectral differences exhibited in the WH07 set  primarily reflect the differences in progenitor composition, especially the systematic increase in the O-shell mass with the mass of the progenitor (factor of $\sim$\,10 increase between s12A and s29A).

The WH07 set exhibit a monotonic increase in the strength and width of the \oidoub\ doublet line from s12A to s29A; in the same order, the H$\alpha$ line weakens and broadens, though only most obviously for the highest mass model s29A; finally, the \caiidoub\ line stays about the same and is never the strongest optical line. Qualitatively, these trends follow expectations and the results from previous calculations (see, for example, \citealt{jerkstrand_04et_12}). The \oidoub\ line strength in type II SN spectra increases with progenitor mass because the O-rich shell captures an increasing fraction of the total decay power absorbed by the ejecta and because \oidoub\ is the main coolant for the O-rich material.  Given the narrow range in ejecta kinetic energy and mass (see Table~\ref{tab_prog}), the variation in line widths reflects the chemical stratification of the progenitor star \citep{dlw10b}:  with increasing progenitor mass, the CO core occupies a growing fraction of the total ejecta mass, which tends to place the O-rich material at larger velocity -- the same feature explains the broad H$\alpha$ in the s29A model. The lack of variation in the \caiidoub\ line strength confirms that it is a poor indicator of progenitor mass (this result is discussed in detail below).

\begin{figure*}
\centering
\includegraphics[width=0.9\hsize]{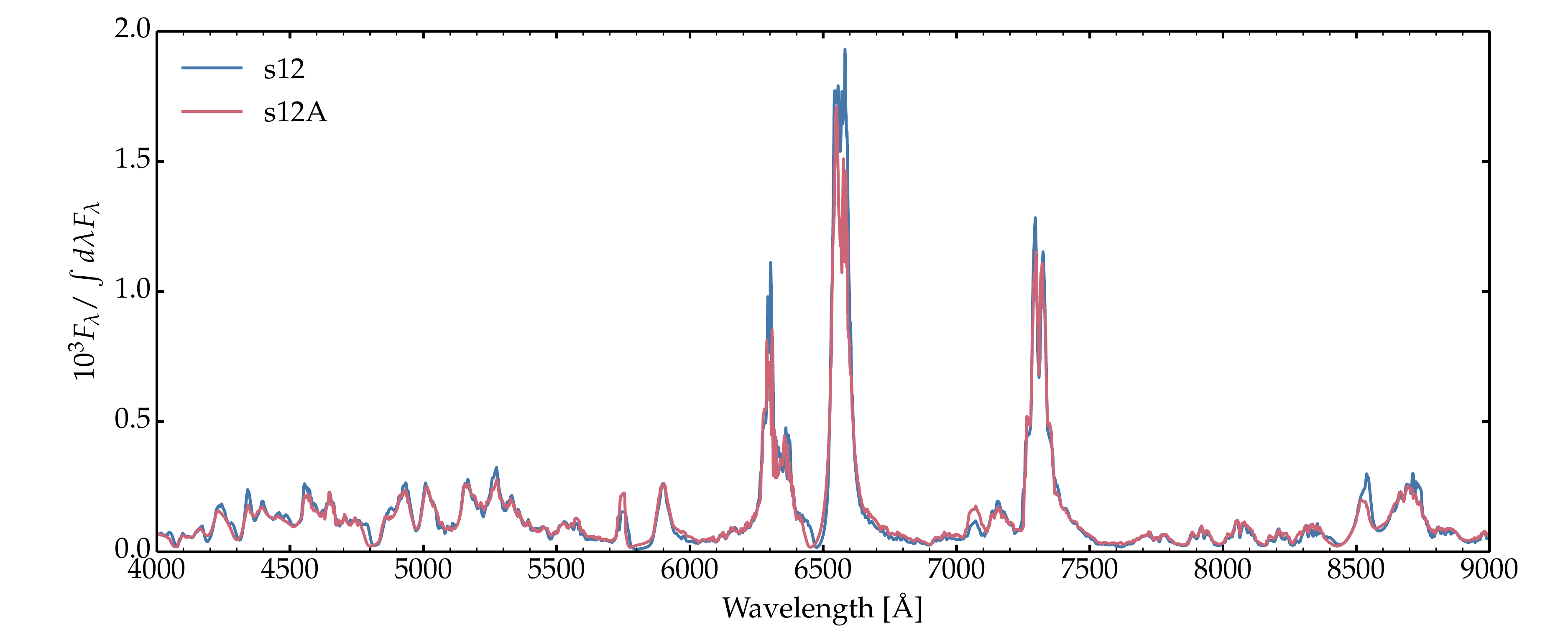}
\includegraphics[width=0.9\hsize]{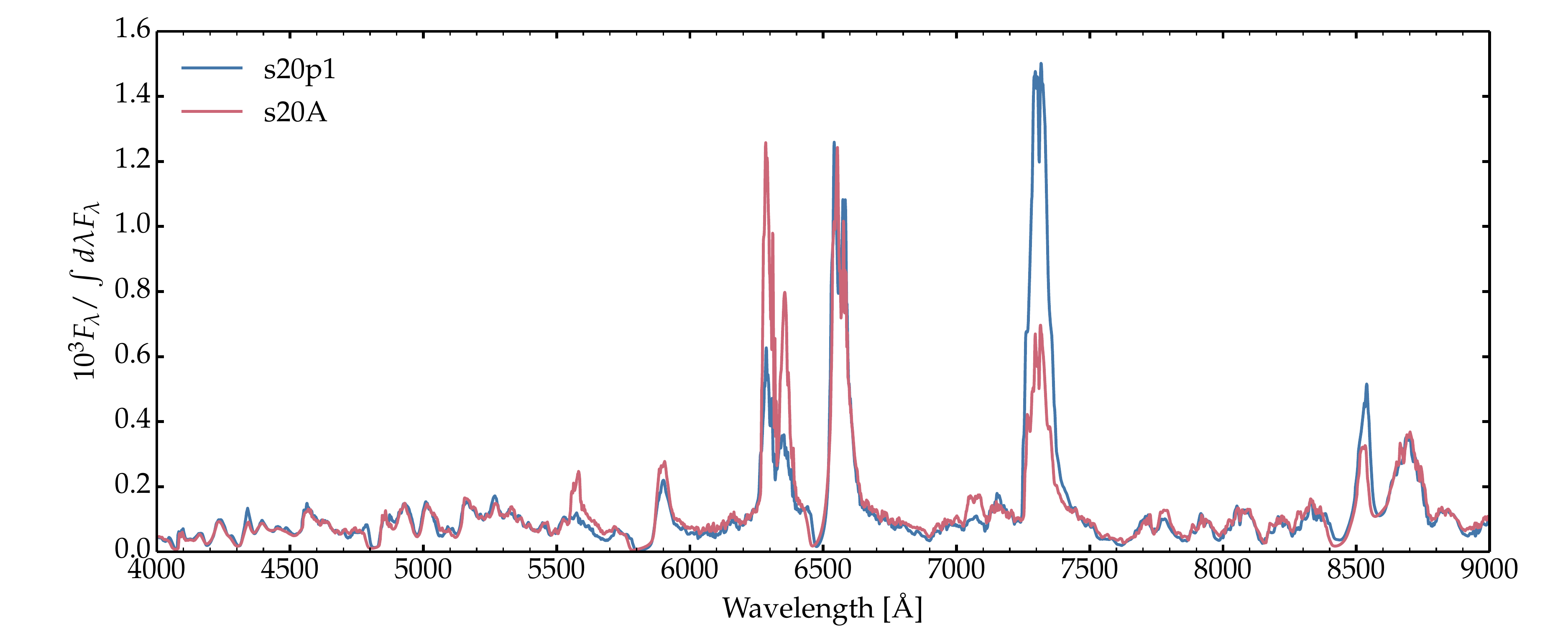}
\includegraphics[width=0.9\hsize]{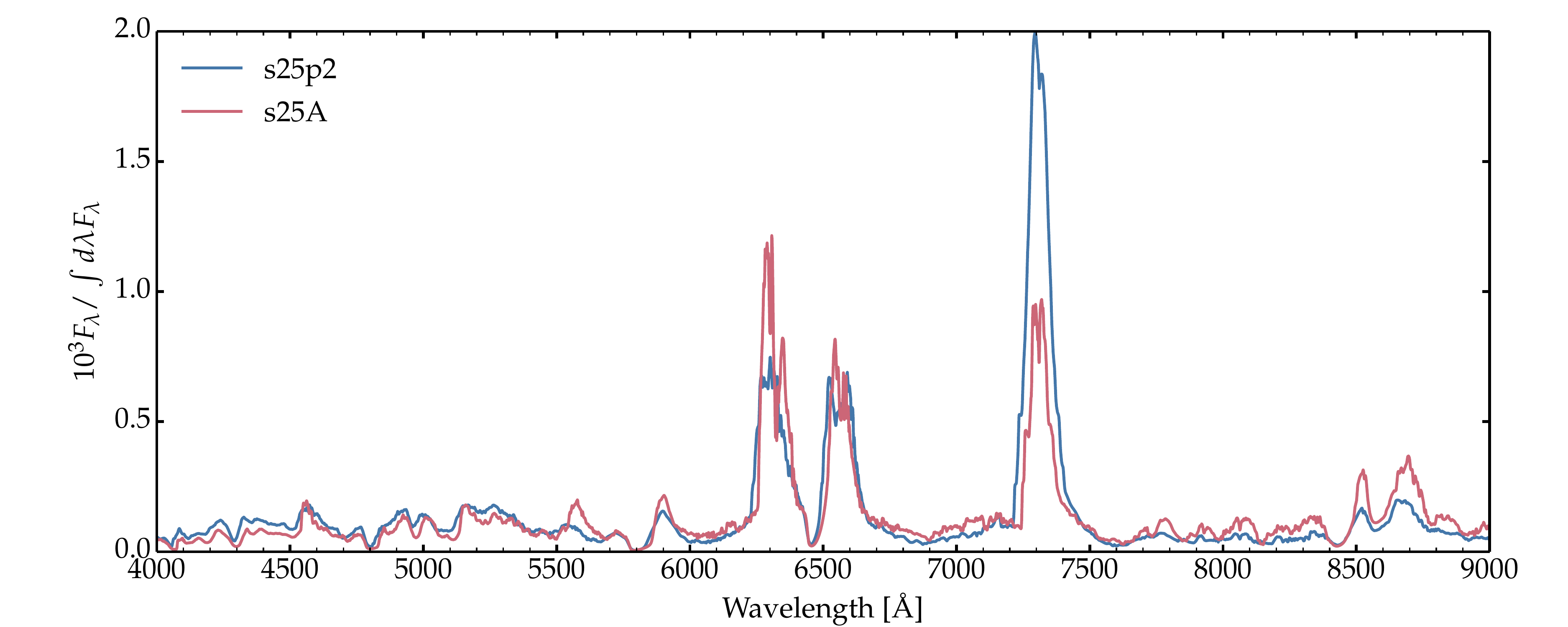}
\vspace{-0.3cm}
\caption{Comparison between the S16 (blue) and WH07 (red) models of comparable main sequence mass, with s12 and s12A (top), s20p1 and s20A (middle), and s25p2 and s25A (bottom). In each panel, the flux is divided by the integral of the flux falling between 1000 and 25000\,\AA\ (and then scaled by a factor of $10^3$), which is similar to forcing all models to have the same total decay power absorbed (in reality, the optical flux differs between models of the same mass because of the differences in \nifs\ or $\gamma$-ray escape; see Table~\ref{tab_prog}). This normalization procedure also allows one to estimate by eye the cooling power of strong lines relative to the total decay power absorbed.
}
\label{fig_comp_same_mass}
\end{figure*}

\begin{figure*}
\centering
\includegraphics[width=0.7\hsize]{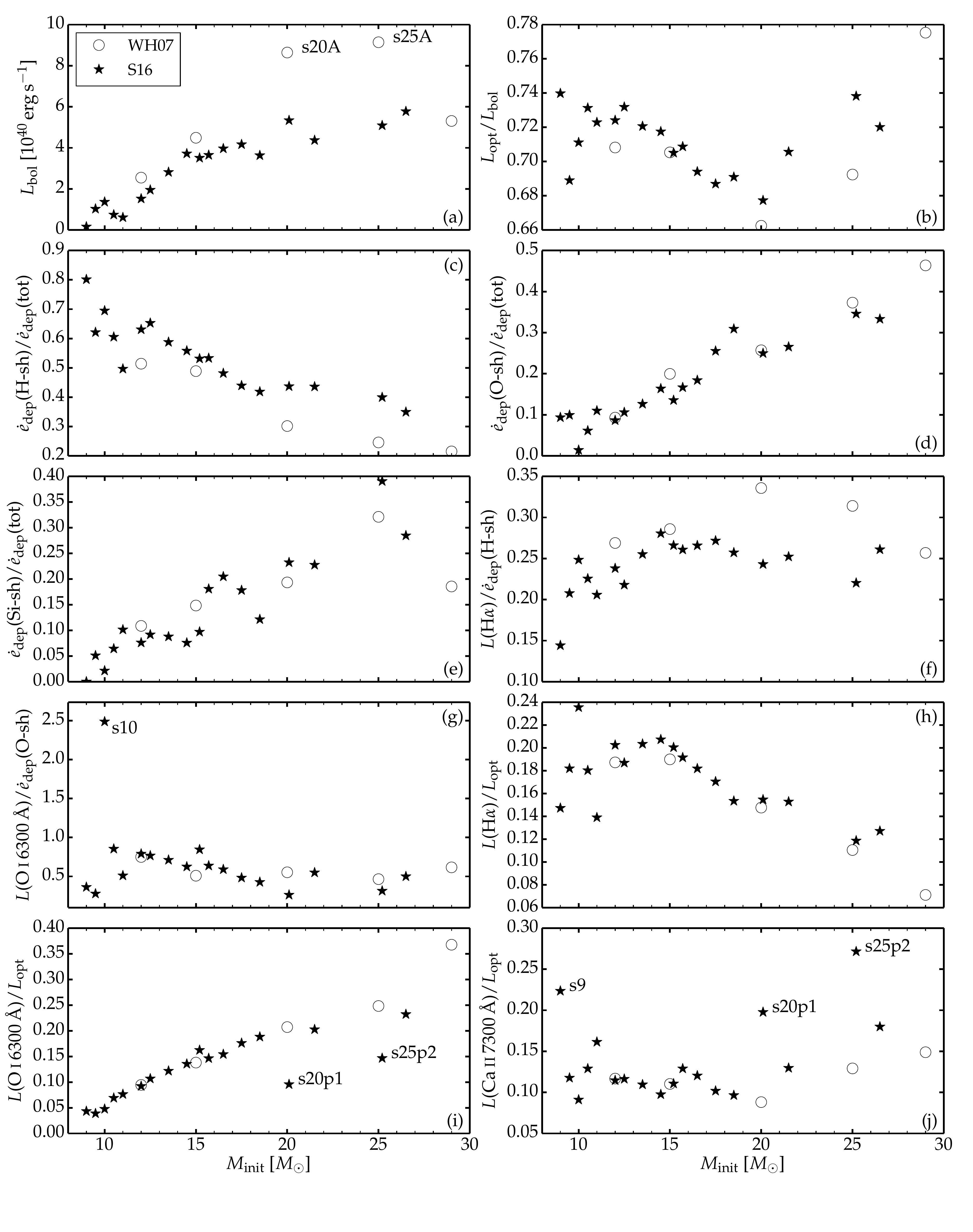}
\vspace{-0.3cm}
\caption{Line fluxes, powers, and their ratios for our grid of radiative-transfer calculations based on the WH07 (circles) and S16 (stars) ejecta models. Each panel in discussed in turn in Section~\ref{sect_quant}. A similar figure, including S16 models scaled in kinetic energy, is presented in Fig.~\ref{fig_line_fluxes_ekin}, in the Appendix.
\label{fig_line_fluxes}
}
\end{figure*}

In the larger S16 model set the spectral evolution shows greater variations and more complicated trends compared to the WH07 model set, with an obvious non-monotonic dependence of the relative line fluxes on progenitor mass. There are several reasons for this. First, the S16 model extends to lower-mass massive stars.  S16 predicts a low explosion energy of 1 to 6 $\times$\,10$^{50}$\,erg for initial masses 9 to 12\,\msun\ (with the s12 model having half the kinetic energy of model s12A). Consequently, the nebular spectra of these models exhibit narrower lines than in the rest of the S16 sample or relative to our selection of WH07 models. Second, S16 predict low \nifs\ masses of 0.004 to 0.03\,\msun\ for these low-energy explosions. This leads to a fainter SN at nebular times, a reduced heating rate, and potentially lower ejecta ionization (Table~\ref{tab_ion}). Third, the lower expansion rate  implies a more compact ejecta at the same SN age, and hence a greater density relative to higher energy explosions. This greater density should favor recombination and hence reduce the ejecta ionization. Finally, the metal yields of lower mass progenitors is smaller, which leads to a very weak \oidoub\ line relative to H$\alpha$.  As a result of these properties, the S16 set of models s9 to s12 stands well apart from the rest of the sample shown in Fig.~\ref{fig_montage_opt}.

For the mass range 12 to 25\,\msun, each WH07 model has a counterpart in the S16 set (s12A and s12, s15A and s15p2, s20A and s20p1, s25A and s25p2). We compare the normalized optical spectra of each model pair in Fig.~\ref{fig_comp_same_mass} -- we  omit the s15A and s15p2 pair since they are similar to each other and only slightly different from the 12\,\msun\ pair. Surprisingly, the optical spectra computed for models s12A and s12 are essentially identical, even though the progenitor evolution and the explosion were computed ten years apart and in a different fashion (see WH07 and S16 for discussion). The shuffling is also done independently for each model and this results in a distinct composition structure. Furthermore, s12A has 1.65 times the ejecta kinetic energy of s12 (and roughly the same $M_{\rm ej}$) but the only indication of this is the offset in the P~Cygni trough associated with H$\alpha$ and H$\beta$. Otherwise, emission lines have essentially the same width. The weak sensitivity to the factor of 1.65 difference in $E_{\rm kin}$ arises because the bulk of the kinetic energy is contained in the outer ejecta layers, which contribute little to the spectrum formation region at 350\,d. Instead, the spectrum forms deeper in the ejecta, where the actual offset in expansion rate for the metal-rich layers is more modest.\footnote{For example, in model s12, the inner 50\% of the ejecta mass contain only about 16\% of the total $E_{\rm kin}$, while 50\% of the total $E_{\rm kin}$ is contained in the outer 20\% of the ejecta mass. Because of this feature, the $E_{\rm kin}$ is better constrained at earlier times from the width of Doppler-broadened lines (which have their own degeneracy with $E_{\rm kin}/M_{\rm ej}$).} Model s12A also has 60\% more \nifs\ than model s12, and thus a 60\% greater luminosity but after normalization, this offset is gone. This offset in \nifs\ mass is too small to drive a visible change in ionization. Spectral line widths at nebular times may also suffer from a degeneracy between explosion energy and mixing. A larger $E_{\rm kin}$ may enhance line broadening, but a similar effect may be caused by a greater mixing of \nifs\ to large velocities. The similar spectral properties at 350\,d between models s12 and s12A suggest that slight differences in progenitor, explosion, or mixing (e.g., as treated via our shuffling technique) properties cannot be easily revealed. Such degeneracies are important limitations of inferences based on nebular phase spectra.

The bottom two panels of Fig.~\ref{fig_comp_same_mass} show a comparison of models s20A with s20p1, and of s25A with s25p2, whose ejecta properties ($M_{\rm ej}$, $E_{\rm kin}$, or yields, including \nifs\ mass) are very similar (Table~\ref{tab_prog}). While some spectral regions (e.g., below 6000\,\AA) and some lines (e.g., H$\alpha$) appear similar in each pair of models, the strong \oidoub\ and \caiidoub\ lines are significantly different. Specifically, the two S16 models exhibit a stronger \caiidoub\ flux and a weaker \oidoub\ flux. This difference is due to the much larger Ca to O abundance ratio in the O-rich shell in the s20p1 and s25p2 models, a consequence of the merging of the Si-rich shell and the O-rich shell during Si burning (see \citealt{DH20_neb}, and references therein).

Due to the merging, the Ca mass fraction in the O-rich shell rises from $\sim$\,10$^{-5}$ to $\sim$\,10$^{-3}$ and the \caiidoub\ doublet becomes a strong coolant of the O-rich material. This occurs even though Ca is still about 1000 times less abundant than O in the O-rich shell. The bulk of the Ca produced in massive stars is located in the Si/S shell (where the Ca mass fraction is about 0.05). Figure~\ref{fig_s20A_s20p1} illustrates these composition properties for models s20A and s20p1, in particular the large Ca abundance in the Si/S shells and the contrast in Ca/O abundance ratios between the O-rich shells of these two ejecta models. Hence, despite the marginal change in the total Ca yield, Si-O shell merging can dramatically alter the nebular phase spectral properties and inferences drawn from them. In model s26p5, the Si-rich and O-rich shells merge, but only partly since Si-rich material contaminates only the inner parts of the O/Ne/Mg shell. Consequently, in that model, the \oidoub\ is not so visibly reduced and appears both strong and broad (bottom curve in Fig.~\ref{fig_montage_opt}).

Turning to the rest of the S16 models, we obtain very similar properties for progenitors with masses between 12.5 and 18.5. Variations are induced in part by the reduction in $E_{\rm kin}$ with increasing progenitor mass, which causes a reduction in line width for all strong lines (e.g., compare s12p5 with s18p5). However, in a relative sense, the \oidoub\ clearly strengthens relative to H$\alpha$ in this sequence. Strikingly, these variations, which are related to the  progenitor mass, are weaker than those arising from Si-O shell merging. The process of Si-O shell merging is indirectly related to progenitor mass (one expects a greater occurrence of this process in higher mass progenitors) but it is an irony that this effect dominates over that caused by the stark contrast in oxygen yields between lower mass and higher mass progenitors (Table~\ref{tab_prog} and Fig.~\ref{fig_prop}).

\subsection{Quantitative description}
\label{sect_quant}

Figure~\ref{fig_line_fluxes} presents a more quantitative description of the results shown in Fig.~\ref{fig_montage_opt}.  For each model, we have measured the bolometric luminosity $L_{\rm bol}$, the optical luminosity $L_{\rm opt}$ (it accounts for the luminosity falling between 3500 and 9500\,\AA), the power deposited $\dot{e}_{\rm dep}$ in the various shells (the total power deposited in the ejecta $\dot{e}_{\rm dep}$(tot) is just $L_{\rm bol}$), and the power emitted in the strongest lines (i.e., \oidoub, H$\alpha$, and \caiidoub).

To measure the power emitted in lines we undertake a formal solution of the radiative transfer equation ignoring all bound-bound transitions except those of the selected ion. The \oidoub\ line flux is then measured from the O\one-only spectrum, the H$\alpha$ line flux from the H\one-only spectrum, and the \caiidoub\ line flux from the Ca\two-only spectrum. Because there are no neighboring lines that could corrupt the measurement, we can integrate the flux over a broad  band (i.e., $\pm$\,10000\,\kms\ for H$\alpha$,  $\pm$\,8000\,\kms\ for the other lines) centered around the rest wavelength of the transition (the mean wavelength is used for doublet lines). This technique allows a direct
 comparison of the line flux with the decay power absorbed in each shell.

To facilitate future discussions we define the H-rich shell as all ejecta locations where the H mass fraction $X_{\rm H}>0.3$, the He-rich shell where  $X_{\rm He}>0.8$, the O-rich shell where  $X_{\rm O}>0.6$ (for models s9 and s9p5, which have a very low O-shell mass, the criterion is $X_{\rm O}>0.2$ because the peak O-abundance is reduced by the minor mixing we apply to avoid having an O mass fraction profile looking like a Dirac delta function), and the Si-rich shell where $X_{\rm Si}>$\,0.1.

The total decay power absorbed is $L_{\rm bol}$ (panel (a) in Fig.~\ref{fig_line_fluxes}) and depends on the \nifs\ mass and the $\gamma$-ray escape fraction. The latter is function of the ejecta density profile (which depends strongly on $E_{\rm kin} / M_{\rm ej}$) and the location of the three \nifs-rich regions (there are three such regions because the original unmixed ejecta has only one and our shuffled-shell approach splits each dominant shell into three equal mass sub-shells).  Our models tend to show a decreasing trapping of $\gamma$-rays towards lower masses (from s25A to s12A, it drops from  90\% to 81\%; from s21p5 to s12p5 it drops nearly monotonically from 95\% to 67\%).  In the higher mass model s29A, the trapping of decay power is only 64\% of the total, primarily because of the reduced mass of the H-rich  shells (i.e., there is less H-rich material to absorb $\gamma$-rays, and the metal-rich core has a lower density, which translates into a greater $\gamma$-ray mean free path). In the S16 models below 12\,\msun\ that have a low explosion energy, the trapping is complete.  Overall, the trapping efficiency is always greater than 60\%, which explains why the distribution of $L_{\rm bol}$ values (top left of Fig.~\ref{fig_line_fluxes}) follows closely the distribution of $M$(\nifs) (top panel of Fig.~\ref{fig_prop}).

The fraction of the model luminosity that falls within the optical range spans the range 66 to 78\% (panel (b) in Fig.~\ref{fig_line_fluxes}), with a representative fraction of 71\%. Most of the remaining flux is emitted in the IR while less than 5\% is emitted in the UV.  The optical luminosity at 350\,d represents at most 70\% of the total decay power emitted, but as low as about 50\% in ejecta with the lowest trapping efficiency in our set (e.g., model s29A). The increase above 20\,\msun\ arises from a relative strengthening of the optical emission lines \oidoub\ and \caiidoub\ (which fall in the optical range), as the metal-rich core increases in mass while the H-rich ejecta layers become progressively lighter. The 10\% increase going from model s20A to model s29A arises from a combination of factors, including the drop by 60\% in decay power absorbed, the drop in optical depth by a factor of three, and the drop from 9.7 to 4.2\,\msun\ of the H-rich envelope mass of the progenitor.

\begin{figure}
\centering
\includegraphics[width=\hsize]{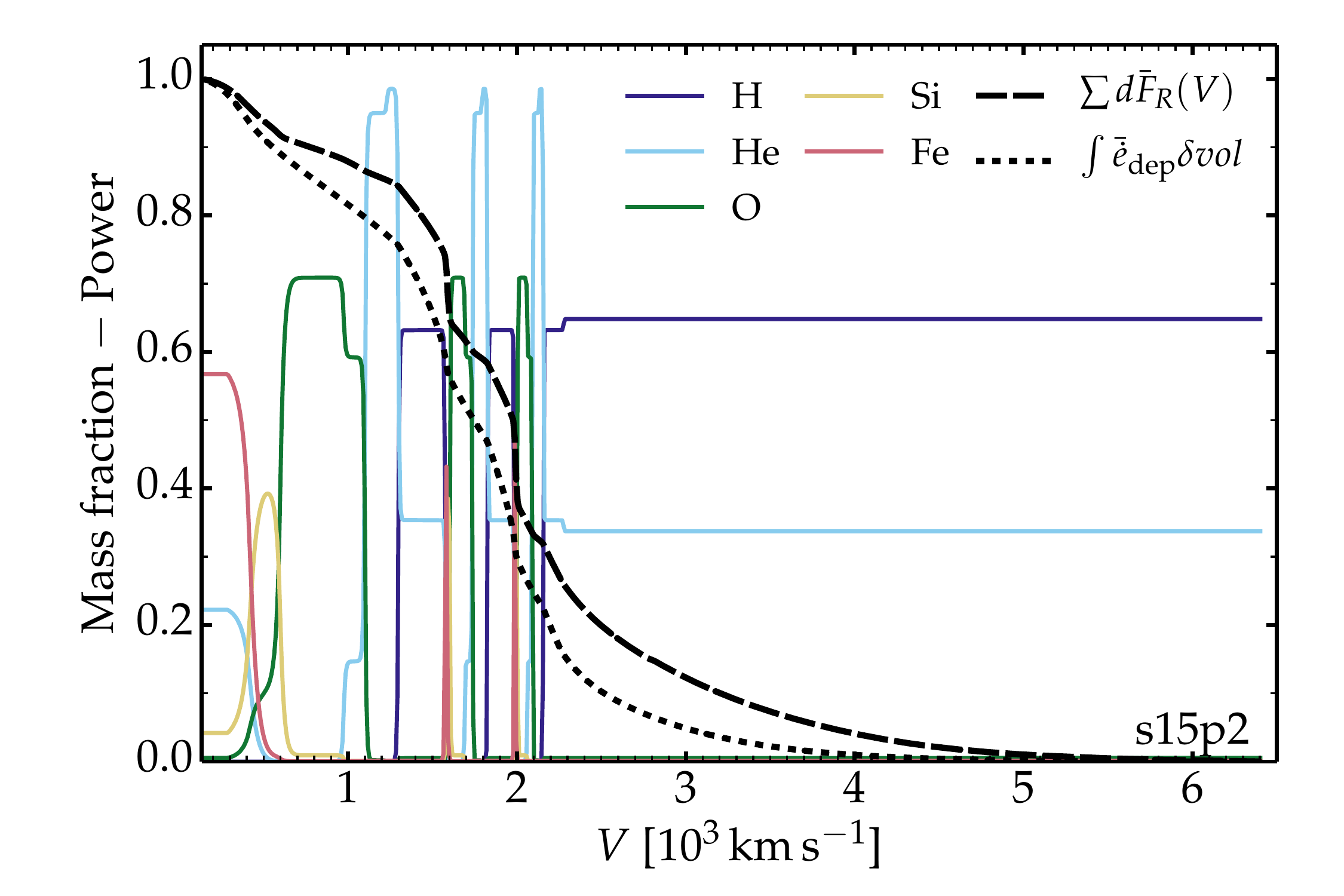}
\vspace{-0.5cm}
\caption{Profiles for the composition of dominant species, the cumulative decay power absorbed, and the cumulative frequency-integrated flux versus depth. For the latter two, the sum is performed inwards from the outermost ejecta location and normalized for visibility.
\label{fig_intdfr}
}
\end{figure}

Panels (c), (d) and (e) of Fig.~\ref{fig_line_fluxes} show the fractional decay power absorbed by the H-rich, O-rich, and Si-rich shells. For the H-rich shell, the power is primarily absorbed within the H-rich regions located below or immediately above $M_{\rm sh}$, hence in regions that coexist with the O-rich and Si-rich shells and that are not too distant from the \nifs-rich material. In other words, the H-rich material at higher velocities contributes less. Because of optical depth effects, the power absorbed in a given shell does not equal the flux escaping from that shell \citep{DH20_neb}. For example, in model s15p2, about 50\% of the total power is absorbed by H-rich material and about 50\% of the total escaping flux emerges from the H-rich material (the equality between the two is a coincidence), but a fraction of the power deposited in the slower-moving H-rich shells is reprocessed by the overlying, faster-moving Fe-rich shells, while the outer H-rich shell, which absorbs about 20\% of the total decay power, releases about 30\% of the total flux (Fig.~\ref{fig_intdfr}).  Nonetheless, these three panels reveal a clear trend, reflecting the composition structure of the progenitors, and give indicative power limits for spectral line luminosities.

As the progenitor mass is increased in both the WH07 and S16 model sets, the fractional power absorbed in the H-rich shells decreases (from 80\% in s9 to 20\% in s29A), On the other hand it increases in the O-rich shell (from nearly zero in s10 up to 50\% in model s29A) and Si-rich shell (from nearly zero in s9 up to 40\% in s25p2). The highest value reached is for model s25p2, and is partially caused by Si-O shell merging in the progenitor which boosts the mass of the Si-rich shell as defined by our criterion $X_{\rm Si}>$\,0.1). These  trends are well understood by considering the variations in the mass of the three shells  \citep{DH20_neb}. The low metal yields and the dominance of the H/He shell in lower-mass massive stars naturally implies that the bulk of the decay power will be absorbed by H-rich material. As the ZAMS mass is increased, the O-rich and Si-rich shells capture a greater fraction of the decay power. Although the mass of the H-rich shell may be the same for a wide range of progenitor mass \citep{d19_sn2p} the  \nifs-rich material is closer to the O-rich and Si-rich shells, which benefit from this proximity to absorb a greater fraction of the available decay power. However for the highest mass models (e.g. s29A) $\gamma$-ray escape from the inner ejecta disfavors the Si-rich shell relative to the massive O-rich shell. The large variations in decay power absorbed are at the origin of the variations in line fluxes, or at least they set the fundamental power limits for the line fluxes.

Panels (f)--(g) of Fig.~\ref{fig_line_fluxes}  relate line flux to decay power absorbed. We find that the H$\alpha$ line flux represents about 25\% of the decay power absorbed in the H-rich shell. Some models exhibit a higher cooling power from H$\alpha$ (i.e., s20A and s25A), and we believe this is caused by the somewhat weaker mixing in those models ($M_{\rm sh} \approx M_{\rm He,c} + 1\,$\,\msun) where less H-rich material is mixed with the metal-rich core (the inner H-rich shells absorb a smaller fraction of the total decay power). In that case, the H$\alpha$ line flux comes from reprocessing of radiation from the inner ejecta. The lower cooling power of H$\alpha$ in model s9 is caused by the strong Ca\two\ cooling in the H-rich layers of that model (as discussed below, this is caused by the low \nifs\ mass, which fosters an ionization shift to Ca$^+$, compared to Ca$^{2+}$ in most other models in our grid; see Table~\ref{tab_ion}). The \oidoub\ line flux relative to the decay power absorbed in the O-rich shell covers a much larger range (from 30 to 90\% for the bulk of models, with one outlier at 250\%!). Model s10 appears as an outlier, and even seems to violate energy conservation. Instead, the apparently large cooling efficiency arises from the formation of the \oidoub\ doublet in both the O-rich shell and the H-rich shell (in the latter, it absorbs power that would otherwise have been radiated by \caiidoub).

The \oidoub\ is a strong coolant of the H-rich layers in model s10 in part because O is neutral there, while Ca is twice ionized (so that \caiidoub\ cannot cool these layers). These ionization patterns arise from the larger $E_{\rm kin}$ of model s10, which causes a lower density for a comparable decay power emitted or absorbed in adjacent models. For example, in model s11, Ca is Ca$^+$ and \caiidoub\ cooling dominates that due to \oidoub\ in the H-rich layers. This effect is striking in lower-mass models because of the low mass of the O-rich shell (which strongly limits the \oidoub\ flux contribution from the O-rich shell). Other effects influencing the \oidoub\ emission is the ionization in the O-rich shell, which can affect the cooling power of O and Mg. Furthermore, Si-O shell merging is one way to quench \oidoub\ emission to the benefit of \caiidoub.

\begin{figure*}
\centering
\includegraphics[width=\hsize]{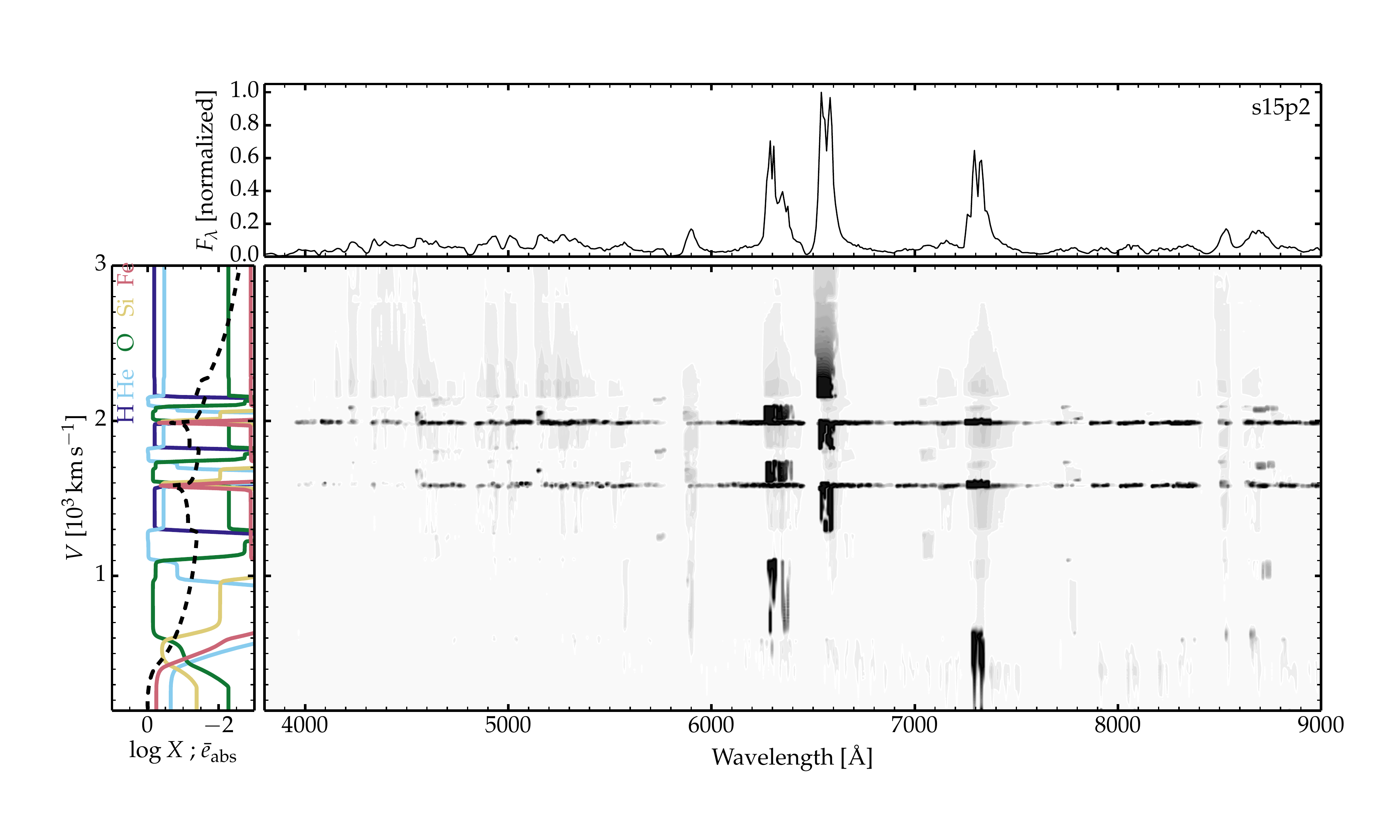}
\vspace{-1cm}
\caption{Illustration of the spatial regions (here shown in velocity space) contributing to the emergent flux in the S16 model s15p2. The greyscale image shows the observer's frame flux contribution $\partial F_{\lambda,V} / \partial V$ (the map maximum is saturated at 20\,\% of the true maximum to bias against the strong emission lines and better reveal the origin of the weaker emission) versus wavelength and ejecta velocity.  The plot at left connects these emission contributions to ejecta shells and the normalized decay-power deposition profile (dashed line) on the same velocity scale. The upper panel shows the corresponding normalized flux $F_{\lambda}$ integrated over all ejecta velocities.
\label{fig_dfr}
}
\end{figure*}

The last three panels (h)-(i)-(j) of Fig.~\ref{fig_line_fluxes} give the line fluxes of H$\alpha$, \oidoub, and \caiidoub\ as a fraction of the total optical luminosity (or flux). This  normalization removes the need for an accurate determination of the \nifs\ mass, the $\gamma$-ray escape, or the distance to the SN -- for observations all that is needed to build such ratios is the reddening. For lower mass progenitors, the fractional H$\alpha$ line flux is between 14 and 24\%, steadily decreasing for initial masses beyond 15\,\msun\ to reach 6\% for model s29A. In lower mass progenitors, the scatter is caused by circumstances in which the radiation leakage in \caiidoub\ is enhanced (models s9 and s11) or inhibited (model s10). For higher masses, the fractional H$\alpha$ line flux decreases because of the decreasing decay power absorbed by the H-rich material and the decreasing optical depth of the outer H-rich ejecta.

With the exception of two outliers (those models with Si-O shell merging prior to core collapse), the fractional \oidoub\ line flux is an increasing function of the progenitor mass, growing from about 4\% in model s9p5 to 25\% in model s25A, with a maximum of 37\% in model s29A. In the absence of Si-O shell merging, the fractional \oidoub\ line flux appears as the best indicator of progenitor mass in both model sets. In contrast to \oidoub, the fractional \caiidoub\ line flux exhibits a lot of scatter, and shows no correlation with progenitor mass -- progenitors with ZAMS masses differing by 20\,\msun\ can exhibit the same flux ratio. The strongest emitters are associated with model s9 (because of the dominance of Ca$^+$ in the H-rich zones and because the H-rich and He-rich shells constitute over 97\% of the ejecta mass)  and models s20p1 and s25p2 (because of Si-O shell merging).

\begin{figure*}
\centering
\includegraphics[width=0.8\hsize]{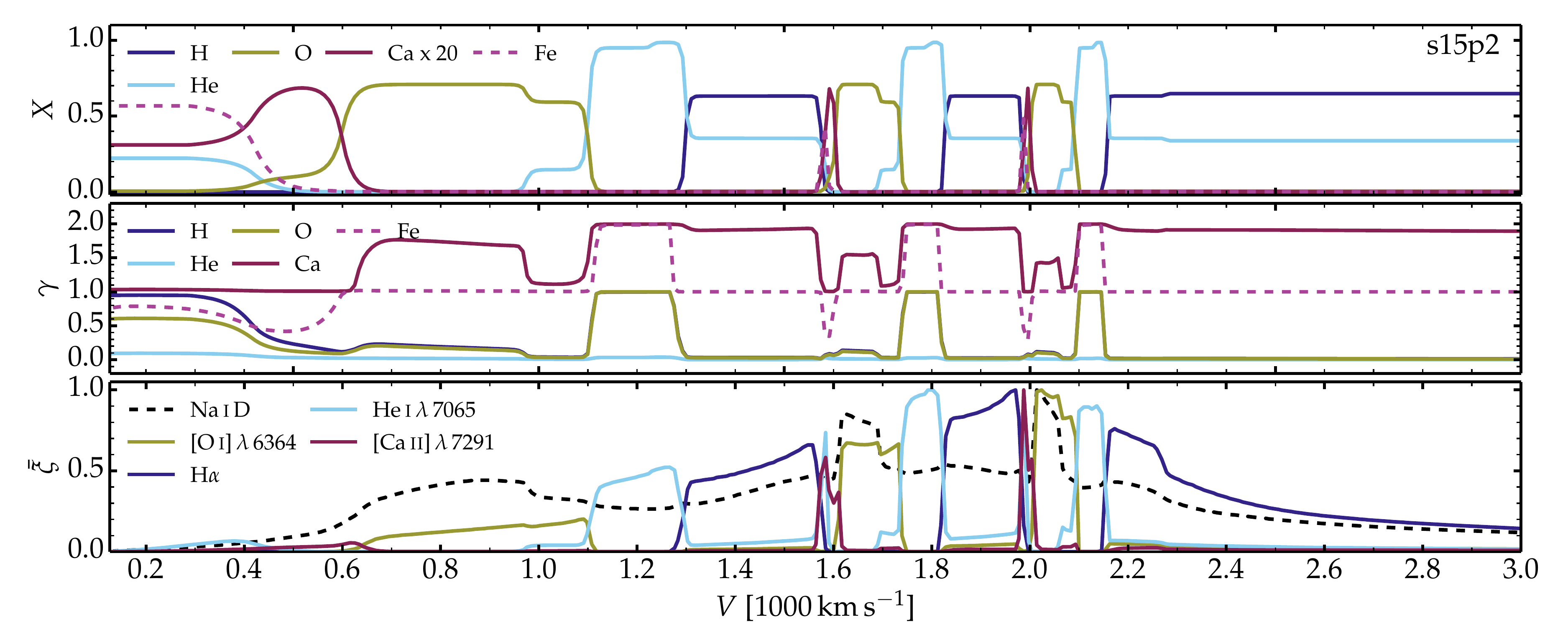}
\caption{Top: Illustration of ejecta and radiation properties for the shuffled-shell model s15p2. From upper to lower panels, we show the profiles versus velocity for the mass fraction of H, He, O, Ca, and Fe, their ionization state (zero for neutral, one for once ionized etc; see also Table~\ref{tab_ion}), and the formation regions for Na\,\one\,D, [O\,{\sc i}]\,$\lambda$\,6364, H$\alpha$, He\,\one\,7065\,\AA, and [Ca\,{\sc ii}]\,$\lambda$\,7291. The quantity  $\int \xi \, d\log V $ is the line equivalent width, which is here normalized to unity for visibility.
}
\label{fig_shuffle_ej_prop_s15p2}
\end{figure*}

\section{Origin of Ca\two\ emission}
\label{sect_ca2}

Below we explore in more detail a variety of features discussed in the preceding sections. We start with the origin of the Ca\two\ emission in SN II nebular spectra. As is well known, the \caiidoub\ is a very strong coolant and may dominate the cooling of the gas \citep{fransson_chevalier_89,li_mccray_ca2_93}. Its importance as a coolant  depends on the Ca abundance, the Ca ionization, and the competition with cooling from other lines.

In their single zone modeling of SN\,1987A, \citet{li_mccray_ca2_93} found that the \caiidoub\ must arise from H-rich material (in which the Ca abundance is given by the LMC metallicity value) rather than from the more abundant and newly synthesized Ca created during the explosion. \cite{jerkstrand_04et_12} also found that the Ca\two\ emission arises primarily from the H-rich material in their modeling of SN\,2004et. In both cases, this H-rich material corresponds to the H-rich material at the base of the progenitor H-rich envelope, or to the H-rich material that was macroscopically mixed down to smaller velocities. An independent confirmation that at least some Ca\two\ emission must arise from H-rich material  is given by the observation of a small common bump in emission seen in H$\alpha$, the \caiidoub\  and Fe\two\ $\lambda 7155$ \citep{spyromilio_clump_93}.

In our simulations for standard-energy type II SNe, we find that in most models the \caiidoub\ forms primarily in the Si-rich shell, hence where Ca is the most abundant.  We found this in previous simulations adopting a simplified ejecta structure \citep{DH20_neb}, in more sophisticated simulations using the shuffled-shell method applied to the s15A model of WH07 \citep{DH20_shuffle}, and we again find this results in the present extended set of models based on the more physically consistent inputs from S16 (few exceptions are discussed below). This is most strikingly revealed by Fig.~\ref{fig_dfr} which illustrates the emission sites for the emergent optical radiation. While some \caiidoub\ emission arises from H-rich zones, the bulk of it arises in the Si-rich zone.  Type IIb/Ib/Ic  SNe also exhibit \caiidoub\ emission \citep[e.g.][]{shivvers_ibc_19},  and in this case it must come from newly synthesized Ca because their progenitor have a low mass H-rich envelope or no H-rich envelope at all \citep[e.g.][]{jerkstrand_15_iib}.

In contrast to the  \caiidoub\ transition,  the \caiitrip\ triplet has a significant contribution from the H-rich shell. In model s12, for  example, approximately 50\% of the triplet emission originates in the H-rich shell. That the  \caiidoub\ and  \caiitrip\ have different emitting regions is not surprising -- they have a different critical density, a different temperature dependence, and the \caiitrip\ has a much larger optical depth.

\begin{figure}
\centering
\includegraphics[width=\hsize]{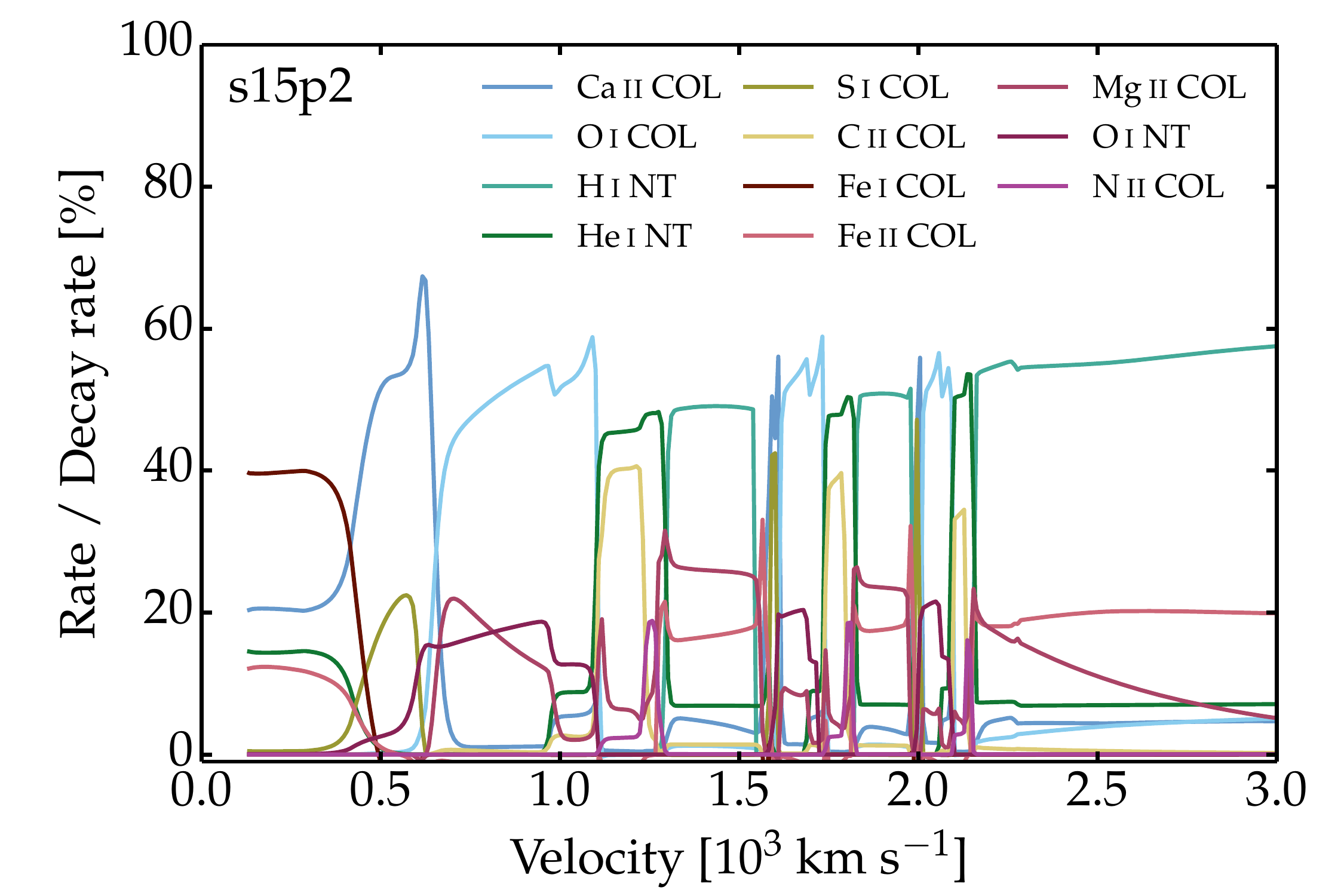}
\vspace{-0.2cm}
\caption{Illustration of the main cooling processes balancing the radioactive decay heating at all ejecta depths in the model s15p2. We show each dominant cooling rate (stepping down from the rate having  the largest peak value at any depth) normalized to the local heating rate. The term ``NT'' stands for non-thermal processes (in this context non-thermal excitation and ionization) and ``COL" stands for collisional processes (i.e., collisional excitation).
\label{fig_cool}}
\end{figure}

Figure~\ref{fig_shuffle_ej_prop_s15p2} reveals more information on the origin on line emission (bottom panel) by connecting it to both the abundance profile (top panel) and the ionization state (middle panel) for H, He, O, Ca, and Fe. In the Si-rich shell, Ca is once ionized and this is where the bulk of the emission is coming from. In all other shells, it is either twice ionized (Ca$^{2+}$ dominates and Ca\two\ emission is negligible) or underabundant (in the C/O shell). In model s15p2, the main deterrent for Ca\two\ emission in the H-rich shell is therefore overionization (Table~\ref{tab_ion}). This is further seen in Fig.~\ref{fig_cool} where the dominant cooling processes are shown versus velocity. As is apparent, Ca\two\ (together with S\one) collisional excitation dominates the cooling of the Si-rich regions, while the main processes cooling the H-rich material are non-thermal excitation of H\one\ and Fe\two\ collisional excitation. This situation holds in essentially all models of the present grid, with only a few exceptions.

\begin{figure*}
\centering
\includegraphics[width=0.8\hsize]{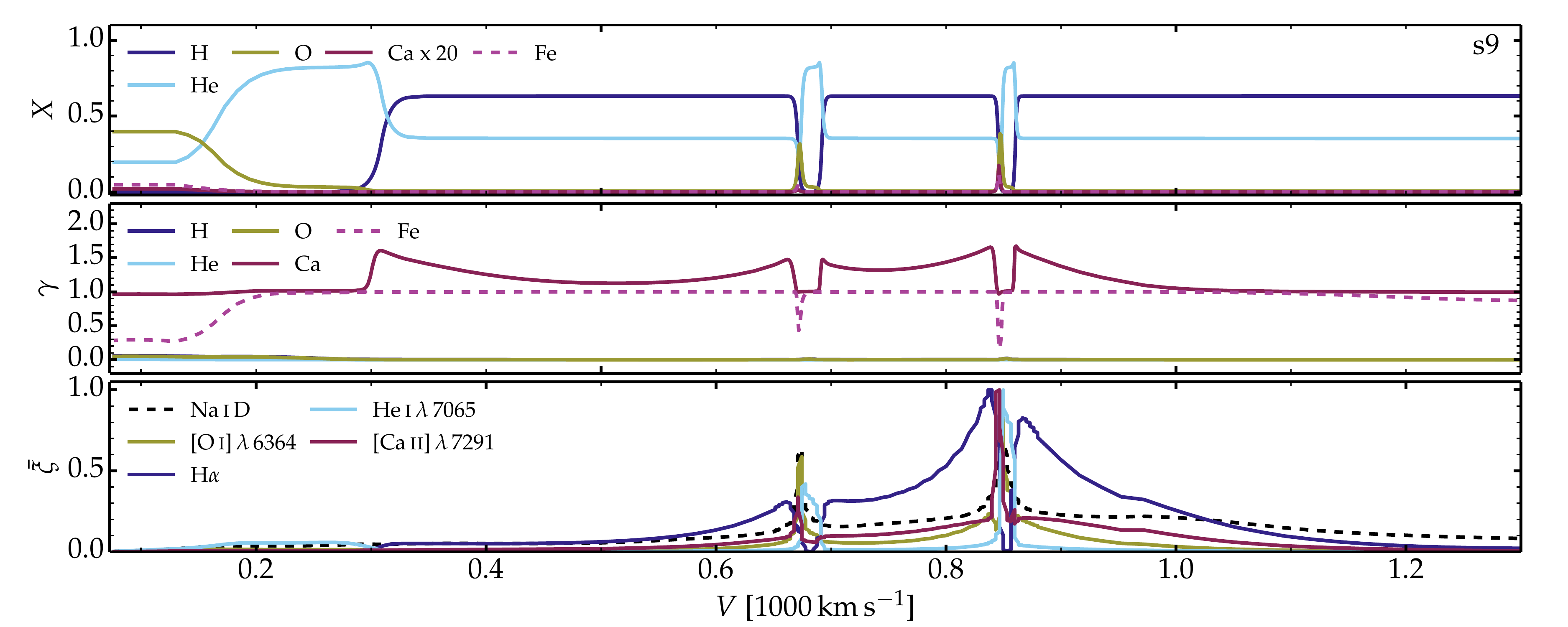}
\caption{Same as Fig.~\ref{fig_shuffle_ej_prop_s15p2}, but now for the s9 model, which corresponds to the lowest mass progenitor model in the S16 set.
\label{fig_shuffle_ej_prop_s9}
}
\end{figure*}

In three of our models \caiidoub\ emission also comes from outside the Si-rich shell,  giving rise to the outliers in the \caiidoub\ line flux plot shown in the bottom left panel of Fig.~\ref{fig_line_fluxes}.  The extended emission arises from two effects. First, in the models that undergo Si-O shell merging in the final stages of evolution prior to collapse, the Ca\two\ emission also arises from the O-rich region contaminated by Si-rich shell material. Hence, a fraction of the power deposited in the O-rich shell is then radiated by \caiidoub, which boosts its strength relative to models without Si-O shell merging. Secondly, a boost to the \caiidoub\ emission can arise if Ca\two\ collisional excitation becomes an important coolant for the H-rich material, as in model s9 (Fig.~\ref{fig_shuffle_ej_prop_s9}).  Because of the low \nifs\ mass (0.004\,\msun), the decay heating in s9 is reduced relative to a more standard explosion model like s12p5. Further, because of the low ejecta kinetic energy ($1.1 \times 10^{50}$\,erg; one tenth of that in model s12p5), the ejecta density is greater at late times. Both effects lead to lower ejecta ionization, and consequently, Ca is roughly singly ionized throughout the ejecta. As a consequence, and because a large fraction of the decay power is absorbed by the H-rich material, the \caiidoub\ forms in both the Si-rich and H-rich layers. An analogous situation affects the O\one\ emission, which comes in part from the H-rich layers in this model. Because of its extended spatial distribution in model s9, the \caiidoub\ is very strong and broader relative to other optical emission lines (see Fig.~\ref{fig_montage_opt} and Fig.~\ref{fig_line_fluxes}).

\begin{table*}
    \caption{Mean ionization state $\gamma$ for a selection of species in the H-rich, He-rich, O-rich, Si-rich, and Fe-rich shells. A value of zero corresponds to neutral, one to once ionized etc.
\label{tab_ion}
}
\begin{center}
\begin{tabular}{l|ccc|c|c|c|c}
\hline
 Model  &    \multicolumn{3}{|c|}{H-rich shell}       & He-rich shell &  O-rich shell & Si-rich shell & Fe-rich shell \\
\hline
        &        $\gamma_{\rm H}$    &     $\gamma_{\rm Ca}$   &      $\gamma_{\rm Fe}$  &       $\gamma_{\rm He}$   &       $\gamma_{\rm O}$  &       $\gamma_{\rm Ca}$  &       $\gamma_{\rm Fe}$    \\
\hline
    s12A  &        0.02    &      1.93   &       1.00  &        0.04   &       0.12  &        1.04  &        0.65    \\
  s15A  &        0.02    &      1.93   &       1.00  &        0.03   &       0.11  &        1.03  &        0.66    \\
  s20A  &        0.02    &      1.89   &       1.00  &        0.03   &       0.10  &        1.01  &        0.57    \\
  s25A  &        0.03    &      1.92   &       1.00  &        0.02   &       0.09  &        1.01  &        0.57    \\
  s29A  &        0.05    &      1.98   &       1.00  &        0.03   &       0.16  &        1.02  &        0.72    \\
\hline
    s9  &        0.00    &      1.09   &       0.95  &        0.00   &       0.02  &        0.99  &        0.38    \\
  s9p5  &        0.00    &      1.26   &       1.00  &        0.01   &       0.04  &        1.00  &        0.29    \\
   s10  &        0.01    &      1.88   &       1.00  &        0.02   &       0.05  &        1.01  &        0.69    \\
 s10p5  &        0.00    &      1.53   &       1.00  &        0.01   &       0.04  &        1.00  &        0.57    \\
   s11  &        0.00    &      1.21   &       1.00  &        0.00   &       0.02  &        0.99  &        0.24    \\
   s12  &        0.01    &      1.85   &       1.00  &        0.02   &       0.07  &        1.01  &        0.71    \\
 s12p5  &        0.02    &      1.96   &       1.00  &        0.03   &       0.10  &        1.04  &        0.75    \\
 s13p5  &        0.02    &      1.96   &       1.00  &        0.03   &       0.12  &        1.06  &        0.64    \\
 s14p5  &        0.02    &      1.95   &       1.00  &        0.04   &       0.12  &        1.06  &        0.66    \\
 s15p2  &        0.02    &      1.91   &       1.00  &        0.03   &       0.12  &        1.01  &        0.59    \\
 s15p7  &        0.02    &      1.89   &       1.00  &        0.02   &       0.08  &        1.01  &        0.61    \\
 s16p5  &        0.01    &      1.84   &       1.00  &        0.02   &       0.07  &        1.00  &        0.43    \\
 s17p5  &        0.01    &      1.77   &       1.00  &        0.01   &       0.06  &        1.01  &        0.45    \\
 s18p5  &        0.01    &      1.71   &       1.00  &        0.01   &       0.05  &        1.00  &        0.36    \\
 s20p1  &        0.01    &      1.82   &       1.00  &        0.01   &       0.04  &        1.00  &        0.61    \\
 s21p5  &        0.01    &      1.82   &       1.00  &        0.01   &       0.05  &        1.00  &        0.46    \\
 s25p2  &        0.03    &      1.96   &       1.00  &        0.03   &       0.08  &        1.02  &        0.73    \\
 s26p5  &        0.03    &      1.93   &       1.00  &        0.02   &       0.07  &        1.01  &        0.60    \\
\hline
\end{tabular}
\end{center}
\end{table*}

\section{\nifs\ Bubble effect}
\label{sect_bubble}

Decay heating causes a time integrated energy injection that can approach the local specific kinetic energy of the \nifs-rich regions. In these regions, decay heating can impact the dynamics for days after shock breakout, at times when homologous expansion would otherwise hold \citep{woosley_87A_late_88,herant_87A_2D_92,basko_56ni_94}. In practice, these \nifs-rich regions can expand and create low-density ``bubbles'' surrounded by a dense wall of swept-up material, thereby triggering RT instabilities and additional mixing. Hence, this bubble effect implies a decrease in density of the \nifs-rich regions and a compression of the swept-up material that builds a thin dense wall around these newly created bubbles. This bubble effect depends on the 3D distribution and abundance of \nifs, as well as on the density and expansion rate of the surrounding material. This process is therefore very non linear and complex. Three-dimensional simulations over days and weeks after explosion suggest that the 3D ejecta can have density variations of about a factor of ten throughout the volume where the bubbles reside \citep{gabler_3dsn_21}. Such inhomogeneities can impact the SN radiation during the photospheric phase \citep{d18_fcl,dessart_audit_rhd_3d_19} as well as the nebular phase properties of core-collapse SNe \citep{li_87A_93,jerkstrand_87a_11}.

In our 1D shuffled-shell method, we can implement this bubble effect by introducing a radial stretching of the zones containing \nifs, and a radial compression of all other zones within $M_{\rm sh}$. For simplicity, we assume that this radial stretching and compression is uniform throughout $M_{\rm sh}$.

We proceed in the following manner. In the original unmixed ejecta model, we initially compute the volume $V_{\rm Ni, 0}$ occupied by the \nifs-rich shell and the volume $V_{\rm sh}$ contained within  $M_{\rm sh}$. We then specify what fraction $\alpha_{\rm Ni}$ of $V_{\rm sh}$ should be occupied by the \nifs-rich bubble in the mixed ejecta at late times (i.e., at 350\,d in the present simulations). Numerical simulations suggest that $\alpha_{\rm Ni}$ can be as large as 0.5, i.e., the bubble takes up half the volume occupied by the progenitor He core.

To conserve mass, the density of the \nifs\ bubble must be divided by a factor $f_{\rm Ni} = (\alpha_{\rm Ni}V_{\rm sh})/V_{\rm Ni, 0}$ and the density of the other shells within $M_{\rm sh}$ must be divided by a factor $f_{\rm other} = (1-\alpha_{\rm Ni})V_{\rm sh} / (V_{\rm sh}-V_{\rm Ni, 0})$. Because $V_{\rm sh} \gg V_{\rm Ni, 0}$ (in model s15A, the \nifs-rich shell represents only 2.5\% of the total volume within $M_{\rm sh}$ immediately after explosion), this treatment implies a large density reduction of the \nifs-rich material but only a modest compression of the rest of the metal-rich regions (for the s15A model with a bubble having $\alpha_{\rm Ni}=$\,0.6, the density in the \nifs-rich bubble drops by a factor of 24, while the rest of the material within $M_{\rm sh}$ is only compressed by a factor of 2.5). While the former is physical, the latter is probably not very realistic since the dense wall around the \nifs--rich bubble is expected to be very dense and the material further away hardly affected by the bubble effect. Hence, in reality, there should be a complicated 3D distribution of compressed and rarefied material within the metal-rich regions of core-collapse SNe.

Having determined the density scalings for the \nifs-rich shell and the other shells within  $M_{\rm sh}$, we proceed from the innermost zone outwards through our shuffled-shell structure and apply a radial stretching or a radial compression to modify the volume of all shells as necessary. When we encounter a \nifs-rich zone of volume $\delta V$ located between radii $r_l$ and $r_{u}$ on the original radial grid, we build a new radial grid and set :
$$
r_u^3  = r_l^3  + \frac{3}{4\pi} f  \delta V  \,\,  ,
$$
where $f$ is equal to $f_{\rm Ni}$ in \nifs-rich zones and $f_{\rm other}$ otherwise. The density of the corresponding zone is scaled by $1/f$ so that the mass of the zone is unchanged. The radial grid and the density beyond $M_{\rm sh}$ are unchanged. Because of homology, the radial stretching or compression is equivalent to a similar distortion in velocity space. With the bubble effect, the \nifs-rich shells cover a larger range of velocities while the rest of the material is confined to narrower velocity regions.

\begin{figure*}
\centering
\includegraphics[width=0.9\hsize]{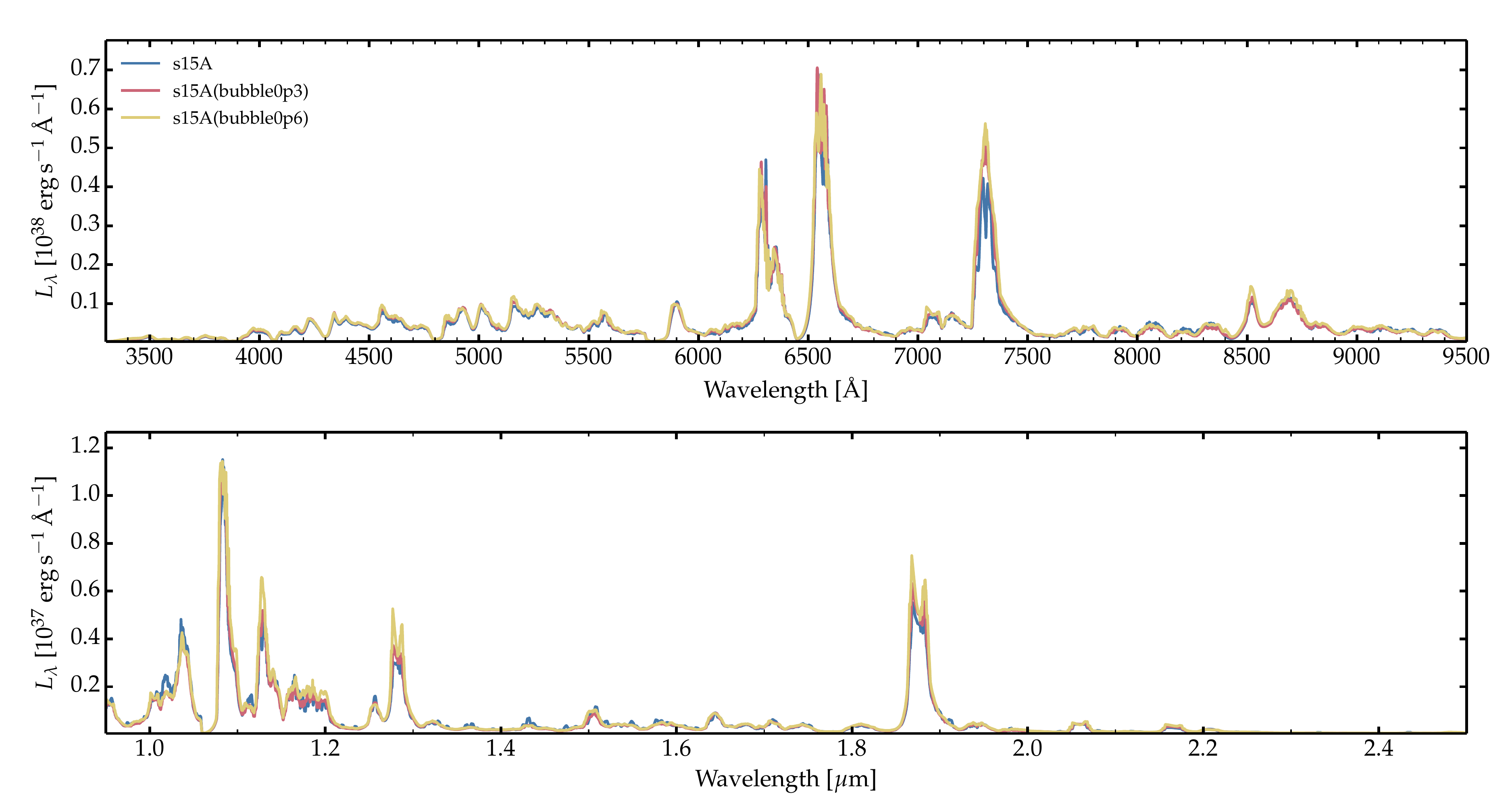}
\includegraphics[width=0.9\hsize]{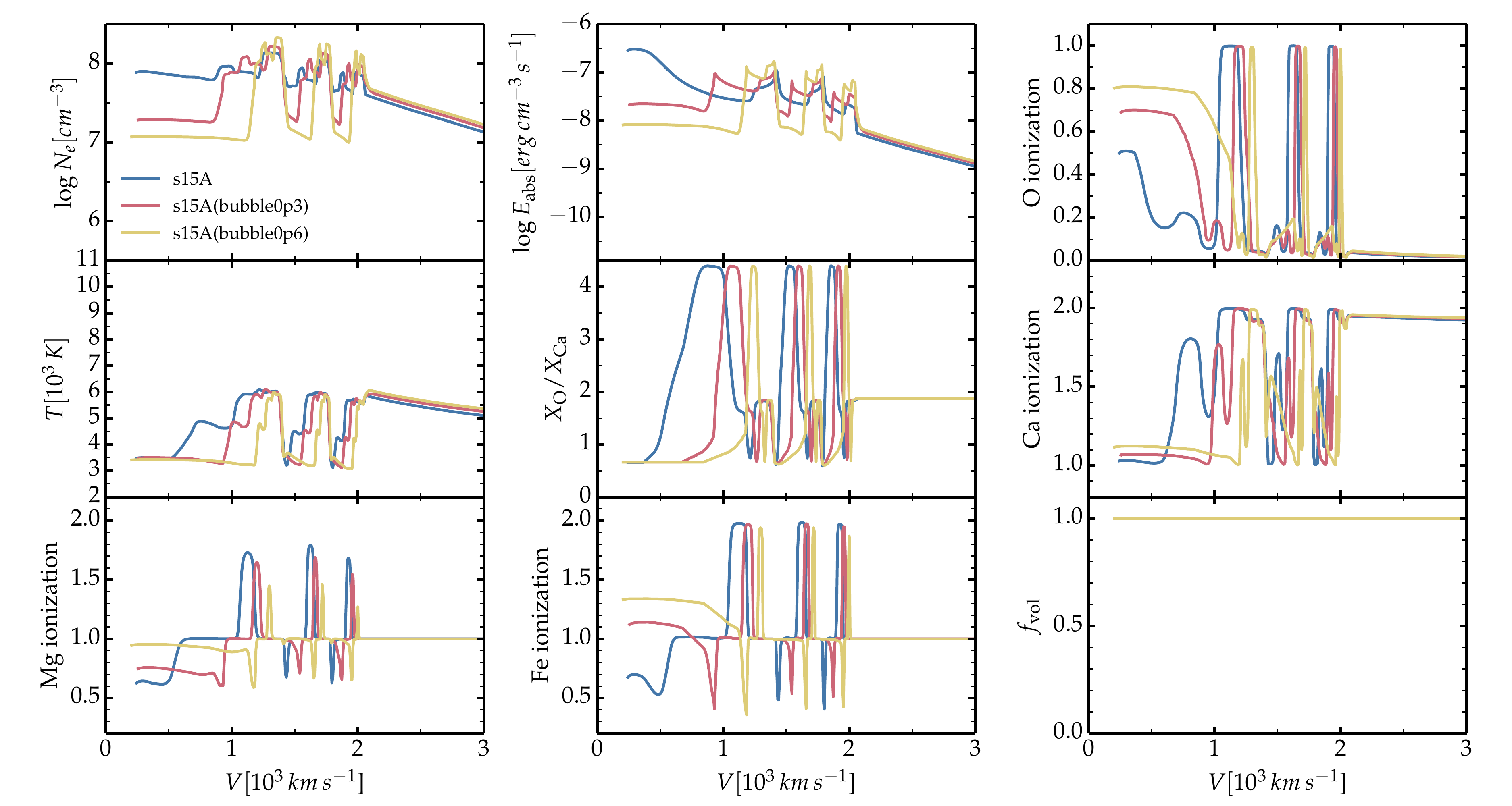}
\caption{Illustration of the influence of the bubble effect applied to the reference model s15A. We show the results for the SN radiation and gas properties in model s15A and for its counterparts in which the \nifs-rich material occupies 30 (model bubble0p3) and 60\% (model bubble0p6) of the volume within $M_{\rm sh}$. [See Section~\ref{sect_bubble} for discussion.]
}
\label{fig_bubble}
\end{figure*}

Using the model s15A with $M_{\rm sh}=$\,5.36\,\msun\ as a reference, we computed two other models to test the influence of the bubble effect adopting $\alpha_{\rm Ni}=$\,0.3 and 0.6 (the \nifs-rich bubble occupies 30 and 60\% of the total volume within $M_{\rm sh}$). The results for the SN radiation and the ejecta properties are shown in Fig.~\ref{fig_bubble}. Interestingly, the impact on the SN radiation is weak, although some slight differences are visible in Ca\two\ lines. Concerning ejecta properties, the bottom panel of Fig.~\ref{fig_bubble} shows that the gas properties are hardly modified by the bubble effect, amounting mostly to a radial or velocity shift of the profiles depicted for the three models, with only an increase in ionization in \nifs-rich regions resulting from the lower density. As said earlier, while the bubble effect leads to a strong reduction of the density in \nifs-rich regions, the adopted uniform compression of the surrounding material is about a factor of two, which is too small to cause an ionization shift, in particular because the ionization is typically low for the dominant species of each zone.

The \nifs\ bubble effect should have a similarly weak effect in other models. In lower-mass models, the \nifs\ mass is small (down by a factor of ten or more relative to s15A), but the $E_{\rm kin}$ is also small (also down by up to a factor of ten relative to s15A) so the bubble effect should be comparable and therefore weak (as for model s15A). In higher mass models, the \nifs\ mass is within a factor of two of the value for s15A so we expect essentially the same behavior as seen in the tests performed for the model s15A.

\section{Influence of a uniform clumping}
\label{sect_fvol}

The \nifs-bubble effect described above is expected to be just one component contributing to the complex 3D structure of core-collapse supernova ejecta. Besides the general density variations discussed in a simplified manner in the preceding section, the \nifs-bubble effect should lead to additional instabilities causing mixing and clumping of the material at the bubble interface with the surrounding medium. The instabilities driven by the bubble effect, and the likely interaction between distinct bubbles, should also build upon the instabilities associated with the explosion itself and the shock propagation in the progenitor envelope \citep{gabler_3dsn_21}. One thus expects a complex distribution of clump densities and composition distributed in a complicated 3D pattern.

To further explore the impact of density variations on the ejecta properties and escaping radiation, we introduce a uniform clumping of the ejecta. Following \citet{d18_fcl},
we assume that the ejecta are composed of clumps that occupy a fraction $f_{\rm vol}$ of the total volume. The clumps are assumed to be small relative to a photon mean free path, and the medium surrounding these clumps is assumed to be void. With these assumptions the initial ejecta model density is simply scaled by a factor of $1/f_{\rm vol}$, while opacities and emissivities  (which are computed with populations and temperature deduced for the clumps) are all scaled by a factor of $f_{\rm vol}$. These assumptions leave column densities unchanged but have a direct influence on processes that depend on the density squared (e.g., free-free), and an indirect influence on the radiation field because of the sensitivity of the kinetic equations to density.

\begin{figure*}
\centering
\includegraphics[width=0.9\hsize]{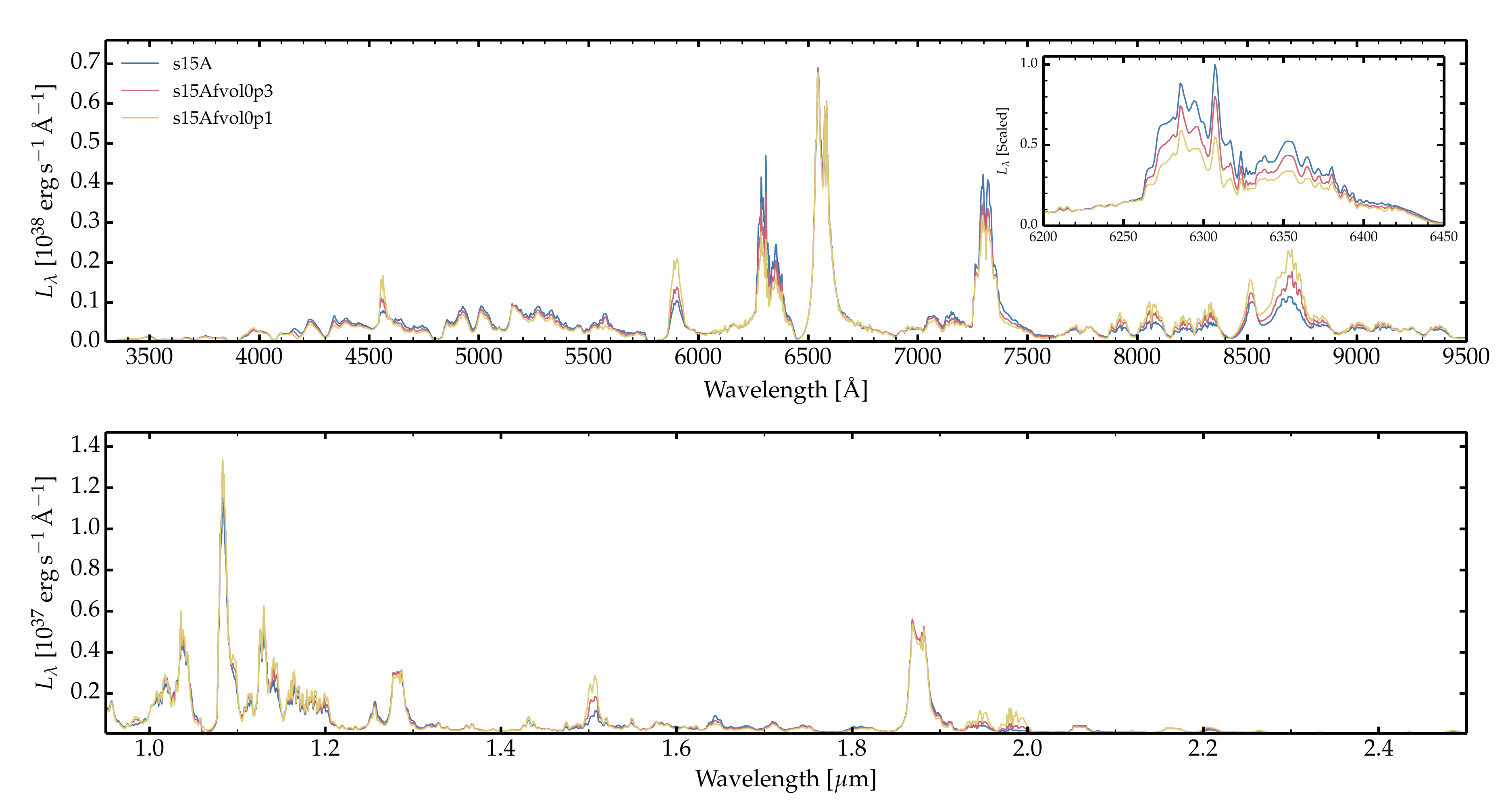}
\includegraphics[width=0.9\hsize]{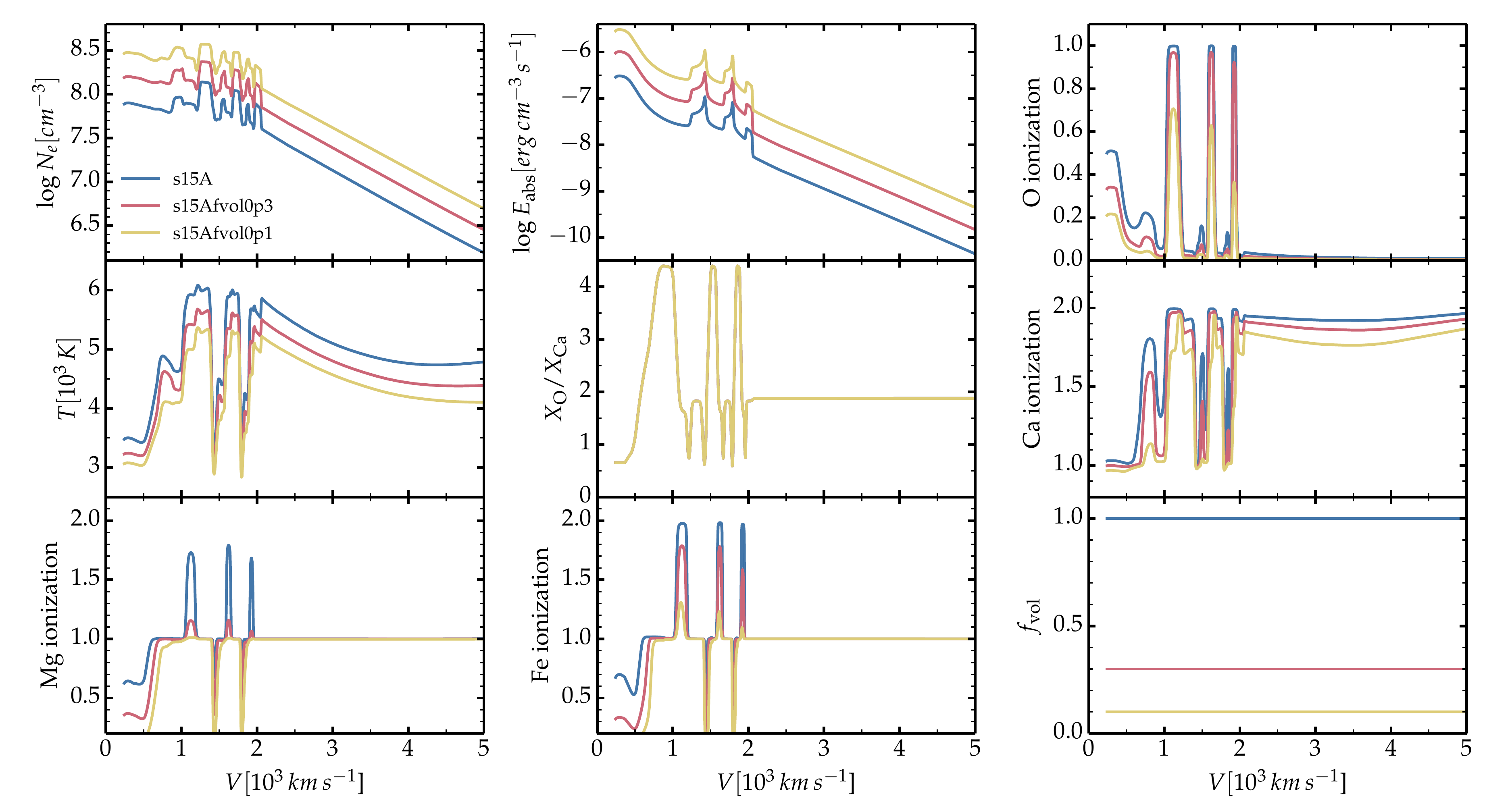}
\caption{Illustration of the influence of clumping, comparing the reference model s15A and its counterparts s15Afvol0p3 and s15Afvol0p1 in which a uniform volume filling factor of 0.3 and 0.1 is used. }
\label{fig_fvol}
\end{figure*}

Figure~\ref{fig_fvol} shows the results for model s15A in the smooth density case and in the clumped case with a uniform volume filling factor of 0.3 and 0.1. These adopted values correspond to density compressions of 3.3 and 10.0, and may be considered representative of what is seen in 3D simulations of core-collapse SNe (see, e.g., \citealt{gabler_3dsn_21}).\footnote{It is hard to be precise here. The 3D density structure of core-collapse SN ejecta is very complex, with a wide range of density compressions and rarefactions obtained in different ejecta regions or in clumps of distinct composition. Hence, there is no ``uniform'' compression factor but instead a wide distribution of clump densities and sizes. Our approach is therefore simplistic and should be considered exploratory.} The impact of this magnitude of clumping on the spectral properties is modest. As expected, clumping causes a reduction in the ionization and the temperature of the gas, but the effect is not strong. It leads to a slightly stronger Ca\two\ NIR triplet (as the Ca$^{2+}$ transitions modestly to Ca$^{+}$). The increased density in the clumped model favors collisional de-excitation, causing a weakening of forbidden lines like \oidoub\ and \caiidoub. The reduction in Mg ionization causes a strengthening of Mg\one\ lines at 4571.1\,\AA\ and 1.502\,$\mu$m. Similarly, the Fe\one\ lines in the red part of the spectrum strengthen with enhanced clumping. For the highest level of clumping, the \oidoub\ does show a sizable reduction in flux (by up to 30\% in the model with a 10\% volume filling factor).

Overall, these changes are modest. A much greater level of clumping would be required to cause a significant ionization shift and a large impact on the spectrum. One limitation for the influence of clumping is that the ionization of the species dominating the cooling in each zone is almost optimal.\footnote{One can contrast this with the high ionization in the ejecta of super-luminous SNe Ic, in which clumping has a strong influence \citep{jerkstrand_slsnic_17,d19_slsn_ic}.}  For example, O is predominantly neutral in the oxygen emitting zones (see Table.~\ref{tab_ion}). By introducing strong clumping, the \oidoub\ line strength decreases because of the enhanced recombination of other species like Na or Mg, which take over some of the cooling through Na\one\,D or Mg\one\,4751\,\AA. In the H-rich layers, Ca$^{2+}$ still dominates and prevents cooling by \caiidoub. The H-rich zones might be significantly clumped if mixed inwards into the metal-rich regions, but clumping of the H-rich material at large velocity is unlikely to be strong since there is no process to cause it.

\begin{figure*}
\centering
\includegraphics[width=0.9\hsize]{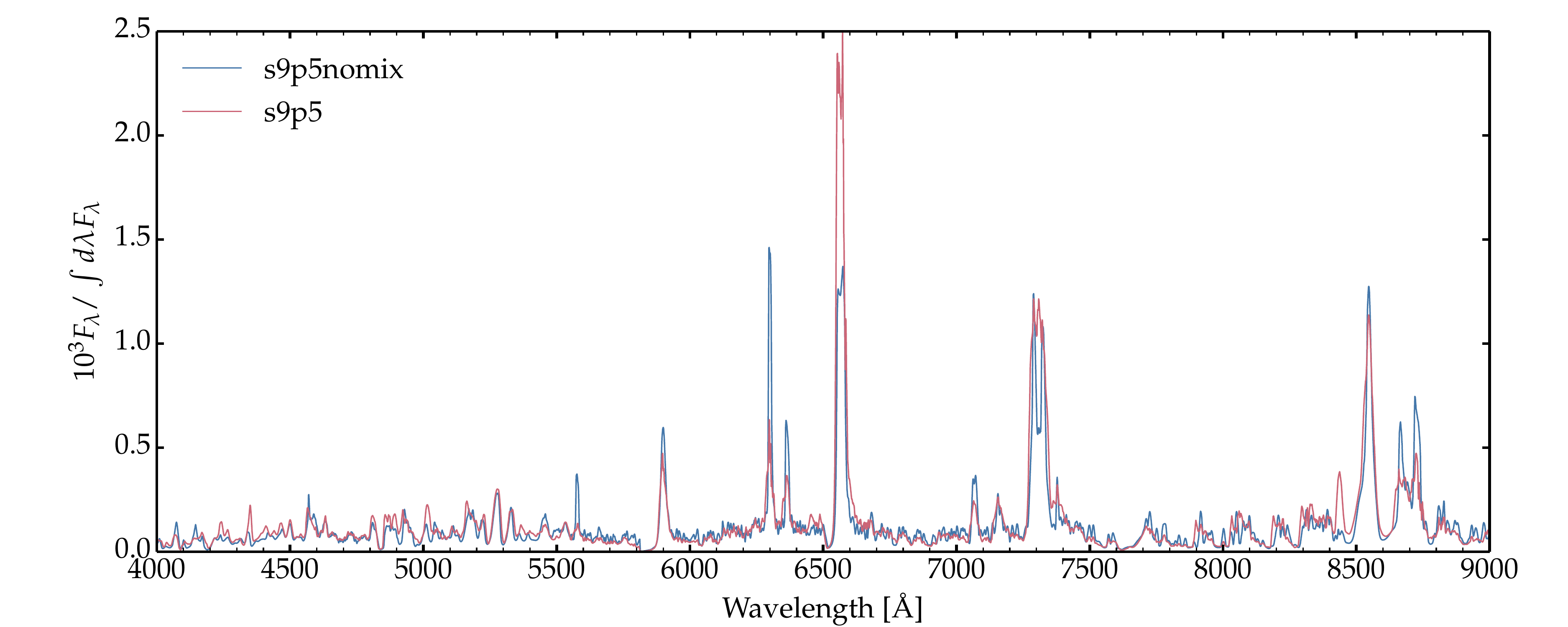}
\vspace{-0.2cm}
\caption{Comparison of model s9p5 with the standard mixing procedure produced by the shuffled-shell technique (see section~\ref{sect_setup}) and the same model in which the original unmixed ejecta has been used without adjustments (model s9p5nomix).
\label{fig_mix_nomix}
}
\end{figure*}

\section{Influence of mixing}
\label{sect_mix}

In \citet{DH20_neb}, we explored the sensitivity of nebular-phase spectra to the adopted ejecta properties, in particular to the level of chemical mixing. Although the mixing procedure was not optimal in that study (since we crafted toy models for simplicity and flexibility), we found that the mixing of \nifs\ was the primary ingredient influencing spectral properties and in particular emission line fluxes. This arises because the location of \nifs\ determines the deposition profile of the decay power, tuning the fraction delivered to other ejecta regions. In the absence of mixing in a SN II, the bulk of the decay power would be absorbed in the inner ejecta (rich in metals) and little in the outer ejecta (rich in H and He). As \nifs\ mixing is enhanced, a greater fraction of the power is deposited further out, biasing against the inner ejecta regions. If only \nifs\ were mixed while other shells remained unmixed, this could strongly impact how the ejecta cools and how the spectra appear.

In nature, one expects a significant mixing of the metal-rich core in core-collapse SNe (say below 2000\,$-$\,3000\,\kms\ in a standard-energy standard-mass SN II), in combination with the mixing of \nifs. Consequently, the near uniform deposition of decay power in the mixed metal-rich inner ejecta implies that the exact distribution of metals in the inner region is unimportant. We confirmed this expectation using our new mixing technique (i.e., the shuffled-shell approach) in \citet{DH20_shuffle}. These metal-rich regions are bathed in a sea of $\gamma$-rays, whose large mean free path allows them to fill the entire volume occupied by the inner ejecta.

The above picture is generally meant in a 1D sense -- a spherical average of an explosion should present a high level of chemical mixing. However, numerical simulations of 3D neutrino-driven explosions exhibit a variety of explosion morphologies, all suggestive of strong mixing. For example, in the simulations of \citet{gabler_3dsn_21}, the model B15 is characterized by a quasi-isotropic distribution of \nifs\ fingers while the model W15 has a few big \nifs\ protrusions but shows very weak \nifs\ mixing along some directions, sometimes covering a wide solid angle.

Our 1D treatment of mixing is probably not accurate since some radial directions may exhibit strong mixing while others may show little. Further, radiative transfer in spherical-averaged ejecta is not the same as performing 2D/3D radiative transfer in realistic 2D/3D geometries. Our approach applies to configurations in which the mixing is quasi-spherical but is inadequate for strongly aspherical composition distributions.

To test the influence of mixing in the present set of simulations, we rerun models s9, s9p5, and s11 using the initial unmixed ejecta conditions from S16. In low energy explosion models, the shock propagation may be initiated sooner after core bounce, cutting short the development of fluid instabilities that are at the origin of much of the chemical mixing (see, however, \citealt{stockinger_9msun_20}). Hence, if weak or absent mixing can occur in a general sense (i.e., be weak along all radial directions), it should be in this type of models. But weak mixing may occur along some radial direction in any SN II, so this test is useful to gauge the implications.

The influence of mixing in the model s9p5 is shown in Fig.~\ref{fig_mix_nomix} -- models s9 and s11 are not discussed since they show a similar behavior. As discussed above, we see that going from the mixed to the unmixed model, the strength of \oidoub\ and \caiidoub\ increases while the strength of H$\alpha$ decreases. This occurs because a greater fraction of the decay power is absorbed by the metal-rich inner ejecta. Emission is biased towards the inner ejecta so that numerous emission lines appear narrower than in the mixed model s9p5. This is exemplified by the \caiidoub\ line which shows distinct emission for each component of the doublet.

\section{Line profile morphology}
\label{sect_prof}

The ejecta velocity in the spectrum formation region controls the width of the emission lines that are present in the spectra. In our model set, the representative ejecta expansion rate (which we take as $\sqrt{2 E_{\rm kin} / M_{\rm ej}}$) covers the range from 1230\,\kms\ (model s9) to 3790\,\kms\ (model s12p5), thus only a factor of three. Since the spectrum forms in the inner ejecta, one may expect a smaller contrast (say a factor of two) in line width between the weakest and the strongest explosions. In addition to its influence on the intrinsic line width, the ejecta expansion rate tunes the importance of overlap with other lines, either from the same atom or ion, or from other elements.  The separation between the components of the O\one\ and Ca\two\ doublets are 3050 and 1330\,\kms, so overlap between the O\one\ doublet components will occur only in the energetic explosion models, while overlap of the Ca\two\ doublet components will occur in all our models.

\begin{figure}
\centering
\includegraphics[width=\hsize]{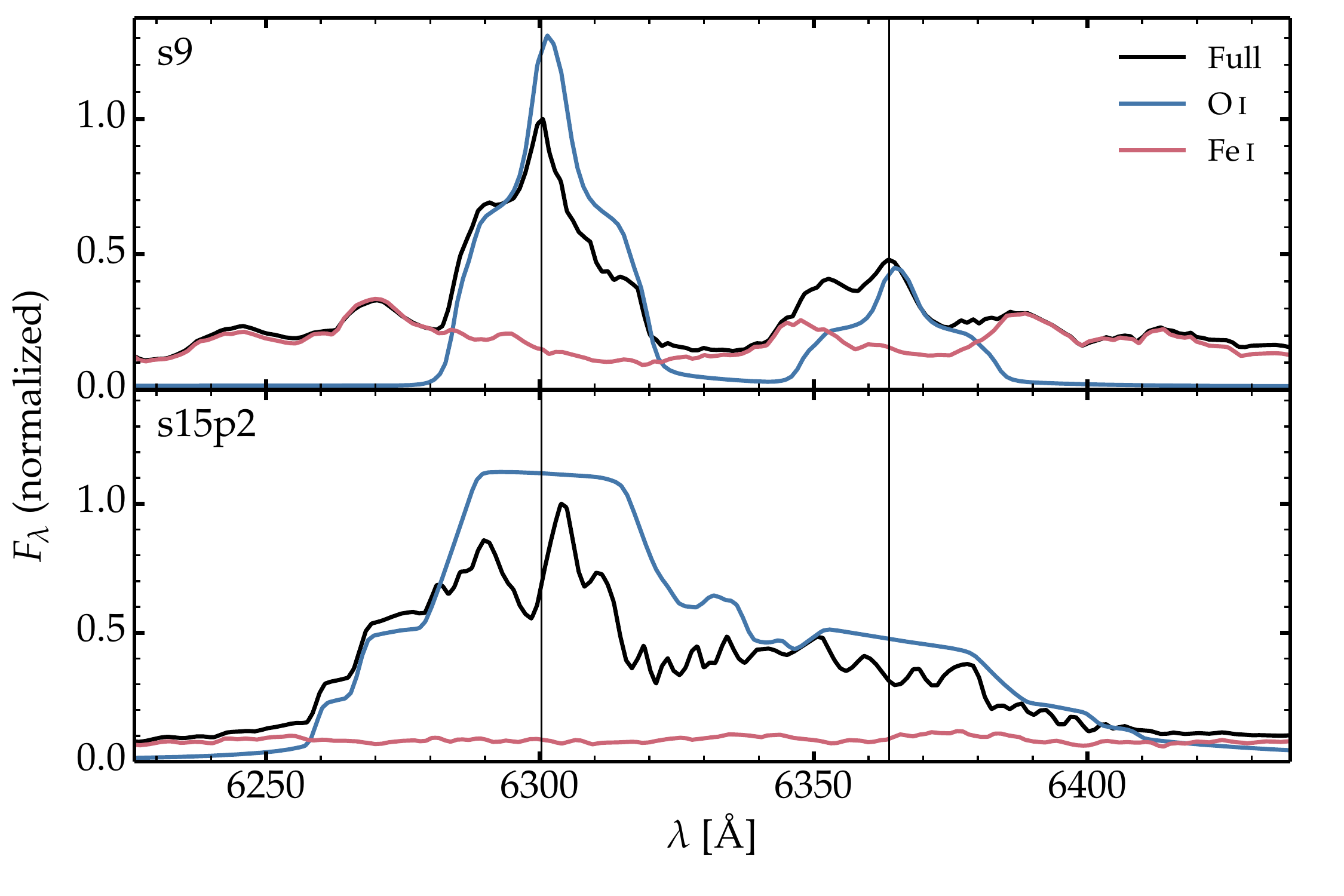}
\includegraphics[width=\hsize]{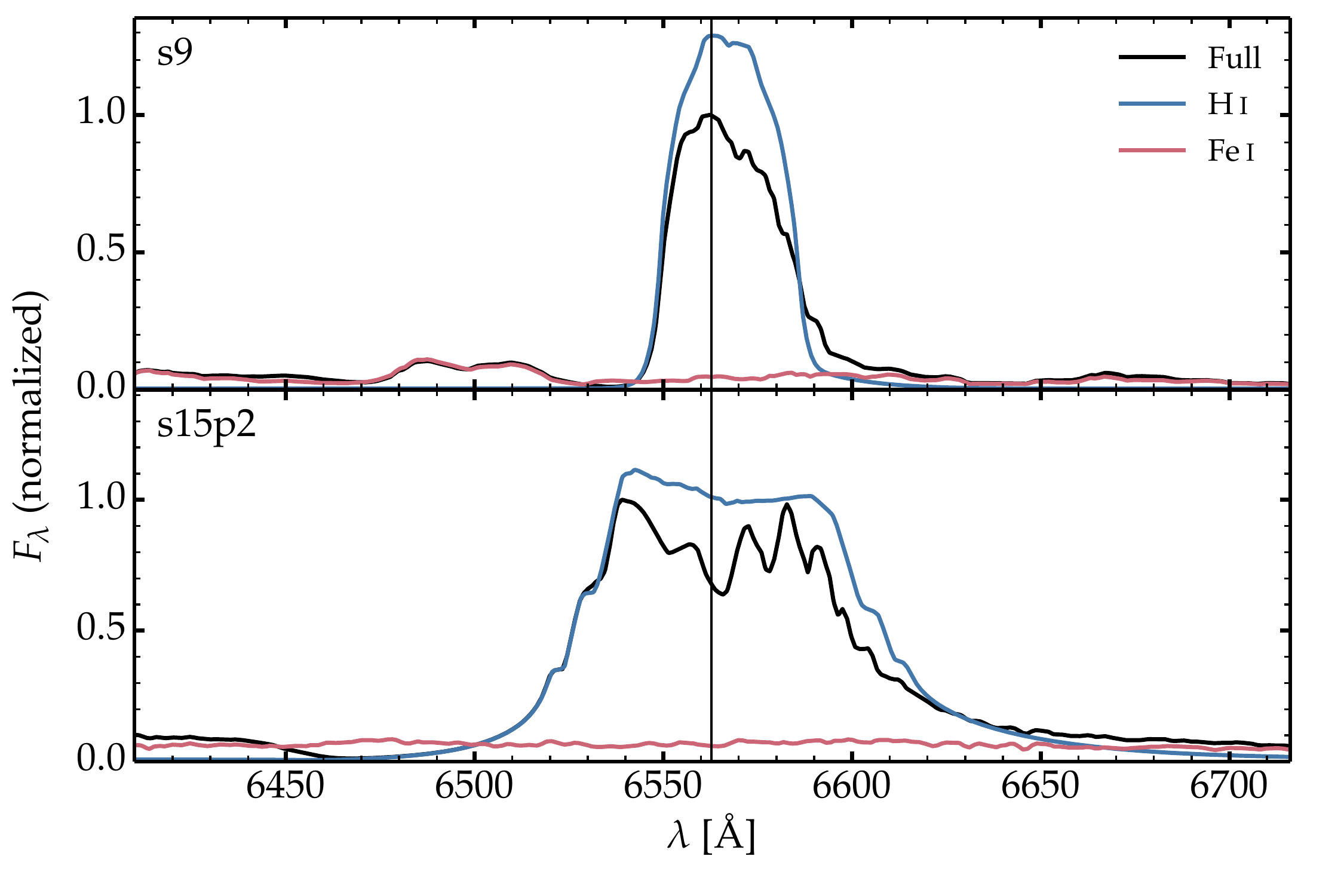}
\includegraphics[width=\hsize]{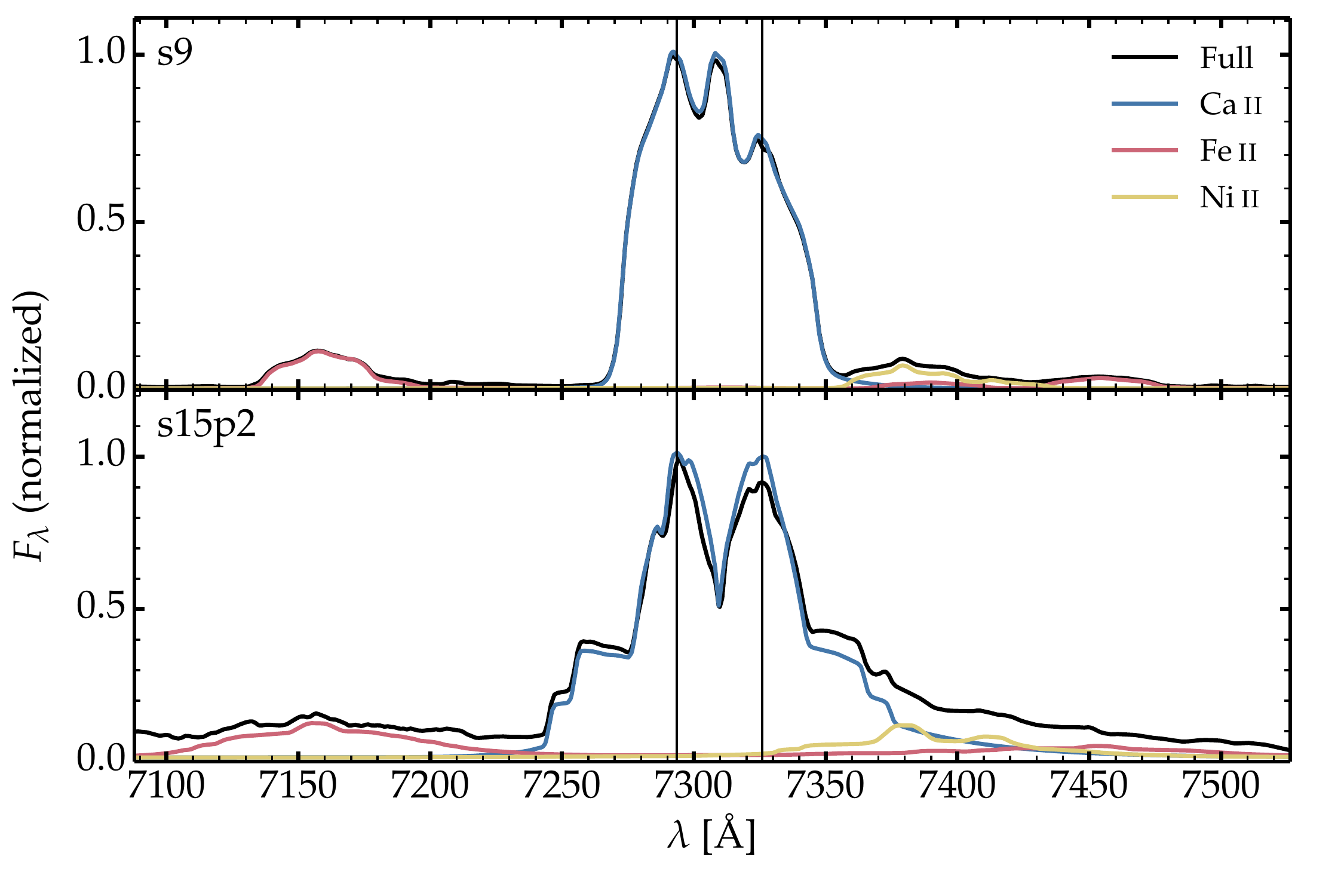}
\vspace{-0.9cm}
\caption{Montage of spectra in the \oidoub\ region (top), the H$\alpha$ region (middle), and the \caiidoub\ region (bottom) for the s9 and the s15p2 models of S16. A thin vertical line indicates the rest wavelength of each component of the doublet lines and H$\alpha$. Colored lines indicate the flux arising exclusively from bound-bound transitions associated with a specific ion (other ions and their opacity are then neglected, which explains the flux difference and the excess flux in places). Notice the presence of [Ni\two]\,7377.8 and [Ni\two]\,7411.6\,\AA\ in the bottom panels.
\label{fig_montage_lines}
}
\end{figure}

Figure~\ref{fig_montage_lines} shows a montage of spectra for the lowest energy explosion model s9 and the more standard energy explosion model s15p2, and zooming on the \oidoub, the H$\alpha$, and the \caiidoub\ spectral regions. In both models H$\alpha$ is optically thick while the other two lines are marginally thin. In model s9, the total Rosseland-mean optical depth is 0.1 (of which 85\% is due to electron scattering), so this model is essentially thin in the continuum, while in model s15p2, this optical depth is 0.4 and will affect spectrum formation. Hence, optical-depth effects cannot be neglected.

With our shuffled-shell mixing technique, the shells of distinct composition are staggered in velocity space so that optically thin emission lines forming in these shells tend to show a boxy profile with steps as one goes from the profile maximum to the blue and red wings.  These steps are easily visible in the two doublets, but less obvious in H$\alpha$ because it forms over a large ejecta volume. The artifact does not affect the total integrated line flux but only its distribution in wavelength.

In all lines shown, the peak emission occurs at or very near the rest wavelength of the line (or its doublet component). In model s9, each component of \oidoub\ is well separated but shows two distinct emissions, one occurring at very low velocity (causing a central peak) and an additional emission at larger velocity and producing a ledge on each side of the line. This \oidoub\ line flux sits on a quasi-continuum produced by a forest of overlapping Fe\one\ lines. The smaller separation of the components of the \caiidoub\ doublet line  produces a profile with three peaks due to overlap in between the two components. In model s15p2, the higher expansion rate of all emitting shells leads to broader features, with overlap even for the \oidoub\ line. Other properties are similar to what we find for model s9.

As evidenced in the top two panels of Fig.~\ref{fig_montage_lines}, the flux due to O\one\ exceeds the total flux. This is suggestive of optical depth effects, primarily due to overlap with Fe\one\ lines in model s9, but probably due to both the influence of Fe\one\ and continuum opacity in model s15p2 (the individual components show a red deficit).  The large H$\alpha$ line optical depth in both models is also at the origin of the extended blue absorption as well as the skewness of H$\alpha$. The extended H$\alpha$ trough is caused by the outer, faster expanding H-rich ejecta along the line of sight, absorbing the overlapping Fe\one\ emission. This reciprocal process (i.e., Fe\one\ also attenuates H$\alpha$ emission nearer line center) likely arises from the interweaving of H-rich and Fe-rich shells, with alternate emission and absorption in selective spectral bands.

In general, our spherically-symmetric ejecta models do not exhibit systematic wavelength shifts, relative to the rest wavelength, of the location of maximum line flux (see Fig.~\ref{fig_montage_opt}), in apparent agreement with the observations of most SNe II (see section~\ref{sect_obs}). This is in contrast with SNe Ibc, which reveal systematic shifts  (see, for example \citealt{taubenberger_ibc_09,milisavljevic_etal_10}).

\section{Comparison to observations}
\label{sect_obs}

The SN II models discussed in the preceding section have a specific set of properties, which depend on the progenitor star (primarily its mass), the final properties of the progenitor at core collapse, and its explosion as SN. While observed SNe occupy a wide parameter space in \nifs\ mass and explosion energy, our set of models define a specific pairing between progenitor mass and SN ejecta. This grid of models nonetheless permits a comparison to a sample of observed SNe. One does not expect a perfect match to any object but if the theoretical framework holds, the distribution of models and observations should compare favorably.

Our selection of observations, their inferred characteristics and associated references are given in Table~\ref{tab_obs}. It includes low-luminosity SNe II-P (SNe\,1997D, 2008bk, and 2016aqf), standard-luminosity SNe II-P (SNe\,1999em, 2004et, 2012aw, 2013ej, 2015bs, and 2017eaw), and the type II-peculiar SN\,1987A. The inferred \nifs\ mass for these observations is generally obtained by assuming full trapping and estimating the fraction of the SN luminosity that falls outside of the observed range. Combined with the uncertainty in SN distance and reddening, these \nifs\ masses are probably uncertain by a factor of two. In our simulations, our synthetic spectra exhibit mostly a bolometric offset and little variation in relative flux between different spectral regions when the \nifs\ mass varies by a factor of a few (see also Section~8 of \citealt{DH20_neb}). Hence, an uncertainty or an offset in \nifs\ mass for both observations and models is of limited impact for the discussion that follows.

\begin{table*}
\caption{
Characteristics of the selected sample of Type II SNe for the comparison to our models at 350\,d.
\label{tab_obs}
}
\begin{center}
\begin{tabular}{
l@{\hspace{4mm}}
|r@{\hspace{4mm}}c@{\hspace{4mm}}c@{\hspace{4mm}}
c@{\hspace{4mm}}c@{\hspace{4mm}}c@{\hspace{4mm}}
c@{\hspace{4mm}}c@{\hspace{4mm}}c@{\hspace{4mm}}
c@{\hspace{4mm}}c@{\hspace{4mm}}
}
\hline
        SN       &     $D$      &  $\mu$   & $t_{\rm expl}$   &    $E(B-V)$    &    $z$   & \nifs\     & References  \\
\hline
                     &     [Mpc]  &     [mag]  &     MJD [d]           &     [mag]        &      &   [\msun] &   \\
\hline
SN\,1997D          &      13.4            &      30.64 &   50362.0      &      0.020 &   4.00(-3)  & 0.002  & B01 \\
SN\,2008bk        &       3.4            &      27.68 &   54546.0         &      0.020 &   9.40(-4)   &  0.0086 & L17, M12 \\
SN\,2016aqf      &      10.8            &      30.17 &   57440.0       &      0.030 &   4.01(-3)     & 0.008 & M20  \\
\hline
SN\,1999em     &  11.5           &      30.30     &   51474.7       &      0.100    &   2.39(-3)  &  0.042 & D06, E03, L12\\
SN\,2004et        &       7.7          &      29.43     &   53270.5     &      0.300 &   9.09(-4)  &  0.109 & S06, V19 \\
SN\,2012aw      &       9.9        &      29.98      &   56002.6      &      0.074   &   2.60(-3)   & 0.074  & D14, J14\\
SN\,2013ej        &      10.2         &      30.04 &   56497.5          &       0.060 &   2.19(-3)  &  0.06 & Y16\\
SN\,2015bs       &     120.2         &      35.40 &   56920.6         &      0.045 &   2.70(-2)  &  0.049 &  A18\\
SN\,2017eaw    &       7.7            &      29.43 &   57885.7        &      0.350 &   1.33(-4)   &  0.075 & V19 \\
\hline
 SN\,1987A       &   0.05            & 18.50          &  46849.8        &      0.150    &   8.83(-4) &   0.069 & B91, P91 \\
\hline
\end{tabular}
\end{center}
{\bf Notes:} A18: \citet{anderson_15bs_18}; B01: \citet{benetti_97d_01}; B91: \citet{bouchet_87A_91}; D06: \citet{DH06}; D14: \citet{dallora_12aw_14}; E03: \citet{elmhamdi_99em_03}; J14: \citet{jerkstrand_12aw_14}; L12: \citet{leonard_99em}; L17: \citet{lisakov_08bk_17}; M12: \citet{maguire_2p_12}; M20: \citet{muller_16aqf_20}; P90: \citet{phillips_87A_90}; S06: \citet{sahu_04et_06}; V19: \citet{vandyk_17eaw_19}; Y16: \citet{yuan_13ej_16};
\end{table*}

\begin{figure}
\centering
\includegraphics[width=\hsize]{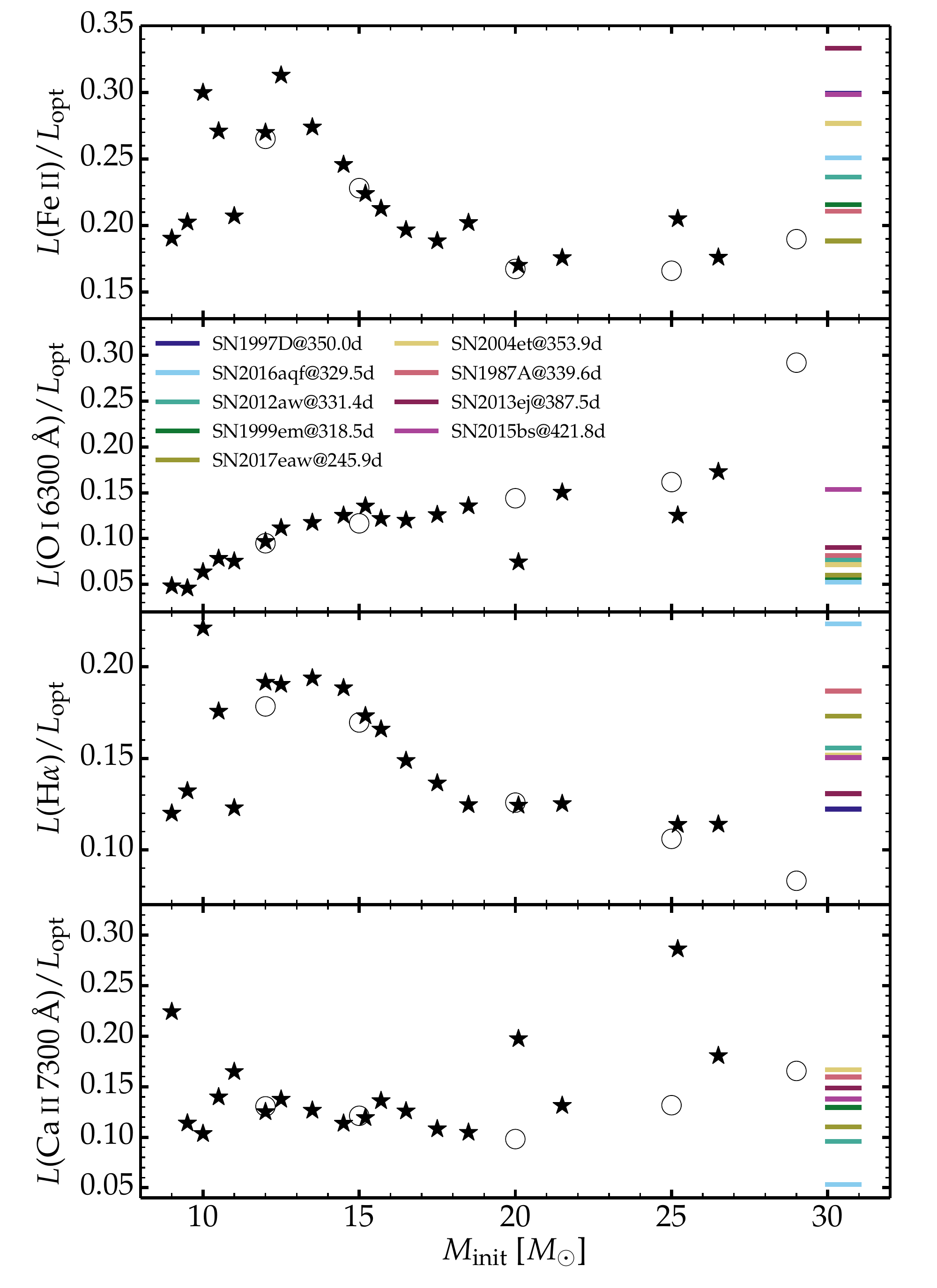}
\caption{Flux measurements for the Fe\two\ forest between 4100 and 5500\,\AA, H$\alpha$, \oidoub\ and \caiidoub, for our models (shown versus initial mass; symbols) and observations (colored bars at right). In all cases, we normalize to the total flux falling between 4000 and 9000\,\AA.
\label{fig_line_flux_obs}
}
\end{figure}

\subsection{Flux comparisons}

\subsubsection{General comments}
The SN luminosity at nebular times is set by the \nifs\ mass (modulo $\gamma$-ray escape). To cancel the offset in \nifs\ mass between our models and observations, we renormalize the spectra, so that they match at a given optical wavelength.  As the \nifs\ mass can also influence the ionization and temperature of the gas we restrict our comparisons to cases where the offset in \nifs\ mass between model and observation is moderate. For standard-luminosity SNe II, the WH07 and S16 models have a \nifs\ mass that is within a factor of two from that inferred for our selection of observations. The offset between model and observations is typically larger for low-energy SNe II, since our selected observations cover a range from 0.002 to $\sim$\,0.009\,\msun, while the \nifs\ mass in our low-energy explosion models is around 0.02 except for the model s9 (with $\sim$\,0.004\,\msun). In some cases, we opted for the better fit even if that meant using a model with a larger offset in \nifs\ mass.

The distribution of \nifs\ within the ejecta (and chemical mixing in general) can also impact line fluxes. The mixing formalism applied in our models is the same irrespective of progenitor mass: the shuffled structure in our models corresponds to a complete mixing of the metal rich core, with some mixing inwards of 1\,$-$\,2\,\msun\ of material from the progenitor H-rich envelope. This level of mixing should hold in most SN II and so our models should be in the ballpark of what is required to match observations. What is lacking in our modeling is the consideration of large scale asymmetries, which  probably affect many, if not all, Type II SNe. As large scale asymmetries require a 2D or 3D treatment they cannot  be addressed in the present work -- they are left to a future study.

Using the set of observations logged in Table~\ref{tab_obs}, we measure the flux falling between 4100 and 5500\,\AA\ to which we refer as Fe flux (it arises primarily from a forest of Fe\two\ lines, with a smaller contribution from Fe\one\ lines -- the latter are more dominant beyond about 6000\,\AA), the \oidoub\ flux, the H$\alpha$ flux, and the \caiidoub\ flux. In contrast to the modeling section~\ref{sect_quant}, we cannot easily estimate  the influence of overlapping lines, and hence we simply measure the flux of an emission feature over a fixed passband. We assume that the spectral range around a given line extends slightly beyond the half-width at half-maximum on both sides of that line, or is bounded by the location of minimum flux if there is a neighboring line (for example, as happens in between \oidoub\ and H$\alpha$).  For the three strongest optical emission lines this ensures that we account for the bulk of the flux associated with the feature while minimizing contamination by neighboring emission lines (for example the flux from Ni\two\ or Fe\two\ lines immediately adjacent to \caiidoub). When treating the models, the measurement is performed in the same manner and is therefore affected by the same bias. These flux measurements, while quantitatively different from those measured with the alternate method of section~\ref{sect_quant}, show the same trends. For the discussion, we assume that the contribution to a given line stems primarily from the associated
transition (e.g, the \oidoub) and  ignore any flux contribution by other species (e.g., Fe\one).

The flux measurements (normalized by the flux falling between 4000 and 9000\,\AA; in cases where the spectrum covers a smaller range, we extrapolate using the mean flux near the spectrum edge) are shown for the selected models and observations in Fig.~\ref{fig_line_flux_obs}. The model and observed distributions of the measurements overlap, although, as discussed in section~\ref{sect_quant}, the non-monotonicity of the trends means that a given fractional flux measurement can be compatible with both a low mass and a high mass progenitor. In a large part, the region of large scatter in models is associated with low mass progenitors, characterized by low metal yields but a wide range in explosion energy (or ejecta expansion rate). Furthermore, one needs to be cautious when comparing to observations, since one match between a SN model and an observation in one plot (for one diagnostics), might correspond to a mismatch in another. So, we will avoid over-interpreting this figure.

Excluding the lower mass progenitors, the fractional Fe and H$\alpha$ luminosity behave in a similar fashion. This suggests the bulk of the Fe emission arises from the H-rich envelope (from Fe at the primordial metallicity) rather than the \nifs-rich layers. This makes sense since the H-rich shells are much more massive than the  \nifs-rich layers (by a factor of  one hundred or more), which implies that it can absorb a large fraction of the decay power. Because of this mass offset, there is about as much Fe in the H-rich layers (where it is primarily in the form of Fe$^+$) of SN II as in the zones that used to be \nifs\ rich (where it is partially neutral). The scatter in the fractional Fe and H$\alpha$ fluxes for the models and observations is similar.


\subsubsection{Ca\two\ emission}

The models show a ``flat'' distribution of \caiidoub\ line flux as a function of progenitor mass, confirming that it is a poor diagnostics of progenitor properties. The observations fall within a narrow range for the normalized \caiidoub\ line flux, between 0.09 and 0.17 (with SN\,2016aqf standing a little off at 0.05). No observed SN in our sample exhibits a  \caiidoub\ line as strong as that obtained in models with Si-O shell merging. This process might still operate, but would require a more moderate contamination of Si-rich material in the O-rich shell. Similarly, we see that no low-luminosity SN in our sample exhibits the strong \caiidoub\ line obtained in model s9. In that model, the large strength arises because Ca is once ionized in the H-rich layers, becoming an important coolant for this material. Hence, this may support our finding that the bulk of the \caiidoub\ line emission in SNe II occurs in the Si-rich material, rather than from the H-rich zones.

\begin{figure*}
\centering
\includegraphics[width=0.495\hsize]{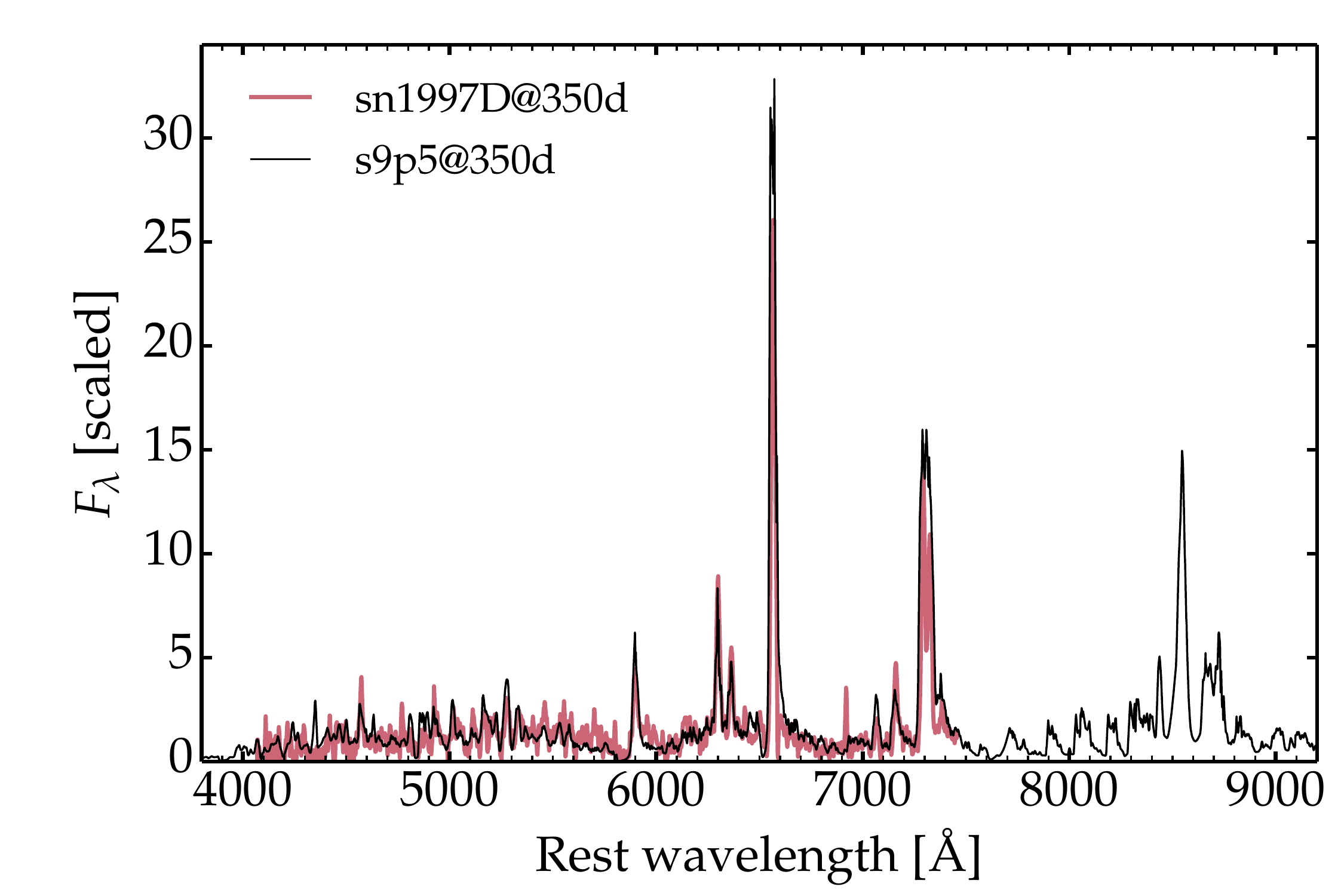}
\includegraphics[width=0.495\hsize]{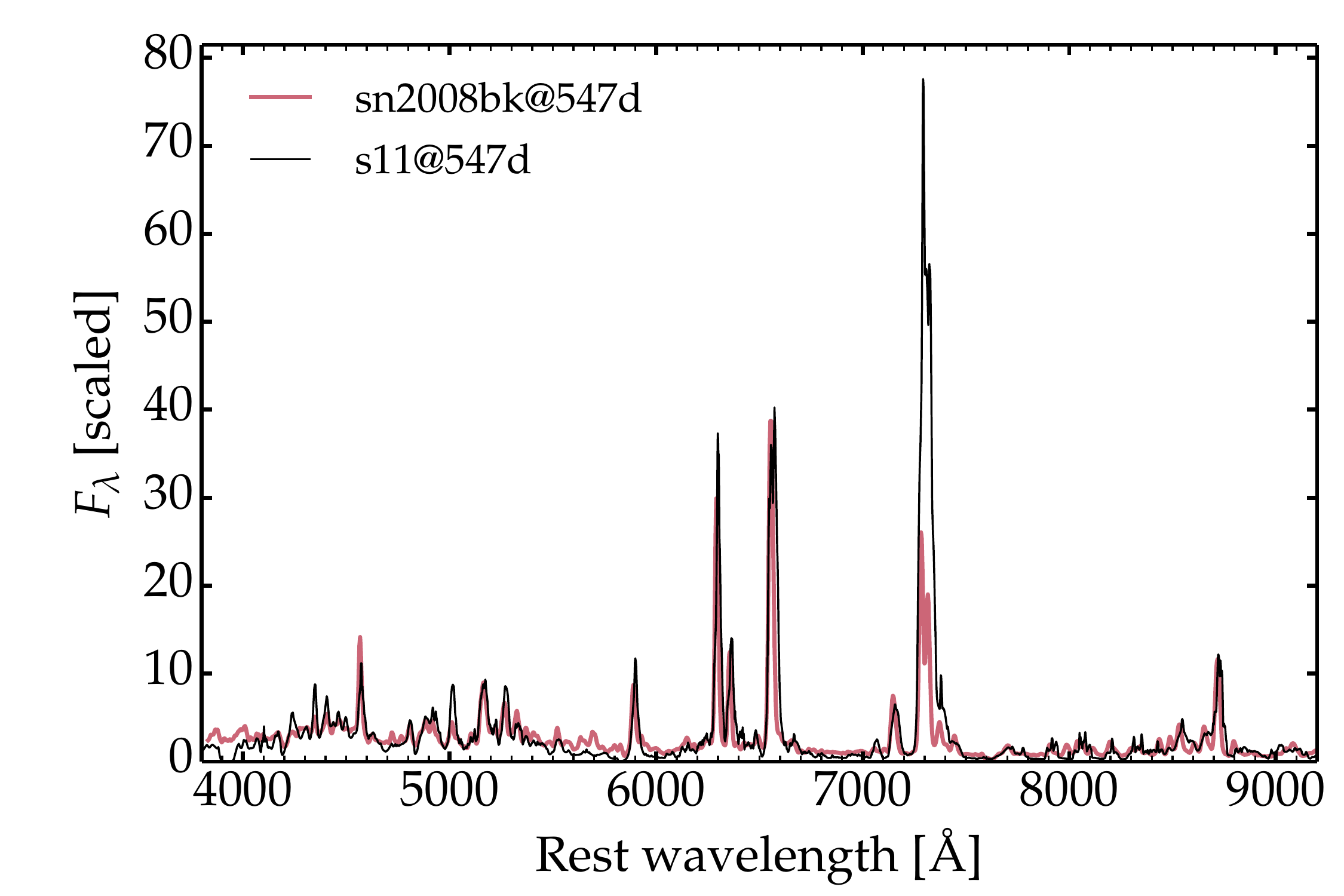}
\vspace{-0.2cm}
\caption{Left: Comparison of model s9p5 ($M(^{56}$Ni$)=0.017$\,\msun) at 350\,d with the observations of SN\,1997D ($M(^{56}$Ni$)=0.002$\,\msun) at 350\,d. Right: Comparison of model s11 ($M(^{56}$Ni$)=0.016$\,\msun) with SN\,2008bk ($M(^{56}$Ni$)=0.0086$\,\msun), both at 547\,d. The fluxes are normalized to unity at 7200\,\AA\ in both panels. The observations are corrected for reddening and redshift.
\label{fig_97D_08bk}
}
\end{figure*}

\begin{figure*}
\centering
\includegraphics[width=0.495\hsize]{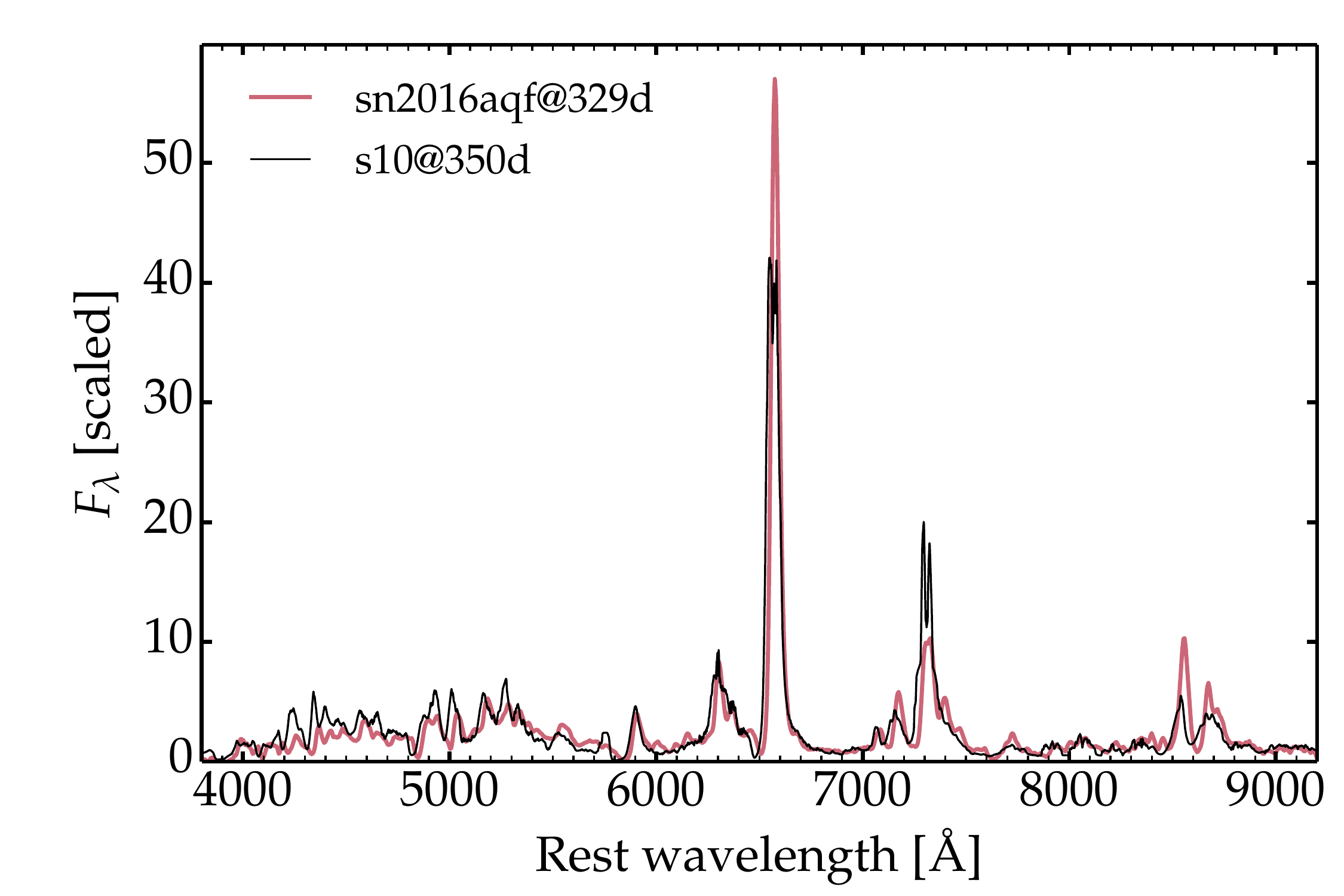}
\includegraphics[width=0.495\hsize]{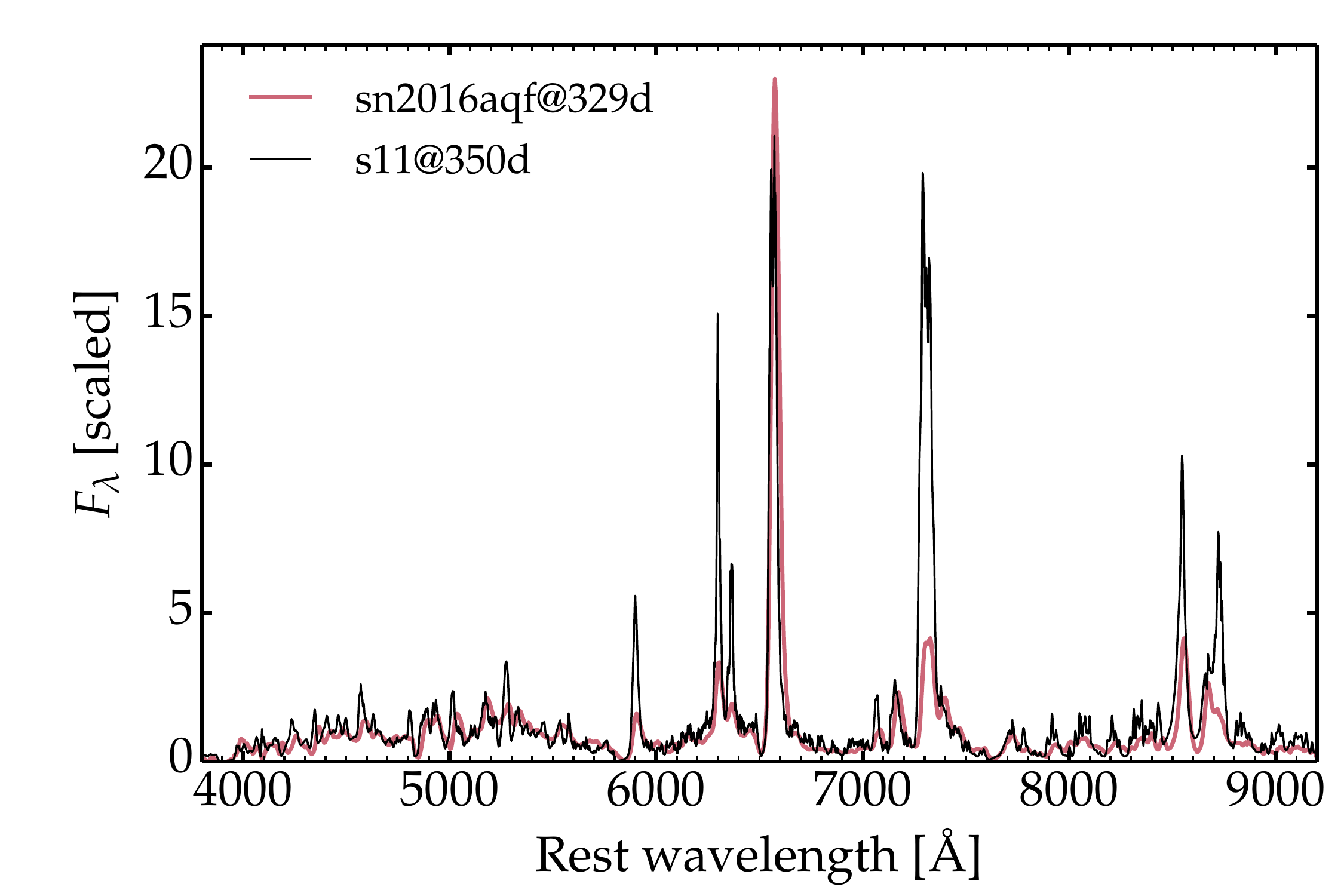}
\vspace{-0.2cm}
\caption{Same as Fig.~\ref{fig_97D_08bk}, but now for a comparison of model s10 ($M(^{56}$Ni$)=0.025$\,\msun) and s11 ($M(^{56}$Ni$)=0.016$\,\msun) to the low-luminosity SN\,2016aqf ($M(^{56}$Ni$)=0.008$\,\msun). The fluxes are normalized to unity at 6800\,\AA\ (left panel) and 5400\,\AA\ (right panel).
\label{fig_16aqf}
}
\end{figure*}

\subsubsection{O\one\ emission}

Probably the most interesting and intriguing result of our studies is that the observed fractional \oidoub\ line fluxes are compatible with our results for low mass progenitors (9\,$-$\,12\,\msun\ mass range). The one exception is SN\,2015bs which is known to require a much higher mass RSG progenitor \citep{anderson_15bs_18}. The match of observed \oidoub\ line flux ratio by lower mass RSG progenitors is perplexing. It implies that \oidoub\ can be a strong coolant even for low or moderate mass progenitors. In \citet{DH20_neb}, we reported this result for our toy models, finding that the cooling efficiency of the \oidoub\ doublet was systematically higher (although exhibiting the same trend) than obtained in the simulations of \citet{jerkstrand_ni_15}. At that time, the offset might have arisen because of the simplistic approach, the simplified composition and the adopted model atoms. Here, we use state-of-the-art explosion models as initial conditions, in particular computed with a large nuclear network. We also apply a more suitable chemical mixing, introducing no artificial microscopic mixing.

There are two obvious physical reasons for this offset. First, in Section~\ref{sect_fvol}, we found that clumping reduces the \oidoub\ line flux by as much as 30\%  for the model with  a 10\% volume filling factor. While this adopted uniform clumping is probably too large, it suggests that adopting a more realistic ejecta structure (i.e., not smooth) will reduce appreciably the current prediction for the \oidoub\ line flux.

Another reason for the offset is our neglect of molecule formation and their potential cooling power. In particular, CO, if it exists, will contribute to the cooling of the O-rich material (specifically the O/C shell, and perhaps the He/C shell if O is abundant there), reducing the amount of power radiated by \oidoub, the primary coolant for these ejecta layers (Fig.~\ref{fig_cool}). In the observations of SN\,2017eaw \citep{rho_17eaw_18}, the flux associated with CO molecules corresponds to a few percent of the optical flux, and therefore represents a sizeable fraction (a few 10\%) of the \oidoub\ line flux. If we assume that only the O/C shell is dominated by CO cooling  \cite[see, e.g.,][]{jerkstrand_12aw_14}, the \oidoub\ flux in the optical will be reduced by about 25\% in model s15p2. Near infrared observations of the fundamental and first overtone CO band are crucial for constraining the amount of CO cooling.

CO fundamental (day 117) and first overtone emission (day 112) was first detected in SN 1987A by \cite{spyromilio_co_87A_88}. Modeling by these authors suggested a CO mass of about $5\times 10^{-5}$\,\msun\  at day 285, while \cite{liljegren_co_20} estimated a much larger mass of $4\times 10^{-3}$\,\msun. Because of the reduction in temperature in the C/O region, \cite{liljegren_co_20} indicates that the O/C region will not be a significant contribution to the \oidoub\ in the optical region.  SiO was also present from $\sim$\,160\,d, with an estimated mass of about $4 \pm 2 \times 10^{-6}\,$\msun\ near day  500 \citep{roche_sio_87A_91}. Observations of the ejecta by the Atacama Large Millimeter/submillimeter Array (ALMA) reveal distinct clumpy and complex torus-like structures for both  CO and SiO \citep{abellan_87A_17} in SN 1987A. CO and SiO molecular formation is probably ubiquitous in Type II SNe after a few hundred days \citep{kotak_04dj_05,kotak_05af_06}.

 Another potential source of error in our models is explosion asymmetries. An offset in \nifs\  from the O-rich material (as in a bipolar explosion) will reduce the energy deposition in O-rich material, and potentially weaken the \oidoub\ flux fraction.

Irrespective of our models there is a very limited range in the observed \oidoub\ flux fractions. The observed ratios differ by less than a factor of three (a factor of two if we exclude SN\,2015bs). This can be compared with our models where the change in the flux ratio is a factor of four, with a factor of two change between progenitor masses of 9 and 12\,\msun.

While determining an accurate progenitor mass is a goal of nebular modeling it is worth mentioning the
weak dependence of the \oidoub\ flux fraction on the oxygen mass. While the oxygen mass varies
by a factor of 50  in our models, and the oxygen mass fraction in the ejecta by a factor of 25, the increase
in the \oidoub\ flux fraction is only a factor of 4. This insensitivity is not surprising -- it arises because the
emitted energy is set by the initial mass of \nifs, and  because \oidoub\ is an important coolant, but only in the
oxygen-rich shells. One can use the absolute  \oidoub\ flux to estimate the oxygen mass, but this requires an accurate temperature estimate, and such an estimate is not easily available from the observations -- it typically requires modeling. In principal [O\one]\,$\lambda$\,$5577$ can be used with the  \oidoub\ to constrain the temperature, but at the epochs considered in this paper it is very weak and badly blended.

\subsection{Specific comparisons to a sample of Type II SNe}

\subsubsection{Comparison to  low-luminosity SNe II 1997D, 2008bk, and 2016aqf}

Figures~\ref{fig_97D_08bk} and \ref{fig_16aqf} present a comparison of models s9p5, s11, and s10 with the observed optical spectra of low-luminosity SNe II 1997D, 2008bk (the model was computed for the same epoch as the observations, that is 547\,d after the inferred time of explosion), and 2016aqf. Unless specifically stated, all models correspond to a SN age of 350\,d. Rather than give a magnitude offset between models and observations, we just state the model \nifs\ mass and that inferred for the observed SN. The two spectra shown are normalized at some wavelength (chosen to improve the visibility and facilitate the comparison; this wavelength is often taken at 6800\,\AA, away from strong lines, in a region dominated by a smooth background flux dominated by Fe\one\ emission).

These low-energy explosions in lower mass progenitors ($E_{\rm kin}$ in the range 2 to $6 \times 10^{50}$\,erg) provide a suitable match to the observations, which are characterized by relatively narrow lines, strong H$\alpha$, weak metal lines. The model s9p5 yields a satisfactory match to the observations of SN\,1997D, although its \nifs\ mass is ten times greater. Alternatively, we could have used model s9, whose \nifs\ mass is only twice that inferred for SN\,1997D. Model s9 yields a very good fit (not shown) to most of the spectral features of SN\,1997D, but strongly overestimates the \caiidoub\ because Ca is mostly Ca$^+$ in the H-rich layers (see Table~\ref{tab_ion}). It is unclear what conditions lead to a modest \caiidoub\ in SN\,1997D. Ca over-ionization is not expected for a low \nifs\ mass. A lower metallicity might explain in part this property. This requires further study.

The model s11 at 350 and 547\,d does a poor job at matching the \caiidoub\ of SN2016aqf and SN\,2008bk, but the rest of the spectra are reasonably well fitted. Model s11 overestimates the strength of \oidoub\ and \caiidoub\ compared to SN\,2016aqf perhaps because this model predicts a significant contribution from the H-rich material. This contribution may therefore not hold (this extra flux at larger velocity also leads to a mismatch in line width, though this could be cured by using a weaker explosion). The emission lines in the observed spectrum of SN\,2016aqf \citep{muller_16aqf_20} also seem redshifted relative to the model, which may point to a large-scale asymmetry of the \nifs\ in that ejecta.

The association of low or moderate mass massive stars exploding as weak explosions with low-luminosity SNe II-P confirms the results from previous studies on SN\,1997D \citep{chugai_utrobin_97D_00}, SN\,2008bk \citep{maguire_2p_12,lisakov_08bk_17}, or SN\,2016aqf \citep{muller_16aqf_20}. The S16 model s9 was also studied in detail by \citet{jerkstrand_9msun_18} and found to compare favorably with the observations of low-luminosity SNe  1997D, 2005cs, or 2008bk. This result seems robust, is corroborated by photospheric phase properties, and appears to hold for the whole class of such events \citep{lisakov_ll2p_18}.

\subsubsection{Comparison to standard-luminosity SNe 2012aw and 2013ej}

\begin{figure}
\centering
\includegraphics[width=\hsize]{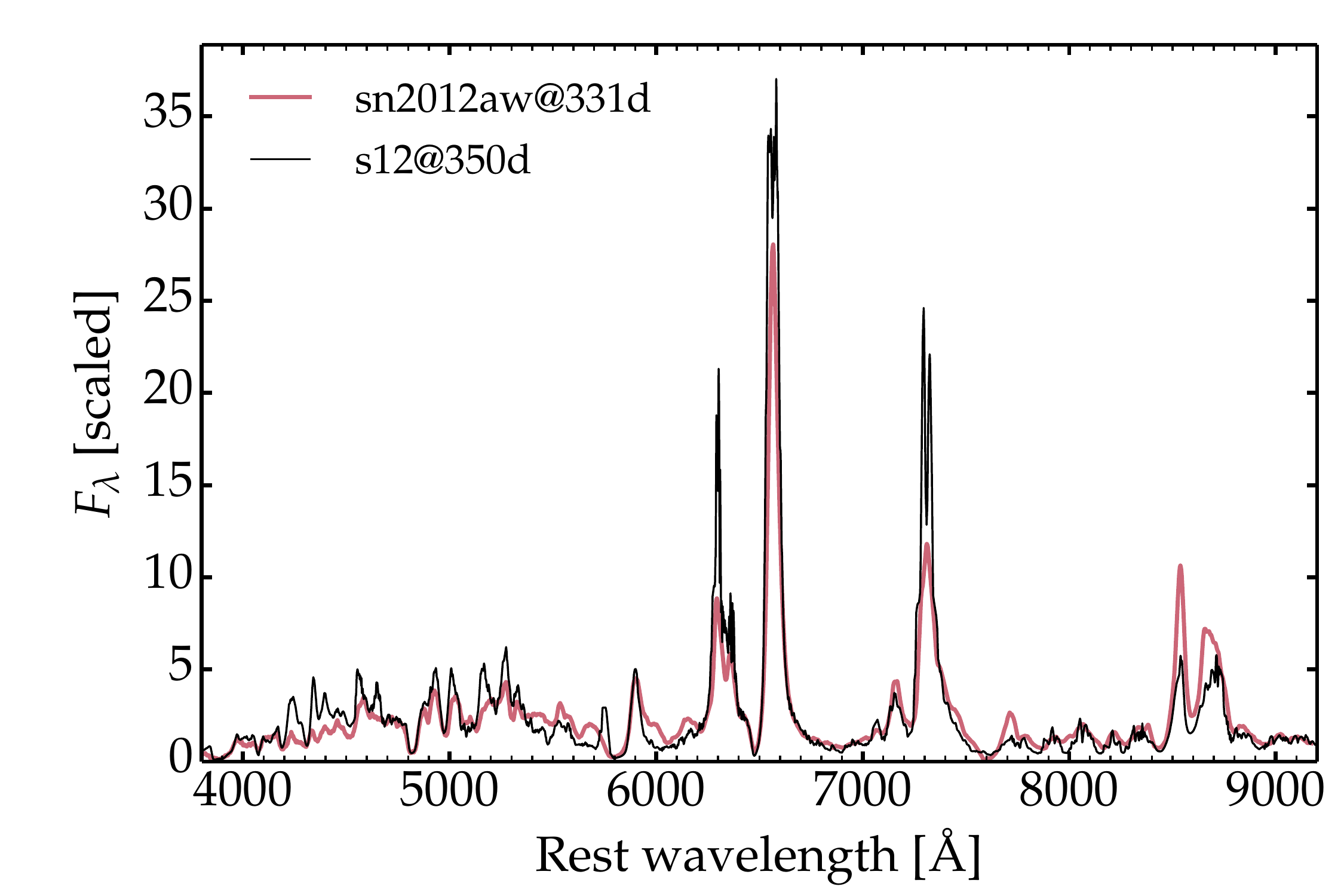}
\caption{Same as Fig.~\ref{fig_97D_08bk}, but now for a comparison of optical spectra between model s12 ($M(^{56}$Ni$)=0.032$\,\msun) and SN\,2012aw ($M(^{56}$Ni$)=0.074$\,\msun). The fluxes are normalized to unity at 6800\,\AA.
\label{fig_12aw}
}
\end{figure}

Figure~\ref{fig_12aw} compares the optical spectra of SN\,2012aw and model s12. The model matches well the Fe\two\ emission forest, the strong O\one, H\one, and Ca\two\ lines. The line profiles are also well matched in width, although the model emission has structure not obviously present in the observations. The model H$\alpha$ lacks a central narrow peak, likely because we do not extend our 1D ejecta models to low-enough velocities (the absence of material at low velocity is an artifact of performing the explosion in spherical symmetry). The model \oidoub\ doublet line has instead narrow peaks at the rest wavelength of each component, while the observed profile is smoother. This arises because the O-emitting material is more randomly distributed in space than adopted in our simplified shuffled-shell structure. The representative velocity of the O\one-emitting volume, as estimated from the full width at half maximum, is however well matched. Probably for the analogous reason, the Fe\one\ emission lines in the 8000\,\AA\ region are also too resolved in the model, something that could be remedied by a more extended volume of emission from Fe\one\ (this Fe\one\ emission comes primarily from the original \nifs-rich material, which is characterized by a lower Fe ionization level compared to the H-rich material in which Fe is once ionized; see Table~\ref{tab_ion}). This is an obvious, though not critical, shortcoming of using a shuffled-shell structure in place of a more realistic and complex 3D distribution of \nifs.

\begin{figure}
\centering
\includegraphics[width=\hsize]{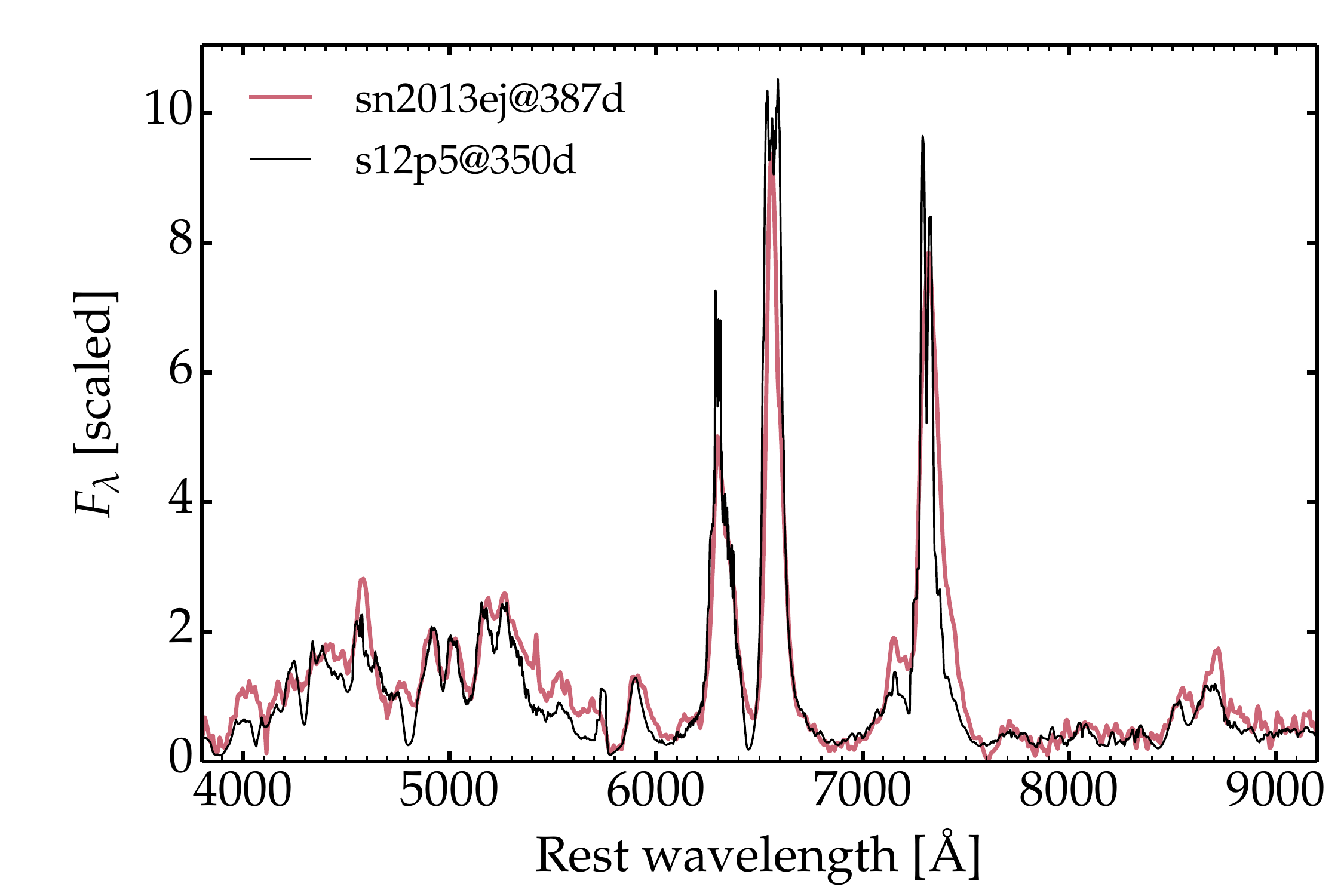}
\vspace{-0.4cm}
\caption{Same as Fig.~\ref{fig_97D_08bk}, but now for a comparison of model s12p5 ($M(^{56}$Ni$)=0.05$\,\msun) to SN\,2013ej ($M(^{56}$Ni$)=0.06$\,\msun). The flux normalization is done at 5100\,\AA.
\label{fig_13ej}
}
\end{figure}

Figure~\ref{fig_13ej} compares model s12p5 with SN\,2013ej. The particularity of this model is its high $V_m$, the largest of our model grid, combined with its modest metal yields, typical of a low-mass massive star. This is conducive to a lower-density ejecta and faster expansion in the emitting region, favoring the formation of forbidden transitions and the production of broad emission lines. These characteristics seem to match well the properties of the optical spectrum of SN\,2013ej. For example, the Fe\two\ emission below 5500\,\AA\ is quite strong relative to the strongest emission lines, which is better explained by a lower mass progenitor like s12p5. The \caiidoub\ lines width is underestimated (but other lines seem well matched in width), which might indicate that some Ca\two\ emission arises from H-rich material at large velocities (something that is prevented in most of our models because Ca is twice ionized in such regions).

Previous studies of SNe\,2012aw and 2013ej have yielded inconsistent estimates for the ZAMS mass
of their progenitors.  Nebular phase modeling of 2012aw by  \cite{jerkstrand_12aw_14} suggest a ZAMS
mass of 15\,\msun\ for the progenitor, while \cite{yuan_13ej_16} suggest 12\,$-$\,15\,\msun\ for 2013ej.
Radiation-hydrodynamics modeling of the bolometric light-curve combined with constraints on the photospheric velocity inferred from Doppler-broadened line profiles have produced higher mass estimates for SN\,2012aw. \cite{dallora_12aw_14} infer a 20\,\msun\ ejecta, pointing to a massive main sequence star of perhaps 25\,\msun. Similarly, \citet{morozova_sn2p_18} constrain a ZAMS mass of 20\,\msun\ and \citet{nikiforova_12aw_21} a ZAMS mass of 25\,\msun. Using nonLTE radiative transfer modeling of the multi-band light curves and optical spectra of SNe 2012aw and 2013ej, \citet{HD19} find that a 15\,\msun\ progenitor (characterized by different pre-SN mass loss rates) with a standard explosion energy can yield a satisfactory match to observations. Given the degeneracy of light curve modeling \citep{d19_sn2p,goldberg_sn2p_19}, it is not surprising that there is a non-unique solution to the progenitor mass with this approach. The same scatter in inferred progenitor mass from light curve modeling holds for both SN\,2012aw and SN\,2013ej.

\subsubsection{Comparison to the Type II-pec SN\,1987A}

\begin{figure*}
\centering
\includegraphics[width=0.495\hsize]{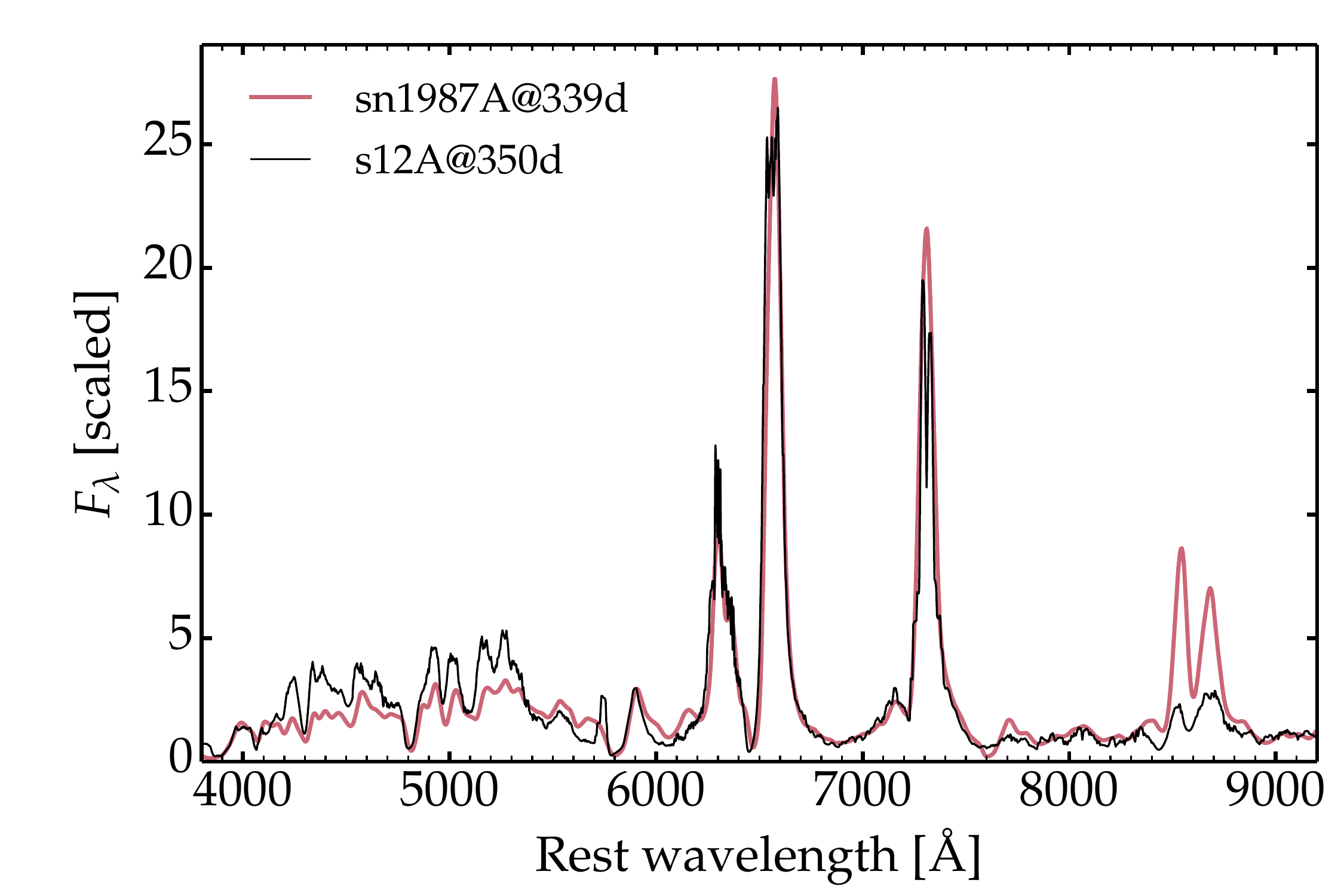}
\includegraphics[width=0.495\hsize]{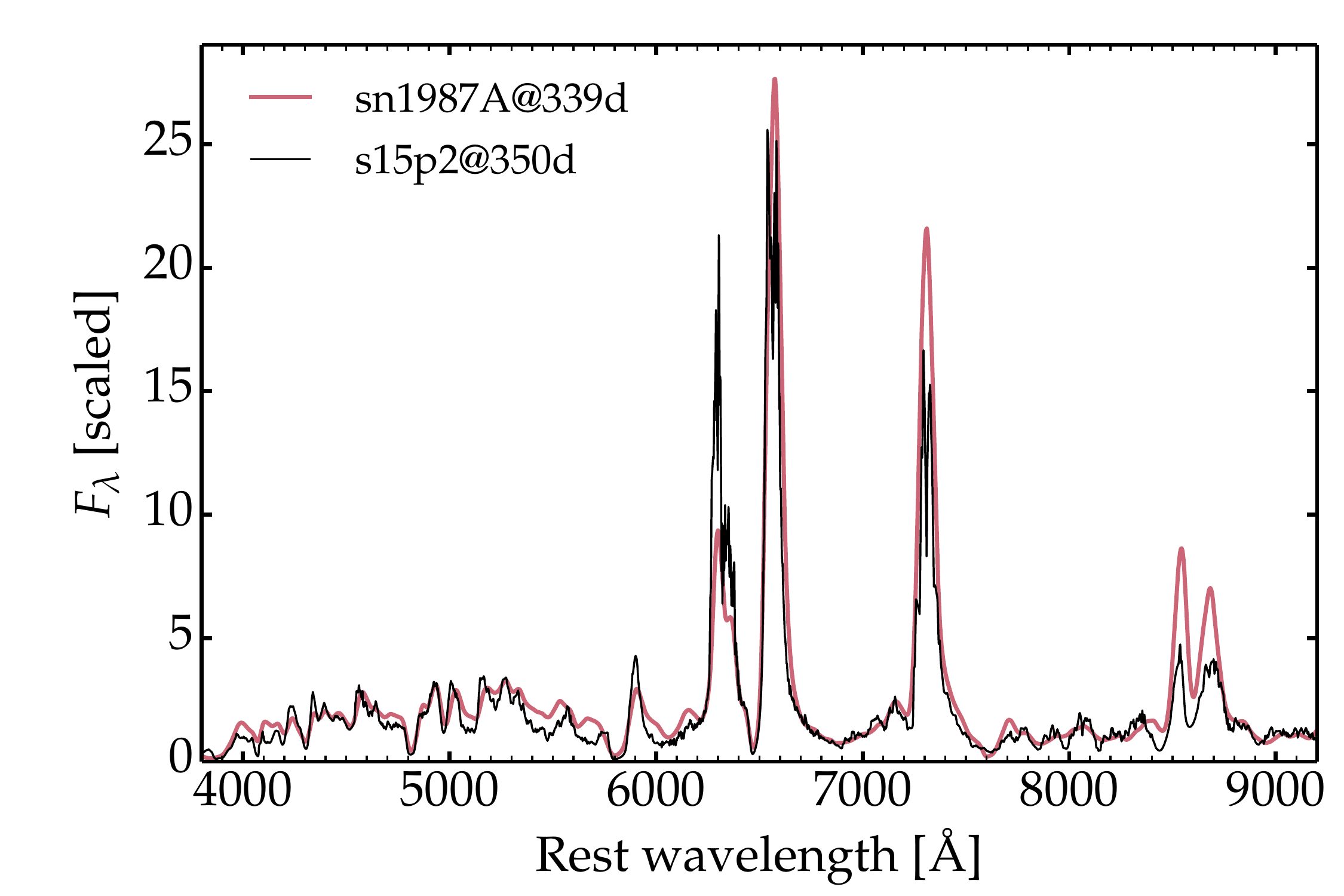}
\caption{Same as Fig.~\ref{fig_97D_08bk}, but now for a comparison of model s12A ($M(^{56}$Ni$)=0.05$\,\msun) and s15p2 ($M(^{56}$Ni$)=0.063$\,\msun) to SN\,1987A ($M(^{56}$Ni$)=0.069$\,\msun). Both models adopt a solar metallicity, not exactly consistent with the LMC metallicity for SN\,1987A. The lower mass progenitor does a
good job at matching the 3 strongest lines,  but severely underestimates the Ca\two\ triplet, and
does a poor job with at matching the Fe emission forest shortward of 6000\,\AA. By contrast, the higher mass model does a better job with the Fe emission, overestimates the O\one\ doublet,  and underestimates the Ca\two\ line strengths. The flux normalization is done at 6800\,\AA.
\label{fig_87A}
}
\end{figure*}

In Fig.~\ref{fig_87A}, we compare models s12A and s15p2 to the Type II-peculiar SN\,1987A.
Although both models offer a reasonable match they also have specific discrepancies. In model s12A, the \oidoub, H$\alpha$, and \caiidoub\ lines are well matched but the Fe\two\ flux below 5500\,\AA\ is overestimated. This probably arises from the greater fraction of decay power absorbed by the H-rich material relative to the O-rich and Si-rich material (which also stems from the greater $V_m$). The Ca\two\ triplet is underestimated by at least a factor of three, perhaps somewhat surprising given that our progenitors have solar metallicity. In contrast, in model s15p2, the Fe\two\ flux is weaker and better matched, but then the strength of \oidoub\ is overestimated. These subtleties highlight the difficulty of inferring accurately the ejecta properties and constraining the progenitor mass.

This comparison is not fully consistent since our models are at solar, rather than LMC metallicity, and were red supergiants (rather than blue supergiants) when they exploded. The actual progenitor of SN\,1987A may have had a mass closer to 16\,$-$\,20\,\msun\ (see, for example, \citealt{jerkstrand_87a_11}; S16), depending on rotation. We do not think the resulting small difference in core structure matters for the spectrum at 1 year. A more important shortcoming in this comparison is the neglect of clumping and CO molecular cooling.

\subsubsection{Comparison to the low-metallicity SN\,2015bs}

\begin{figure*}
\centering
\includegraphics[width=0.495\hsize]{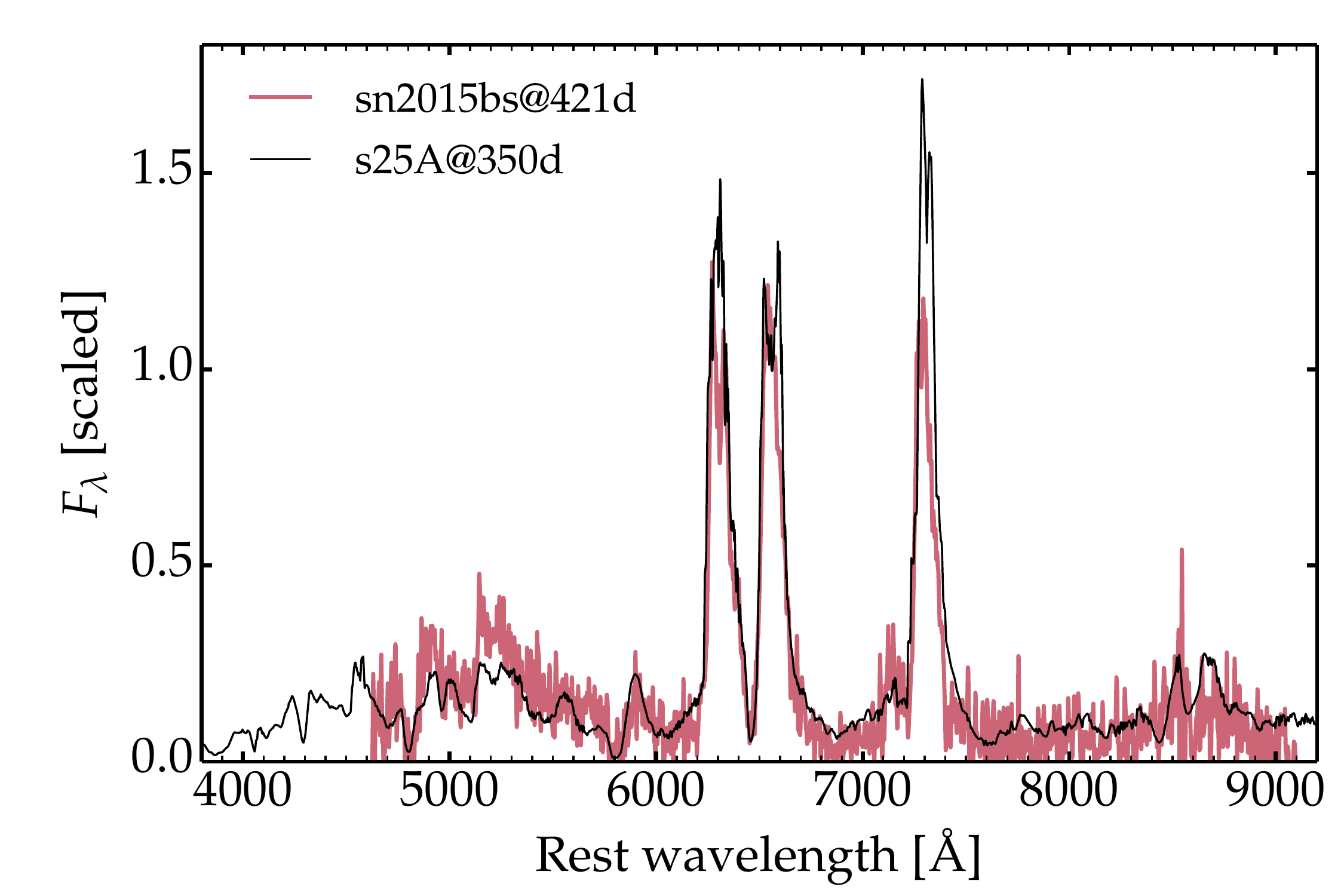}
\includegraphics[width=0.495\hsize]{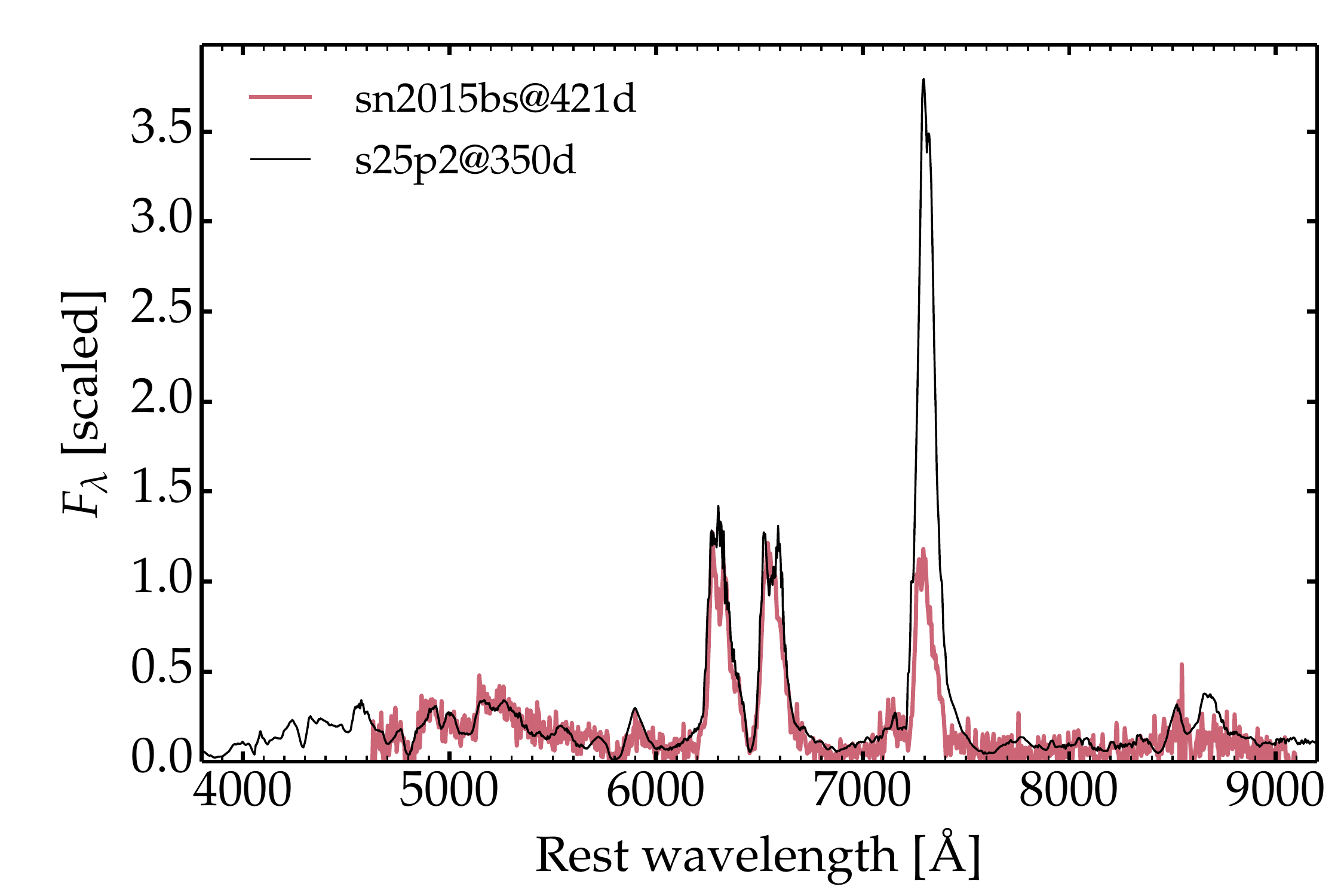}
\vspace{-0.2cm}
\caption{Same as Fig.~\ref{fig_97D_08bk}, but now for the comparison between model s25A ($M(^{56}$Ni$)=0.157$\,\msun; left) and s25p2 ($M(^{56}$Ni$)=0.112$\,\msun; right) and observations of SN\,2015bs ($M(^{56}$Ni$)=0.048$\,\msun). Both models adopt a solar metallicity, not exactly consistent with the one tenth solar metallicity inferred for SN\,2015bs \citep{anderson_15bs_18}. The flux normalization is done at the H$\alpha$ rest wavelength in both panels. Some gaussian smoothing has been applied to the observed spectrum to reduce the level of noise.
\label{fig_15bs}
}
\end{figure*}

Figure~\ref{fig_15bs} presents the spectral comparison for models s25A and s25p2 at 350\,d with the observations of SN\,2015bs at 421\,d after explosion. SN\,2015bs is remarkable in two ways, first because of its location in a very low metallicity environment of about a tenth solar, and second because of its exceptional nebular spectrum exhibiting the strongest \oidoub\ line flux of all SNe II for which such data exist \citep{anderson_15bs_18}. The later can only be matched by a higher-than-standard progenitor mass. Here, we show that model s25A reproduces satisfactorily the entire optical spectrum of SN\,2015bs (although the spectrum is noisy and prevents a proper comparison of weak lines). As expected, the model s25p2, characterized by Si-O shell merging prior to explosion, exhibits a huge \caiidoub\ flux, in conflict with observations. We thus confirm the previous conclusions reached with more simplistic models \citep{DH20_neb}, as well as the previous work from \citet{anderson_15bs_18}.

\section{Conclusions}
\label{sect_conc}

We have presented nonLTE radiative transfer  calculations based on state-of-the-art explosion models from WH07 and S16. Rather than trying to adequately adjust the models to match a specific observation, we simply compared the WH07 and S16 models to observations, without any scaling of their yields, ejecta mass, or ejecta kinetic energy. By retaining the global properties of the original models, our work provides the nebular-phase observables characterizing the ab initio explosion models. Although this approach lacks flexibility, it allowed us to construct a comprehensive grid of nebular phase spectra for the explosion of stars with an initial mass between 9 and 29\,\msun.

 Mixing was treated with a shuffled-shell technique \citep{DH20_shuffle}, with the ansatz that the metal-rich core is fully mixed macroscopically and that 1\,$-$\,2\,\msun\ of material from the H-rich envelope is mixed into the metal rich core. While this 1D shuffling implies an adjustment to the initial ejecta models of WH07 and S16, it treats the chemical mixing that must somehow be accounted for to reflect the 3D mixing witnessed in observations \citep{abellan_87A_17} and systematically present in 3D simulations of the neutrino-driven explosion mechanism of massive stars \citep{gabler_3dsn_21}.

The general trend of increasing explosion energy, metal yields, and \nifs\ mass with increasing progenitor mass leads to nebular phase spectra of increasing brightness, line width, and line strength. The low energy explosion models from 9\,$-$\,11\,\msun\ stars match some of the well known low-luminosity SNe II-P and confirm the widespread notion that these faint SNe II-P arise from the lowest mass massive stars undergoing Fe-core collapse. Their low metal yield is an intrinsic characteristic of both the progenitor and the ejected material rather than the result of fall back in a high mass progenitor. This results in weak and narrow \oidoub, but a very strong narrow H$\alpha$ testifying for the large fraction of decay power absorbed by H-rich material. Allowing for binarity, the progenitors of low-luminosity SNe II-P are expected to extend below 9\,\msun\ \citep{zapartas_bin_17}. Stars with an initial mass above about 12\,\msun\ that produce an ejecta kinetic energy of about 10$^{51}$\,erg are broadly compatible with standard-energy SNe II-P like SN\,2012aw or 1987A. The ab-initio explosion models of S16 therefore depict a landscape of progenitors and explosions that are broadly consistent with observations of SNe II-P.

However, within this general picture, our results reveal diversity and scatter, even for models that differ little in progenitor mass. This scatter takes its origin from the somewhat chaotic pattern of kinetic energy, \nifs\ mass, and envelope composition exhibited by the S16 models. In the present grid of models, one ingredient that impacts drastically the spectral properties is the Ca ionization in the H-rich material. In general,  Ca$^{2+}$ dominates and this relatively high ionization quenches the \caiidoub\ line emission from the H-rich layers. In that case, \caiidoub\ arises primarily from the Si-rich layers. However, Ca$^+$ may dominate in H-rich layers when both the explosion energy and the \nifs\ mass are low (as in model s9 from S16), causing  very strong \caiidoub\ line emission. Observations suggest that this is rarely the case, and hence indicate that \caiidoub\ emission is generally impaired by Ca overionization. Although not explored in this work, a  low primordial metallicity could contribute to inhibit metal-line emission from the H-rich envelope.

A second ingredient is Si-O shell merging in the progenitor prior to explosion which may boost the Ca abundance in the O-rich shell by a factor of 100. \caiidoub\ then becomes an important  coolant for the O-rich material, weakening \oidoub\ emission and strengthening \caiidoub\ emission. The magnitude of the effect, seen in models s20p1 and 25p2, is very strong, and typically incompatible with observations. However, one cannot exclude that it occurs in a milder form, thus allowing a higher mass progenitor to mimic the spectral appearance of lower mass progenitor stars, which are characterized by a weaker \oidoub.

Most of the erratic behavior concerns \caiidoub\ emission. In contrast, the flux associated with Fe\two\ emission shortward of 5500\,\AA, the \oidoub\ or the H$\alpha$ line flux exhibit a more consistent evolution with progenitor mass. Excluding the lower mass progenitors (which are easily identified -- see above), models exhibit a strengthening \oidoub\ and a weakening H$\alpha$ with increasing progenitor mass, reflecting the corresponding trend of increasing (decreasing) O (H) mass. While these evolutions are robust, there are not generally strong enough to separate two models unless they differ in progenitor mass by several \msun. The most vivid example of this evolution is the extreme case of model s29A, our highest mass progenitor with a light residual H-rich envelope.

In general, our models produce a \oidoub\ line flux that is stronger than previously obtained with \sumo\ \citep{jerkstrand_ni_15}, as already reported in \citet{DH20_neb}. Generally we can explain the nebular-phase spectra of observed standard-energy SNe II with a progenitor in the mass range 12 to 15\,\msun. This offset might indicate that we are missing a coolant from the O-rich zone. One possibility is molecular cooling, which is presently ignored. Clumping is also found to reduce the \oidoub\ line flux. Another possibility is the slight contamination of Ca into the O-rich shell, which only occurs (but then strongly) in two of the higher mass progenitor models. A combination (and perhaps all) of these effects probably act at some level in Type II SN ejecta.

We continue to find that SN\,2015bs is the current best evidence that higher mass RSG stars of 20\,$-$\,25\,\msun\ can explode. One could wonder whether SiO and CO  cooling could affect this inference. This is unlikely because the O/Si and C/O shells, where these molecules may form, represent a smaller fraction of the O-rich shell mass in higher mass progenitors: a growing fraction of the decay power will thus be absorbed by the O/Ne/Mg shell, in which such molecules do not form (see bottom panel of Fig.~\ref{fig_prop} for the O/C shell mass in our model set).

Metal line emission from the H-rich material suggests that the \oidoub\ and \caiidoub\ lines could be stronger (weaker) at higher (lower) metallicity, in particular in lower-mass progenitors because they are characterized by low metal yields. This applies in particular to Fe\one, Fe\two, and Ca\two\ line emission.

Our model grid suggests that at 350\,d, about 70\% of the total decay power absorbed is radiated between 3500 and 9500\,\AA. The fraction of the decay power emitted that escapes the ejecta varies between models, being a few percent at most in weaker explosion and at most 35\% in stronger explosions.

To improve the consistency of our ejecta models, we have explored the impact of the \nifs-bubble effect and the influence of clumping on SN radiation. For the present set of type II SN ejecta models, we find that these processes play a modest role, probably because the ejecta regions concerned by these processes have a modest ionization, or because the effect is to weak. A significant improvement for future work will be to use 3D explosion models as initial conditions for our radiative-transfer calculations, such as those of \citet{gabler_3dsn_21}.

A major drawback of the current work is the assumption of spherical symmetry for the progenitor evolution, explosion phase, and the radiative transfer. For the former, little can be done. However, 3D explosion models exist and can provide critical insights for radiative transfer. Furthermore, it is possible with \cmfgen\ to address multi-D effects by performing 1D simulations along multiple radial directions of a 3D explosion model. We can also post-process such simulations with 2D polarized radiation transfer to study the influence of asymmetry on line profile morphology and polarization (see, for example, \citealt{D20_12aw_pol}).

\begin{acknowledgements}

This work was supported by the "Programme National de Physique Stellaire" of CNRS/INSU co-funded by CEA and CNES. Hillier thanks NASA for partial support through the astrophysical theory grant 80NSSC20K0524. This work was granted access to the HPC resources of  CINES under the allocation  2019 -- A0070410554 and 2020 -- A0090410554 made by GENCI, France. This research has made use of NASA's Astrophysics Data System Bibliographic Services. HTJ acknowledges funding by the Deutsche Forschungsgemeinschaft (DFG, German Research Foundation) through Sonderforschungsbereich (Collaborative Research Centre) SFB-1258 \say{Neutrinos and Dark Matter in Astro- and Particle Physics (NDM)} and under Germany's Excellence Strategy through Cluster of Excellence ORIGINS (EXC-2094)---390783311.

\end{acknowledgements}


\appendix

\section{Results for S16 models with a scaled kinetic energy}

\begin{figure*}
\centering
\includegraphics[width=0.8\hsize]{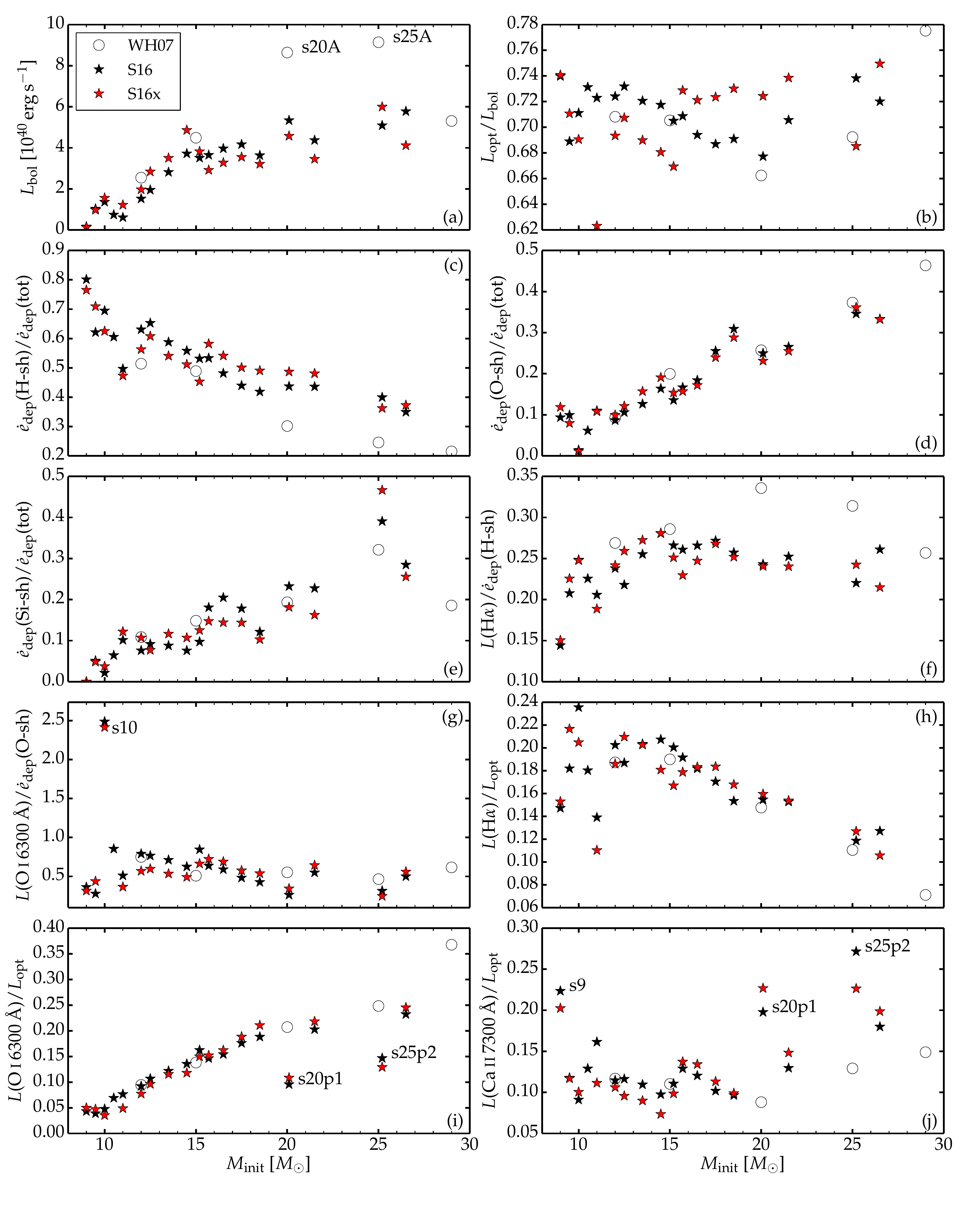}
\vspace{-0.3cm}
\caption{Same as Fig.~\ref{fig_line_fluxes}, but now for a larger set of models including the 23 models discussed in the main text, as well as the S16 models that were scaled in kinetic energy by a factor of 2 or 0.5 (filled red stars, models labelled S16x).
\label{fig_line_fluxes_ekin}
}
\end{figure*}

Figure~\ref{fig_line_fluxes_ekin} presents results for the measurements of various powers and line luminosities for a larger set of models. Besides the selected 23 models from WH07 and S16, we include 18 model counterparts from the S16 set in which the ejecta kinetic energy was scaled up or down by a factor of two. In practice, models with a lower (higher) kinetic energy relative to the models of similar initial mass were scaled in velocity by a factor $\sqrt{2}$ ($1/\sqrt{2}$). The radius was also adjusted to retain the same SN age, and the density was scaled so that $R^3 \rho$ remains unchanged. Such scaled versions are artificial. For example, we retain the same chemical composition from explosive nucleosynthesis although the explosion energy was modified. Nonetheless, they offer a means to gauge the sensitivity of spectral signatures to controlled variations in ejecta properties.

The factor of two variation in kinetic energy that was artificially introduced leads to small variations in the results, as was anticipated from the earlier discussion in Section~\ref{sect_qual} (see, for example, the spectral differences between model s12 and s12A in Fig.~\ref{fig_comp_same_mass}).

\section{Comparison of models s20A and s20p1}
\label{sect_app_prog}

\begin{figure*}
\centering
\includegraphics[width=0.495\hsize]{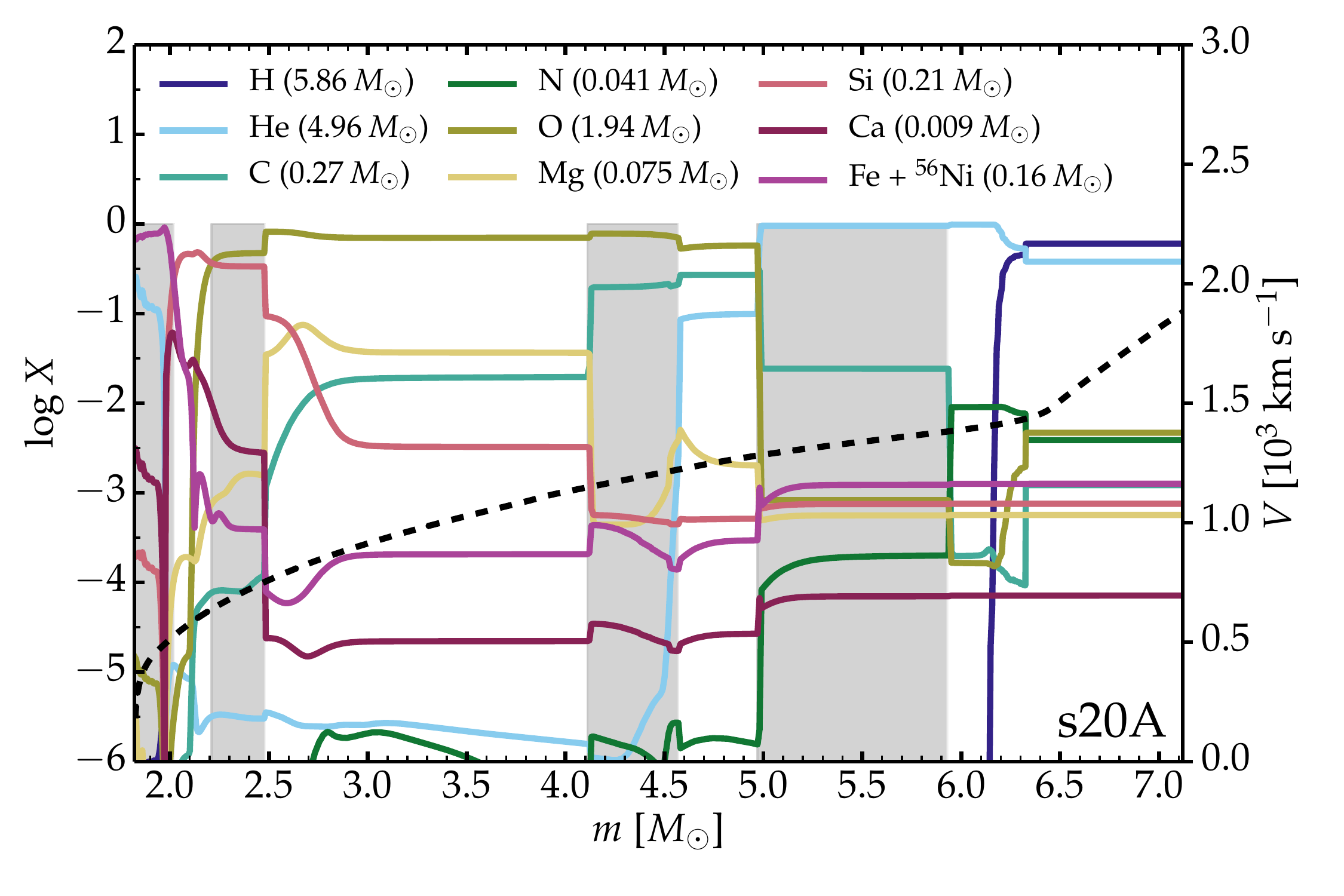}
\includegraphics[width=0.495\hsize]{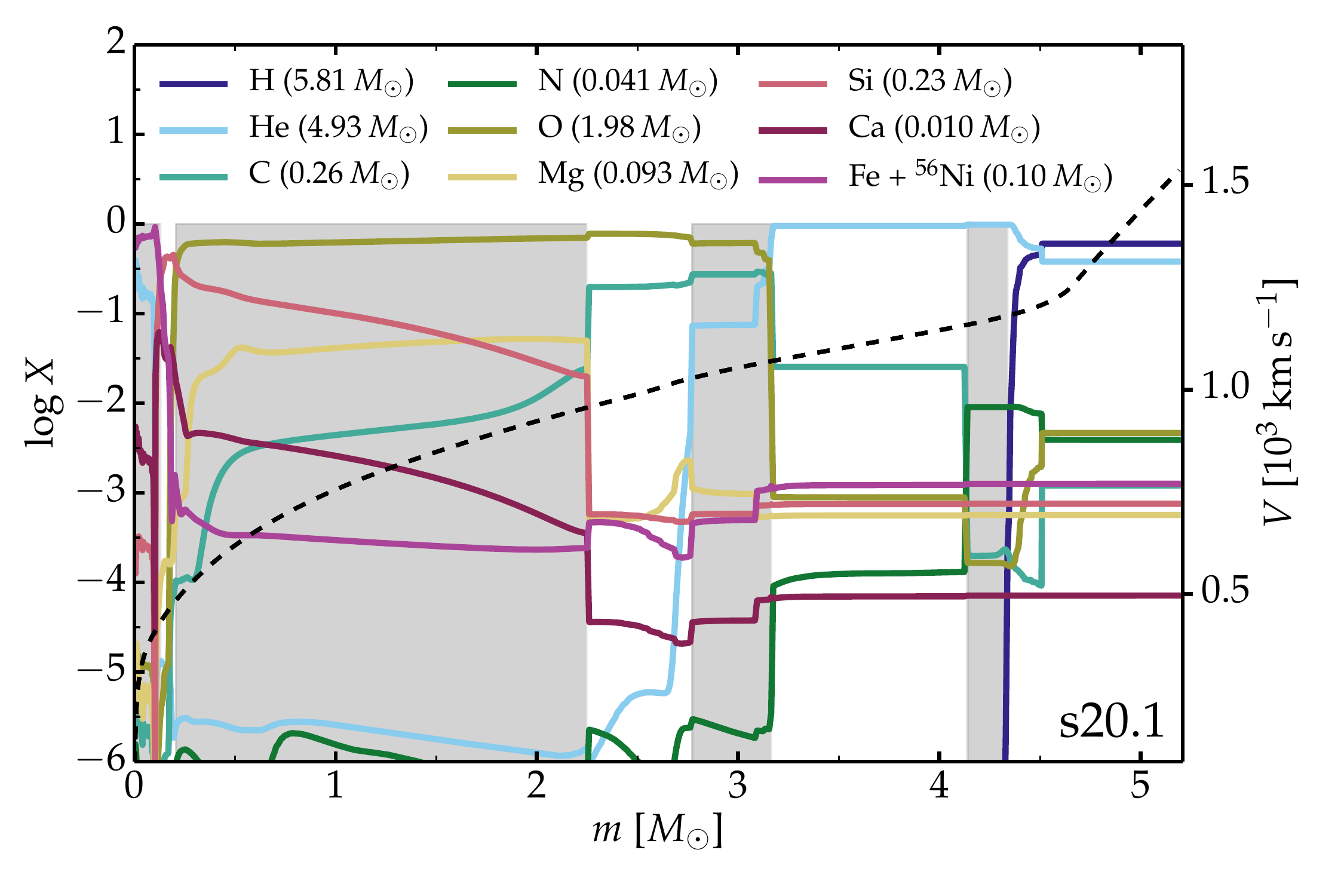}
\vspace{-0.2cm}
\caption{Chemical composition versus Lagrangian mass in the s20A  model (left) of WH07 and the s20p1 model (right) of S16 at  350\,d (the original mass fraction of \nifs\ is shown together with that of Fe). The origin of the x-axis is at the neutron-star center in the left panel, or at the neutron star surface (at a Lagrangian mass of 1.82\,\msun) in the right panel. Si-O shell merging is present in the s20p1 model, but absent in the s20A model.
\label{fig_s20A_s20p1}
}
\end{figure*}

Figure~\ref{fig_s20A_s20p1} compares the chemical composition of the inner ejecta in the models s20A from WH07 and s20p1 from S16. The latter underwent a Si-O shell merging in the last stages of evolution prior to core collapse, while the former was not affected by this phenomenon. The corresponding spectral properties are discussed in the main text, and shown in the middle panel of Fig.~\ref{fig_comp_same_mass}. With Si-O shell merging, the \caiidoub\ doublet becomes a strong coolant for the O-rich material, which quenches the \oidoub\ line flux.

\end{document}